\documentclass{ar-1col-S2O}
\usepackage{lmodern}
\usepackage{rotating}
\usepackage[comma]{natbib}
\usepackage{url}
\usepackage[unicode]{hyperref} 
\usepackage{amsmath}
\usepackage{amssymb}
\usepackage{multirow}
\usepackage{longtable}
\usepackage{booktabs}    
\usepackage{pdfpages}
\usepackage{bibunits}
\defaultbibliography{references}
\defaultbibliographystyle{ar-style2}
\usepackage[unicode]{hyperref}
\hypersetup{
    hidelinks
}
\renewcommand{\figureautorefname}{Figure}

\newcommand{\mainfig}[1]{\hyperref[#1]{Figure~\ref{#1}}}
\newcommand{\SMsec}[1]{\protect\hyperref[#1]{Supplemental Text, Section~\ref{#1}}}
\newcommand{\SMsup}[1]{\protect\hyperref[#1]{Supplemental Section~\ref{#1}}}
\newcommand{\SMtab}[1]{\protect\hyperref[#1]{Supplemental Table~\ref{#1}}}

\def\aap{A\&A}
\def\apj{ApJ}
\def\aapr{A\&A Rev.}
\def\apjl{ApJ}
\def\mnras{MNRAS}
\def\araa{ARA\&A}
\def\aj{AJ}
\def\qjras{QJRAS}
\def\physrep{Phys. Rep.}
\def\prd{Phys. Rev. D}
\def\nat{Nat}
\def\aaps{A\&A Supp.}
\def\apss{Ap\&SS}
\def\apjs{ApJS}
\def\pasj{PASJ}
\def\pasp{PASP}

\jname{Annu. Rev. Astron. Astrophys.}
\jvol{64}
\jyear{2026}
\doi{10.1146/annurev-astro-043024-121502}
  
\begin{document}

\markboth{Whitaker \& Bezanson}{Galaxy Quenching}

\title{Quenching of Star Formation in Massive Galaxies} 

\author{Katherine E. Whitaker$^{1,2}$ and Rachel Bezanson$^3$ 
\affil{$^1$Department of Astronomy, University of Massachusetts, Amherst, Massachusetts, USA; email: kwhitaker@astro.umass.edu}
\affil{$^2$Cosmic Dawn Center (DAWN), University of Copenhagen, Denmark}
\affil{$^3$Department of Physics and Astronomy and PITTPACC, University of Pittsburgh, Pittsburgh, Pennsylvania, USA}}

\begin{abstract}
The shutdown of star formation -- quenching -- marks a pivotal transition in the lives of massive galaxies, which dominate the present-day stellar mass density. This review synthesizes our current understanding of the mechanisms that trigger and maintain quiescence. We discuss the nuances of how quiescent systems are identified across cosmic time and summarize the evolving physical properties of the growing massive population, including their stellar populations, chemical enrichment histories, and gas and dust reservoirs, highlighting several key results:
\parbox{0.75\textwidth}{
\begin{itemize}
\item Quiescent galaxies can be identified with empirical color selections, but evolving specific star formation rate thresholds offer a more robust physical distinction from star-forming systems.
\item The earliest massive quiescent stellar populations show rapid formation histories and high metallicities, with enhanced $\alpha$-elemental abundances often distinct from local analogs.
\item Nascent studies of gas and dust in quiescent galaxies reveal diverse multiphase reservoirs and outflows, pointing to fast ejective and slow regulatory modes of galaxy quenching.
\item In situ processes establish galaxy central density, while assembly continues via (minor) mergers post-quenching, reshaping all massive galaxies and disrupting rotation in most cases.
\end{itemize}}
We distill observations into two broad modes by which massive galaxies form and quench: one involves a rapid, early shutdown driven by supermassive black hole outflows on short timescales; the other proceeds gradually through gas exhaustion, virial heating, or preventative feedback, each leaving distinct observational signatures. Together, these pathways offer a testable framework for modeling the formation and evolution of massive galaxies, which will be informed by future studies of their stars, gas, dust, and dynamics.
\end{abstract}

\maketitle

\tableofcontents

\begin{bibunit}
\section{INTRODUCTION}
  
The discovery of elliptical galaxies dates back to the early 20th century, a transformative era in astronomy when astronomers debated and ultimately confirmed that galaxies exist beyond the Milky Way. \citet{Hubble1926} introduced a landmark morphological classification system grouping galaxies into ellipticals, spirals, and irregulars. Elliptical galaxies, in particular, were defined by their smooth, featureless light profiles and a lack of spiral structure or prominent dust lanes \citep{Sandage1961}. Mid- to late-20th century spectroscopy further revealed that these systems were dominated by old, red stellar populations with little or no evidence of recent star formation \citep{Faber1972, Burstein1984}. This, combined with their generally large masses and overabundance in dense galaxy clusters \citep{Oemler1974,Dressler1980}, 
positioned them as key pieces in understanding the diversity and evolution of galaxies in the Universe. The key to understanding their origin necessitates looking toward earlier epochs, observing massive galaxies while they are actively growing and ultimately shutting down their star formation, before signatures of their formation are obscured by subsequent merging. Owing to the expanding Universe, observing starlight in the rest-frame optical from their epoch of formation mandates near-infrared (NIR) facilities (to sample the rest-optical light at $z\gtrsim1$). Advancements to detect these relatively faint sources and resolution to distinguish them from the plentiful NIR bright sources have only reached sufficient sensitivity within the past few decades, with the first distant quiescent galaxies having been noted by \citet{Dunlop1996} and \citet{Spinrod1997}. Thus, the story of galaxy quenching leads us to the early Universe ($\gtrsim$10 Gyr in the past), where much of the action occurred -- it is a cosmic epoch that has come into view with remarkable clarity over the last two decades. 
 
Although galaxy quenching is observationally implicit by virtue of the existence of `red and dead' elliptical galaxies, it is a postdictive addition to cosmological simulations in order to resolve two longstanding discrepancies between the $\Lambda$ Cold Dark Matter ($\Lambda$CDM) framework and observations. First, theoretical dark matter halos are overabundant at the lowest and highest masses relative to the observed stellar mass function (SMF) \citep{Behroozi2013, Moster2013}.
Feedback from active galactic nuclei (AGN, i.e., accreting supermassive black holes, SMBHs), stellar feedback, and gravitational effects have been invoked to stall star formation and reconcile observations with theoretical models \citep[e.g.,][]{Benson2003, Croton2006}. Second, the typical age of a stellar population scales with the stellar mass of the host galaxy, with the most massive galaxies hosting the oldest stars \citep{Thomas2005}.  At face value, this latter observation is in tension with the hierarchical framework established by the $\Lambda$CDM model, as the most massive galaxies should have assembled at moderately late times.  However, the distinction to be made here is the difference between in situ star formation and ex situ assembly \citep[e.g.,][]{Oser2010,Naab2009,vanDokkum2015}.  Whereas a massive galaxy is thought to form and quench rapidly in the early Universe, it continues to grow via minor mergers and accretion \citep[e.g.,][]{Newman2012} and, thus, fully assemble at late times. Hence, the leading story of galaxy quenching retells a galaxy's reconstructed formation history (nature) coupled with the impact of the surrounding environment (nurture).  

\begin{marginnote}
\entry{$\Lambda$ cold dark matter ($\Lambda$CDM)}{cosmological model in which dark energy and cold dark matter drive expansion and structure formation}
\entry{Active galactic nuclei (AGNs)}{luminous emission from gas accreting onto a supermassive black hole.}
\entry{Supermassive black holes (SMBHs)}{central black holes having mass of 10$^{6}$–10$^{10}$ M$_{\odot}$}
\end{marginnote}

Establishing the physical mechanisms that shut down star formation in massive galaxies has been a central challenge for decades. Although many theoretical models have been proposed, no single mechanism has achieved consensus or is likely universally applicable. This review outlines proposed quenching mechanisms (\autoref{sec:quenching_mechanisms}), 
then focuses on what can be inferred observationally from the properties of massive galaxies that have already quenched and uses this evidence to synthesize a broader narrative of galaxy evolution. Although we draw on theoretical insights when relevant, we do not provide an exhaustive treatment here, instead referring readers to dedicated reviews. From an observational perspective, we review the selection techniques used to identify quiescent populations (\autoref{sec:selection}), highlight their physical properties (\autoref{sec:anatomy}), and trace how their demographics evolve with redshift (\autoref{sec:demographics}); then we integrate this evidence into a unifying picture of quenching and its likely physical drivers (\autoref{sec:framework}).

\section{OVERVIEW OF QUENCHING MECHANISMS}
\label{sec:quenching_mechanisms}

Stars form from molecular gas that cools and settles through energy loss from the thermal radiation of warm and hot gas residing within the surrounding halo.  Both theoretical \citep[e.g.,][]{Keres2005,DekelBirnboim2006} 
and observational studies \citep[e.g.,][]{Rubin2012} 
suggest that this bountiful reservoir is continually replenished through accretion along cosmological filaments. When the conditions for star formation are no longer met---whether owing to suppression of gas accretion, inefficient cooling, or the disruption of star formation itself---the result is quenching. It remains unclear whether any galaxy is completely devoid of star formation; the local elliptical prototype NGC 4472 may be the most extreme known case, with multiwavelength studies constraining its current star-formation rate (SFR) to $\lesssim 0.01\,M_\odot\,\mathrm{year}^{-1}$ \citep[e.g.,][]{Temi2009}.
Given its stellar mass of $\sim 3\times10^{11}\,M_\odot$ \citep{Cappellari2013}, it has a corresponding specific star-formation rate (sSFR) of $\lesssim 3\times10^{-14}\,\mathrm{year}^{-1}$. In practice, quenched galaxies are defined in relative terms, forming substantially fewer stars than counterparts of similar mass at the same epoch (i.e., those lying well below the star-formation main sequence or SFMS). Quenched galaxies also form fewer stars relative to their own past histories, as captured by the sSFR (the ratio of the current SFR to the current stellar mass). Although multiple quenching mechanisms may act concurrently, their ultimate impact on a galaxy depends critically on their characteristic timescales. 

\begin{marginnote}
\entry{Star-formation rate (SFR)}{rate at which mass is converted into stars, typically in $M_\odot\,\mathrm{year}^{-1}$.}
\entry{Specific star-formation rate (sSFR)}{star-formation rate per unit stellar mass (SFR/$M_\star$); inverse of the stellar mass–doubling timescale.}
\entry{Star-formation main sequence (SFMS)}{tight, redshift- dependent relation between stellar mass and star-formation rate for star-forming galaxies.}
\end{marginnote}

Ultimately, the fate of the cold gas, the fuel for new star formation, governs how galaxy quenching unfolds. As we discuss in \autoref{sec:framework}, galaxies quench either rapidly through mechanisms that primarily expel the gas on short ($\lesssim$Gyr) timescales or much more slowly, through processes that either heat or restrict gas replenishment over several gigayears. Additional mechanisms act primarily to maintain quiescence once star formation has ceased, whereas mergers and accretion can contribute to late-time mass growth. As noted by \citet{ManBelli2018}, quenching thus encompasses the physical processes that both halt in situ star formation and preserve quiescence. Hereafter, we distinguish these modes explicitly, noting that in some cases a single mechanism can accomplish both.  

\begin{figure}[h]
\includegraphics[width=0.92\linewidth]{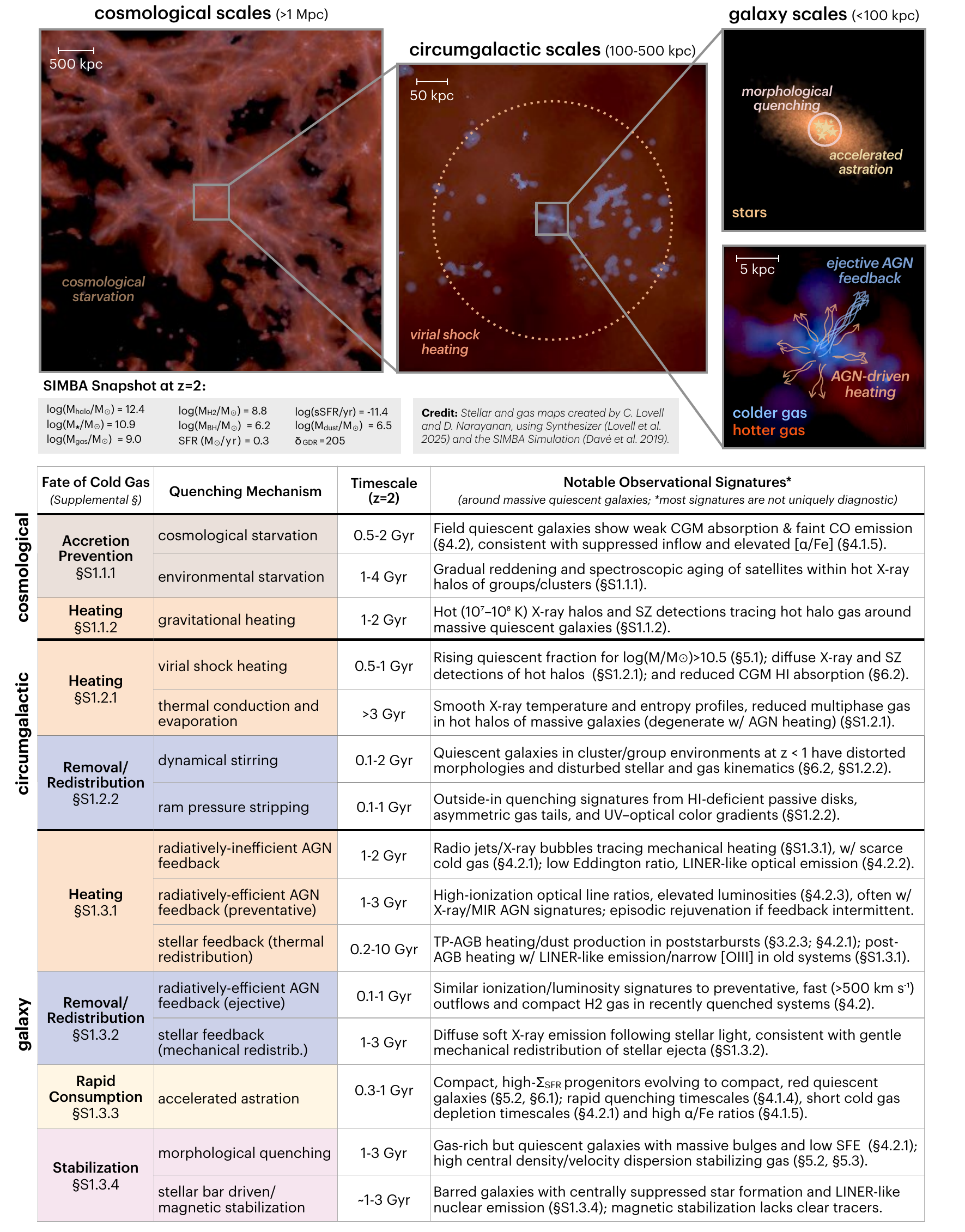}
\caption{Schematic overview of quenching mechanisms, across cosmological (5 Mpc), circumgalactic (500 kpc), and galactic (50 kpc) scales. Shown is a $z=2$ massive quiescent galaxy from a hydrodynamic simulation \citep{Dave2019} with hotter gas (\emph{red}), colder gas (\emph{blue}), and stellar maps (\emph{yellow}; visible only at $<$100 kpc) rendered using \texttt{synthesizer} \citep{Lovell2025}. The table lists each mechanism, its quenching timescale at $z\sim2$, and key observational signatures; full descriptions appear in \SMsec{sec:supplement_quenching_mechanisms}. Abbreviations: AGB, asymptotic giant branch; AGNs, active galactic nuclei; CGM, circumgalactic medium; ICM, intracluster medium,
LINER, low-ionization nuclear emission-line region; SFE, star-formation efficiency; SFR, star-formation rate; SZ, Sunyaev–Zel’dovich
effect; TP, thermally pulsing; UV, ultraviolet.
}
\label{fig:quenching_mechanisms}
\end{figure}

In this section, we separate quenching mechanisms that act on different spatial scales, ranging from subkiloparsec to many hundreds of megaparsecs, largely inspired by \citet{ManBelli2018}. We define cosmological scales as distances $>$1 Mpc (inclusive of galaxy clusters), circumgalactic medium (CGM) scales as $\sim$100--500 kpc, and galaxy scales as $<$100 kpc. \textbf{\autoref{fig:quenching_mechanisms}} summarizes quenching mechanisms by the physical scale on which they operate, using stellar mass (50 kpc box) and hot/cold-gas maps (all three scales) from the SIMBA simulation \citep{Dave2019}, rendered with \texttt{Synthesizer} \citep{Lovell2025} for a massive quiescent galaxy at $z=2$. This figure aims to provide a holistic overview of quenching mechanisms, highlighting key observational signatures and characteristic timescales at $z\sim2$. Current empirical constraints remain strongest at small scales, likely reflecting observational bias. Full descriptions of each mechanism are given in \SMsec{sec:supplement_quenching_mechanisms}.

\begin{marginnote}[]
\entry{Circumgalactic medium (CGM)}{diffuse gas around galaxies that regulates accretion, outflows, and baryon cycling.}
\entry{Star-formation history (SFH)}{the time evolution of a galaxy’s star-formation rate.}
\end{marginnote}

\section{SELECTING QUIESCENT GALAXIES}
\label{sec:selection}

Characterizing the physical properties of quiescent galaxies first requires identifying robust samples. Below, we summarize the main selection methods used in extragalactic surveys, following their historical development alongside our evolving understanding of galaxy quenching. These criteria are inherently phenomenological, identifying galaxies with suppressed recent star formation or dominated by evolved stellar populations as detailed star formation histories (SFHs) are hard to constrain observationally.

Over the past two decades, deep NIR imaging has dramatically improved our ability to detect evolved stellar populations at high redshift.  Extragalactic surveys increasingly rely on longer wavelength selection bands to approximate mass completeness. For example, $K$-band selection (rest-frame 1 $\mu$m at $z=1$) traces light from long-lived, low-mass stars, reducing outshining biases. At $z\gtrsim3$, rest-frame optical selections relied on shallow Spitzer Space Telescope data until the advent of the James Webb Space Telescope (JWST). Even today, additional blind spots can remain, impacting scientific analyses; stellar mass and surface brightness limits must be carefully assessed in any dataset \citep[e.g.,][]{Marchesini2009}. 

No single definition of a quiescent galaxy is all-encompassing. Quiescence implies a suppression, not cessation, of star formation---relative to both galaxies of similar mass and epoch and the galaxy’s own past growth. Because the SFMS evolves rapidly with redshift \citep[e.g.,][]{Leslie2020}, any operational definition must be time-dependent. 

A further complication is the implicit binary division of galaxies between star-forming and quiescent. Quenching is not instantaneous: Some mechanisms act on rapid timescales ($\lesssim$100 Myr), while others unfold over many gigayears, inevitably producing transitional objects. 
This ambiguity is amplified at high redshift, where evolutionary timescales are compressed. 

Finally, all empirical diagnostics of star formation rely, directly or indirectly, on short-lived O and B stars and, thus, become unreliable at low levels. Beyond measurement uncertainties, other processes can mimic star formation: Hot evolved stars contribute UV flux, complicating the attribution of blue colors to young stars (\autoref{subsubsec:UVJ}); old stars heat dust and affect far-infrared (FIR) emission (\autoref{subsec:sSFRs}); and low-level AGNs or shocks can bias traditional spectral line tracers such as H$\alpha$ (\autoref{subsec:stellarages}). For these reasons, treating any individual diagnostic as a ground truth and using it to evaluate others has limited utility.

Given these caveats, the selection of quiescent galaxies must be understood as a best-effort attempt rather than a definitive classification. This likely explains why no community-wide standard has yet emerged. In this section, we review common empirical and model-based approaches for selecting quiescent galaxies in photometric catalogs, highlight their limitations, 
and offer best-practice recommendations for future studies (see the sidebar titled Best Practice Recommendations for Quiescent Galaxy Selection).

\subsection{Classification Schemes Based on Quiescent Spectral Shapes} 

The overall shape of a galaxy’s spectral energy distribution (SED) encodes the cumulative effects of stellar age, dust attenuation, and recent SFH. Rather than relying directly on derived physical quantities such as sSFR, this section summarizes approaches that classify quiescent galaxies using observable spectral features or statistical patterns in broadband colors and spectra, ranging from simple color–color selections to more data-driven techniques that extract dominant spectral components.

\begin{marginnote}[]
\entry{Spectral energy distribution (SED)}{a galaxy’s flux versus wavelength, used to infer stellar mass, SFR, dust, and age.}
\end{marginnote}

\begin{textbox}[h]\section{BEST PRACTICE RECOMMENDATIONS FOR QUIESCENT GALAXY SELECTION}
\subsection{Rest-Frame Color Selection} 
Rest-frame color techniques have historically served as the most efficient means of identifying quiescent populations. A community-wide adoption of a standard, redshift-independent UVJ criterion is recommended for studies to $z\sim3$ (or the redshift at which rest-frame J-band is no longer empirically constrained) \citep{Whitaker2012}, defined as 
$(U - V) > 1.3$, $(V - J) < 1.6$, and $(U - V) > 0.8 \times (V - J) + 0.7$. 
The classic UVJ selection identifies quiescent galaxies older than $\sim$300 Myr \citep{Belli2019}; relaxing the lower U–V bound included younger systems \citep{Park2023}, albeit at the risk of adding bursty star-forming galaxies \citep{Mintz2026}. Conversely, the upper V–J bound prioritizes purity over completeness \citep{Lustig2023}, although the nature of this ``maximally red'' wedge remains poorly understood. At $z>3$, alternative color selections are required \citep[e.g.,][]{Antwi-Danso2023,Long2024}, UVJ criteria must be supplemented (or replaced) with sSFR to remove contaminants \citep{Baker2025a} or data-driven methods trained on spectroscopic samples \citep{Jafariyazani2024}. While UVJ retains value for quick-look analyses and assessing selection biases, a transition toward SED-based approaches is preferable for self-consistency.

\subsection{Specific Star Formation Rate Selection} 
Advances in the sophistication and efficiency of SED modeling support using either a fixed offset from the redshift-dependent SFMS or the physically motivated inverse-Hubble timescale, $0.2/t_{\mathrm{H}}$.
A sSFR value $>2\sigma$ ($\gtrsim0.6$ dex) below the SFMS is a logical baseline, although the exact offset remains subjective. A tiered scheme is useful: galaxies 3$\sigma$ ($\sim$0.9 dex) below the SFMS (equivalent to $0.2/t_{\mathrm{H}}$ at $z\gtrsim1$) define a conservative quiescent sample, while those 2–3$\sigma$ below may be treated as transitional. Systems near any boundary warrant caution, as they may reflect prior-sensitive SFHs such as slow quenching \citep{Carnall2018}, rejuvenation \citep{Akhshik2021}, or bursty SFHs \citep{Looser2023}.

\subsection{Understanding Targeting Selection and Completeness} 
Robust identification of quiescent samples benefits from detection in deep NIR imaging (ideally rest-frame $\gtrsim$1$\mu$m) to minimize outshining effects and trace stellar mass. However, all samples depend heavily on data quality and characterization of the full multi-wavelength SED; often wavelength sampling is more important than broadband depth. Injection–recovery tests are essential for establishing reliable mass and surface brightness completeness in any given dataset \citep[e.g.,][]{Marchesini2009,Pozzetti2010}.
\end{textbox}

\subsubsection{Color selections}  
\label{subsubsec:UVJ}
Photometric color selection has long been a popular way to separate quiescent from star-forming galaxies, enabled by the advent of wide-area deep NIR surveys \citep{Franx2000,Gawiser2006}.
Current approaches build on early \textit{observed-frame} techniques used to identify red sources \citep[e.g.,][]{Franx2003,Daddi2004} or the low-redshift red sequence \citep{Strateva2001,Bell2004,Faber2007}, with a modern comparison presented in \citet{Shahidi2020}. 
These methods rely on rest-UV–optical colors that trace the Balmer/4000$\mathrm{\AA}$ break but are also highly sensitive to dust reddening, leading to significant contamination from dusty star-forming galaxies. This limitation can be mitigated by applying dust corrections to optical colors \citep[e.g.,][]{Schawinski2014} or, with improved photometric redshifts, by directly computing rest-frame colors. 

\begin{figure}[h]
\includegraphics[width=\linewidth]{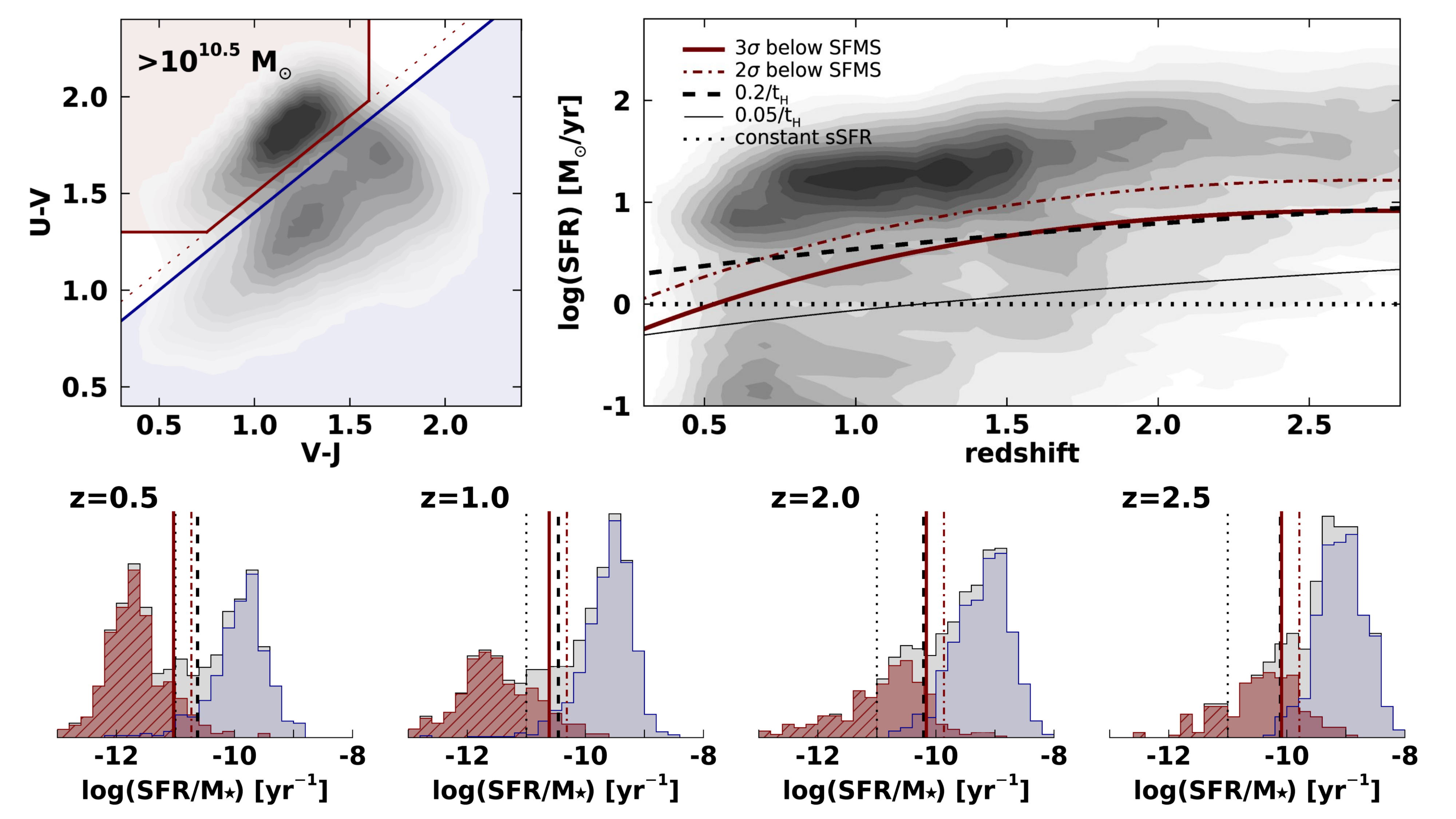}
\caption{There is no single correct way to select quiescent galaxies. Two common approaches are rest-frame color–color diagrams (top left) and redshift-dependent log(sSFR) thresholds (top right). We use 3D-HST photometric catalogs \citep{Skelton2014} with SPS modeling from \citet{Leja2022}. The bottom panels show log(sSFR) distributions for UVJ-selected quiescent (red) and star-forming (blue) galaxies, with boundary cases in grayscale. A fixed log(sSFR/yr) = –11 cut is too strict at $z>1$, whereas a 0.2/t$_\mathrm{H}$ criterion follows a boundary $\sim$3$\sigma$ below the SFMS ($\sigma$$\approx$0.3 dex).  Galaxies 2–3$\sigma$ below the SFMS are often in transition from star-forming to quiescent.
Hashed red regions mark parameter space where sSFR constraints are systematics-limited. Abbreviations:
SFR, star-formation rate; SFMS, star-formation main sequence; sSFR, specific star-formation rate. 
}
\label{fig:ssfr}
\end{figure}

Rest-frame color selections, typically defined using pairs of UV–to-NIR colors, are motivated by theoretical expectations and remain simpler, more computationally efficient, and less model-dependent than full SED fitting (though see \autoref{subsubsec:color_caveats} for caveats). These methods are broadly applicable and reproducible across datasets and naturally exploit the strongest spectral features that separate quiescent from dusty star-forming galaxies. Traditionally, one color straddles the Balmer/4000\,\AA\ break, long used in low-redshift color–magnitude studies to define the red sequence \citep[e.g.,][]{Baldry2004}, while a second rest-optical-to-NIR color distinguishes truly passive populations from dust-reddened star-forming galaxies, whose SEDs rise steeply at longer wavelengths rather than turning over. Commonly used rest-frame color selections include the UVJ 
\citep[e.g.,][]{Labbe2005,Williams2009,Whitaker2011}, UVI \citep{Wang2017}, and NUVrJ \citep{Arnouts2007,Davidzon2017,Moutard2018} diagrams. Although we refer hereafter primarily to UVJ selection (top left, \textbf{\autoref{fig:ssfr}}), these alternatives perform similarly well. 

\begin{marginnote}[]
\entry{UVJ}{Rest-frame $U\!-\!V$ vs.\ $V\!-\!J$ color diagram used to separate quiescent and star-forming galaxies.}
\end{marginnote}

To date, no clear community standard exists for rest-frame color selection, in part because the boundaries are typically tuned empirically rather than derived from stellar population theory. Early definitions evolved with redshift, largely reflecting the data quality available at the time: the first wide-area NIR surveys lacked both the depth and photometric redshift accuracy to reveal a clear bimodality in rest-frame colors at $z>2$ \citep{Williams2009}.  Subsequent improvements in photometric redshifts, especially with the addition of NIR medium-band filters, established a distinct quiescent population at $z\geq2$ \citep{Whitaker2011,Straatman2014} and revealed little physical motivation for redshift-dependent UVJ boundaries (even though such cuts remain widely used). A recommended redshift-independent UVJ selection is defined in the sidebar titled Best Practice Recommendations for Quiescent Galaxy Selection. That said, the increased scatter near the boundary at high redshift reflects genuine diversity in SFHs, which is an unavoidable outcome given that stellar ages must be less than a few gigayears. Further discussion of UVJ trends is provided in \SMsec{sec:supplement_selection}.  

\subsubsection{Data-driven classification methods} 
\label{subsubsec:PCA}
A powerful statistical tool used to reduce the dimensionality of galaxy spectra while preserving the most informative features is principal component analysis (PCA). In the context of selecting quiescent galaxies, PCA enables pattern recognition in SEDs that distinguish passive stellar populations from those undergoing active star formation \citep{Wild2007,Wild2016,Chen2012}.
In this sense, it is a more sophisticated extension of traditional color selection techniques.  By transforming the spectra into a set of orthogonal components ranked by variance, PCA captures subtle variations in features sensitive to stellar age and recent star formation activity, such as the Balmer/4000$\mathrm{\AA}$ break and Balmer absorption lines. Efficient and objective classification is possible when these components can be labeled, as quiescent galaxies occupy a distinct region in PCA-derived spectral space \citep[e.g.,][]{Wild2016,Zhang2026}. This data-driven approach complements traditional color–color (\autoref{subsubsec:UVJ}) or sSFR-based (\autoref{subsec:sSFRs}) selection methods and is especially valuable when working with the high-dimensional and noisy datasets typical of extragalactic surveys. More recently, machine learning–based classifiers have emerged as a scalable extension of these approaches, and when trained on spectroscopic samples, they can achieve higher completeness and purity than traditional color selections \citep[e.g.,][]{Humphrey2023,Jafariyazani2024}.

\begin{marginnote}[]
\entry{Principal component analysis (PCA)}{reduces the dimensionality to identify dominant orthogonal components and decompose spectra or SEDs.}
\end{marginnote}

\subsubsection{Complications and caveats with color selections}  
\label{subsubsec:color_caveats}

Color techniques are widely used as their efficiency generally outweighs known shortcomings---i.e., the $\sim$10--30\% contamination rates revealed by FIR detections, spectroscopy, or detailed modeling \citep{Diaz-Garcia2019,Schreiber2018,Leja2019}. That said, the reliability of these selections depends critically on data quality---the depth, number, and wavelength coverage of filters---which directly affects the accuracy of photometric redshifts and, thus, the inferred rest-frame colors \citep{Whitaker2010,Brammer2011,Bezanson2016}. For example, increasing spectral sampling via medium-band filters reduces rest-frame color uncertainties more effectively than deeper broadband observations \citep{Whitaker2011,Straatman2014,Esdaile2021}.  

The adopted methodology for measuring the rest-frame colors is also important.  When fitting linear combinations of template sets (e.g., EAZY) \citep{Brammer2008}, templates must span the full color range of the galaxy sample \citep[e.g.,][]{Marchesini2010,Muzzin2013}, given that many codes adopt model photometry in lieu of interpolating between observed fluxes directly \citep[e.g.,][]{Taylor2009}. By contrast, rest-frame colors inferred from physical SED models (e.g., \texttt{prospector}, \texttt{BAGPIPES}) often show reduced scatter in the UVJ plane because the fits impose physically motivated priors \citep[e.g.,][]{Carnall2018,Leja2019,Leja2020}. This partially reflects methodology, as individual posterior draws or best-fit models exhibit broader distributions than rest-frame colors marginalized over the full posterior \citep[e.g.,][]{Wang2024}. Such reduced scatter can improve internal consistency but may also reflect model assumptions rather than purely empirical constraints.

Finally, new color-selection techniques have been proposed to improve sample purity and completeness at $z{>}3$ \citep{Antwi-Danso2023,Long2024}, although their benefits at lower redshift, particularly for massive galaxies, remain more limited. Additional discussion is provided in \SMsec{subsection:uvj_modifications}. Looking ahead, expanded rest-frame far-ultraviolet and mid-infrared (MIR) coverage via new facilities may further optimize color selections \citep{Leja2019}. However, the community is increasingly shifting toward full SED modeling.

\subsection{Classification Schemes Based on Specific Star Formation Rate Indicators}
\label{subsec:sSFRs}

Classification based on the SFR of a galaxy is the most physically intuitive way to separate active star-forming and passive systems. In practice, however, the consistent use of SFR indicators for the full population is difficult as tracers probe different timescales and suffer distinct systematics \citep{Kennicutt2012}. UV and H$\alpha$ trace recent star formation but are highly sensitive to dust, whereas IR-based indicators mitigate extinction but require deep, broad wavelength coverage rarely available at high redshift. More fundamentally, all SFR---and thus sSFR---estimates are model dependent: The choice of priors and parameterizations can strongly influence inferred values and ultimately the classification of quiescent galaxies \citep{Pacifici2012,Carnall2019,Leja2019,Kaushal2024}. 

\subsubsection{Evolving specific star-formation rate thresholds}
As the average sSFR evolves with redshift (\textbf{\autoref{fig:ssfr}}), any definition of quiescence tied to sSFR must follow accordingly. It is therefore not immediately obvious how to define a quiescent galaxy consistently across cosmic time. \textbf{\autoref{fig:ssfr}} illustrates that two commonly adopted approaches effectively identify quiescent populations: (1) a fixed offset below the SFMS, typically 3$\sigma$ ($\sim$0.9 dex) below the ridgeline, with 2$\sigma$ (0.6 dex) being more inclusive of recently quenched systems, and (2) an evolving sSFR threshold equal to 20\% of the inverse Hubble time \citep{Franx2008,Tacchella2015,Pacifici2016}, where $t_H$ is numerically similar to the mass-doubling timescale in the early Universe.  These two definitions are broadly equivalent at $z>1$, though they diverge at lower redshift. Both evolving sSFR definitions yield samples broadly consistent with UVJ color selections (colored histograms in \textbf{\autoref{fig:ssfr}}). By contrast, a stringent fixed threshold of ${\rm sSFR}<10^{-11}$ yr$^{-1}$ performs well at low redshift but fails to identify most quiescent galaxies at $z\sim2$. Such a strict cut, equivalent to $\sim$10\% of the inverse Hubble time at $z\sim0$, is thus not advisable. Galaxies with rest-frame colors near the UVJ  boundary typically fall between the star-forming and quiescent sSFR distributions, with most lying 2-3$\sigma$ below the SFMS, consistent with their interpretation as  transition systems. 

Adopting a meaningful sSFR-based selection, as summarized in the sidebar titled Best Practice Recommendations for Quiescent Galaxy Selection, requires robust sSFR measurements. This becomes particularly challenging below $\sim10^{-11}$ yr$^{-1}$, where galaxies clearly have suppressed star formation, but the absolute rates are highly uncertain. In this low-sSFR regime, SFR indicators often yield discrepant results \citep{Salim2007,Utomo2014,Feldmann2015,Leja2019}. Furthermore, degeneracies with stellar age, dust attenuation, metallicity, and assumptions about SFHs can bias both the inferred stellar masses and SFRs and, therefore, the derived sSFRs \citep{Pacifici2012,Pacifici2023,Leja2019}.

\subsubsection{Estimating star-formation rates in quiescent galaxies} 
Until recently, SFRs were commonly estimated from a combination of rest-frame UV and FIR luminosities \citep{Kennicutt2012,ForsterSchreiber2020}. Deep Spitzer 24~$\mu$m imaging in extragalactic fields \citep[e.g.,][]{LeFloch2005,Elbaz2011,Wuyts2011} popularized using the MIR flux as a proxy for FIR dust emission, but this approach carries important caveats. Warm dust emitting at MIR wavelengths can be heated by older stars, contaminating SFR estimates \citep[e.g.,][]{Fumagalli2014,Hayward2014,Utomo2014}. Although the effect is negligible in star-forming galaxies, most FIR flux in quiescent systems originates from dust heated by $\gtrsim$100-Myr-old stars (see \SMsec{sec:supplement_anatomy}), leading to SFR overestimates of up to an order of magnitude \citep{Utomo2014}. Recent JWST results at $z\sim0.7$ further demonstrate that the interstellar medium (ISM) of quiescent galaxies can exhibit significant MIR emission from dust and polycyclic aromatic hydrocarbons (PAH) even in the absence of ongoing star formation \citep[e.g.,][]{BlanquezSese2023}. In some cases, for example, at $z=4.7$ \citep{Ji2024_MIRI}, this MIR emission is remarkably extended, reaching spatial scales $\gtrsim$3$\times$ further than the stellar light. These findings show that dust heating by old stars, and dust distribution itself, need not track the stellar profile. Although AGN complicate MIR interpretation \citep{Stern2005}, the growing Mid-Infrared Instrument (MIRI) samples from JWST will soon enable much clearer separation of stellar, dust, and AGN contributions across cosmic time.

\begin{marginnote}
\entry{Interstellar medium (ISM)}{gas and dust within a galaxy, including its molecular, atomic, and ionized phases.}
\entry{Polycyclic aromatic hydrocarbons (PAHs)}{large carbonaceous molecules producing MIR emission features that trace dust and ISM conditions.}
\entry{Stellar population synthesis (SPS)}{models combining stellar evolution tracks and spectra to infer galaxy properties from integrated light.}
\end{marginnote}

Given the shortcomings of traditional IR-based SFR tracers, advances in stellar population synthesis (SPS) modeling \citep[see reviews in][]{Conroy2013,Iyer2026} have greatly improved our ability to characterize the physical properties and SFHs of quiescent galaxies. Modern SPS codes include flexible, nonparametric SFH models within Bayesian frameworks, incorporating variable dust laws, metallicity treatments (often fixed but with evolution available from semianalytic or cosmological libraries), and updated stellar evolution prescriptions \citep[e.g.,][]{Pacifici2012,Leja2017,Carnall2019,Wang2023}. For quiescent galaxies, SED modeling offers key advantages for estimating sSFRs: (a) accounting for contamination from dust-heated old stars; (b) providing self-consistent estimates of stellar mass, age, dust, and SFH; and (c) enabling self-consistent population-level comparisons and selections. Importantly, modern implementations increasingly incorporate physically motivated priors, tightening SFH constraints while clarifying where model assumptions bias inferred properties \citep{Carnall2019a,Leja2019a}.

\subsubsection{Complications and caveats with stellar population synthesis modeling techniques}

Despite major advances in SED modeling and data quality, fundamental uncertainties in SPS modeling continue to limit precise inference of quiescent formation histories. A central challenge is the age--metallicity--dust degeneracy: Dust reddening can mimic older ages or higher metallicities, especially without high signal-to-noise ratio (SNR) spectroscopy of key features \citep[e.g.,][]{Worthey1994,Gallazzi2005}. This longstanding problem now intersects with the need to model nonsolar abundance patterns at younger ($\leq$1 Gyr) ages typical of high-redshift quiescent galaxies. Although $\alpha$/Fe-variable models extend to $\sim$0.1 Gyr \citep[e.g.,][]{Thomas2003,Johansson2012}, they rely on index-based response functions rather than full-spectrum synthesis, and they lack the abundance flexibility required for JWST spectroscopy. Recent work has begun to address this need by constructing full-spectral models with variable $\alpha$-element abundances \citep{Park2025}, though additional stellar abundances may influence the spectra as well \citep[e.g.,][]{Choi2019}. Deriving accurate mass-to-light (M/L) ratios is further complicated by bursty SFHs \citep{Narayanan2024,Wang2025} and by potential environment- or redshift-dependent initial mass function (IMF) variations \citep[e.g.,][]{Conroy2012IMF,Steinhardt2022}. 

Additional uncertainties arise from poorly understood stellar phases, including thermally pulsing asymptotic giant branch (TP-AGB) stars, horizontal branch (HB) morphology, and blue stragglers \citep{Maraston2005}, which can bias age and mass estimates, especially in recently quenched or composite SFH systems \citep{Lu2025}.   
Empirical constraints on these stellar phases remain limited, and many SPS models include only simplified TP-AGB and HB treatments while omitting blue stragglers entirely \citep[e.g.,][]{Bruzual2003,Villaume2015}.  TP-AGB light can substantially impact NIR SEDs of $\sim1$ Gyr populations; early work found substantial contributions \citep[e.g.,][]{Maraston2006,vanDerWel2006}, whereas composite SEDs at $z$$\sim$1-2 suggested a weaker effect \citep[e.g.,][]{Kriek2010}. Spectroscopy from the Near Infrared Spectrograph (NIRSpec) on JWST now reveals clear longer-wavelength TP-AGB features \citep{Lu2025}, and models with enhanced TP-AGB emission \citep[e.g.,][]{Maraston2005,Maraston2013,Noel2013} provide better matches to these data. Wholesale inclusion of these models could have a profound impact: Inferred ages decrease by $\sim$0.1--1 Gyr, and stellar masses decrease by $<$0.3 dex relative to models that underpredict TP-AGB flux. These results underscore the pressing need for updated, empirically calibrated TP-AGB prescriptions \citep{Conroy2009}. 

\begin{marginnote}
\entry{Asymptotic giant branch (AGB)}{late stellar phase with strong winds, dust production, and enrichment of the surrounding ISM.}
\entry{Horizontal branch (HB)}{core-helium-burning phase of low-mass stars that influences a population’s UV and optical colors.}
\end{marginnote}

Finally, the complex growth histories of massive quiescent galaxies, often shaped by mergers and episodic star formation, remain difficult to capture with traditional SFH modeling, especially parametric ones \citep{Cochrane2024}. High-redshift measurements of rest-frame optical and NIR absorption features are still sparse \citep[e.g.,][]{Beverage2024,Beverage2025}, limiting direct application of spectral indices calibrated in the local Universe \citep[e.g.,][]{Gallazzi2005}. Although SPS modeling provides critical insight into quiescent galaxy evolution, substantial uncertainties persist in the derived physical properties. 

\section{ANATOMY OF A QUIESCENT GALAXY}
\label{sec:anatomy}

The goal of this section is to critically evaluate the observational methods used to probe the stellar populations (\autoref{subsec:stellar}) and ISM (\autoref{subsec:ism}) of massive galaxies that have already undergone quenching. We caution that the vast majority of measurements are luminosity weighted and, therefore, primarily sensitive to galaxy centers. Although we focus on global properties, which is where the bulk of progress has been made, resolved analyses provide essential insight into formation pathways. We return to the role of spatially resolved studies in \autoref{sec:summary}.

\subsection{Stellar Populations of Quiescent Galaxies}
\label{subsec:stellar}

Observations of stellar populations provide the direct archaeological record of how massive galaxies formed and subsequently quenched. In this section, we summarize the key observational constraints derived from spectroscopy and broadband SED modeling, progressing from measured spectral diagnostics to inferred physical properties such as stellar mass, ages, SFHs, chemical abundances, and dust attenuation.

\subsubsection{Observed spectral features}
\begin{figure}[h]
\includegraphics[width=0.97\linewidth]{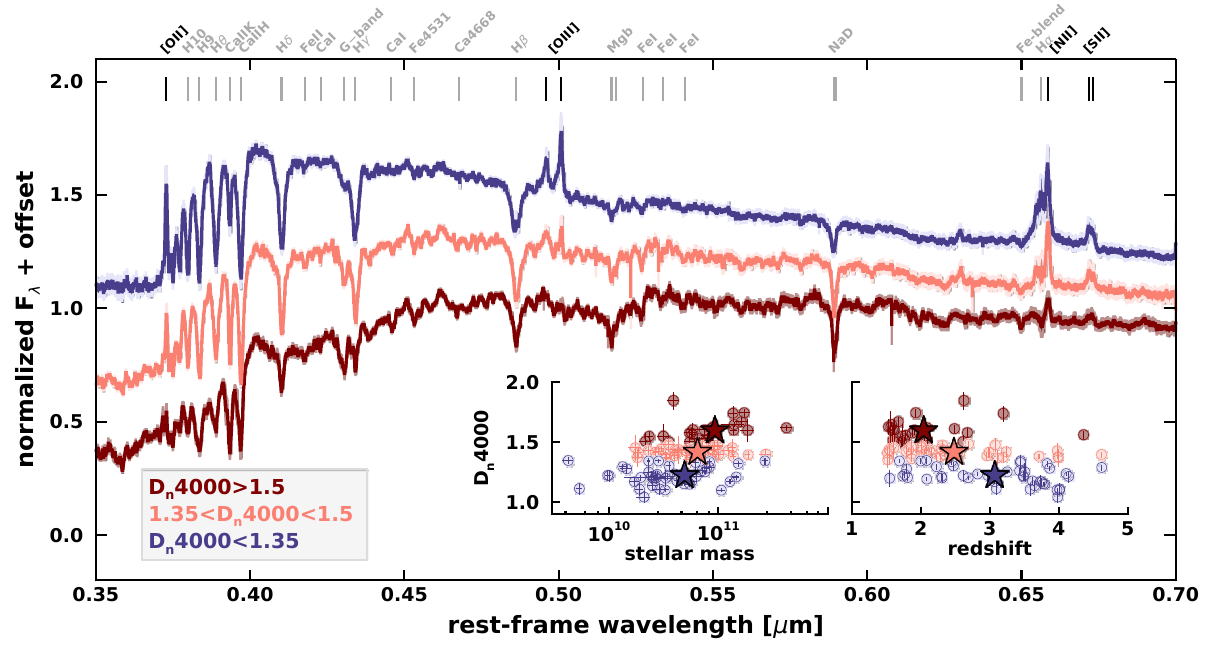}
\vspace{-0.2cm}
\caption{Massive quiescent galaxies show strong spectral variations, evident in stacked spectra binned by $\mathrm{D_{n}4000}$. Systems with strong breaks and high $\mathrm{D_{n}4000}$ (\emph{maroon}) are typically older and more metal rich, while lower values identify recently quenched galaxies (\emph{navy}) with pronounced Balmer absorption. The latter also show increasingly strong emission lines often linked to AGN activity. These stacks, from 140 quiescent galaxies at $1<z<5$ with spectroscopy from the Near Infrared Spectrograph on the \emph{James Webb Space Telescope} are normalized by median flux at $\lambda_{\mathrm{rest}}$ of 0.44–0.48$\mu$m. Absorption line markers are shown in gray, emission in black, with H$\alpha$ appearing in both. Inset panels indicate that higher $\mathrm{D_{n}4000}$ corresponds to higher stellar masses at lower redshift. Although selection biases may limit the representativeness of this sample, the trends naturally reflect the downsizing paradigm. Figure adapted with
permission from \citet{Ito2025}.}
\label{fig:deepdive}
\end{figure}

\textbf{\autoref{fig:deepdive}}, adapted from \citet{Ito2025}, compiles publicly available JWST/NIRSpec medium-resolution grating spectra of galaxies with suppressed star formation at cosmic noon, empirically illustrating the strong variations in spectral features. The observed SED shapes motivate the simpler rest-frame color techniques (\autoref{sec:selection}), but spectroscopy now affords the use of higher spectral resolution features to infer their physical properties more robustly.  

In addition to the Balmer/4000$\mathrm{\AA}$ break (often quantified by the $\mathrm{D_{n}4000}$ spectral index), several other spectral features are commonly used to study quiescent galaxies. Prominent Balmer absorption lines, particularly H$\delta$, are key indicators of recent quenching, and are especially strong in post-starburst systems dominated by A and late B-type stars. Characteristic features of older metal-rich populations include the CaII H\&K lines, the G-band (4304$\mathrm{\AA}$), the Mgb triplet near 5175$\mathrm{\AA}$, and FeI absorption around 5270 and 5335$\mathrm{\AA}$. Crucially, quiescent galaxies generically lack strong H$\alpha$ emission, which probes HII regions around young, massive (and short-lived) stars and therefore traces active star formation. However, both H$\alpha$ and the less robust SFR tracer [OII]$\lambda$3727, can be powered by shocks or AGN, complicating interpretation; such emission is common in quiescent spectra from $z{\sim}1$ \citep[e.g.,][]{Maseda2021} to $z{\sim}3$ \citep[e.g.,][see \textbf{\autoref{fig:deepdive}}]{Ito2025}.  As shown in \textbf{\autoref{fig:deepdive}}, emission line strength increases for younger quenched systems with lower $\mathrm{D_{n}4000}$.  Suppressed UV continuum, particularly at $\lesssim$2800$\mathrm{\AA}$, is an additional hallmark of passive stellar populations.

A central theme in studying quiescent galaxies is identifying the spectral features that trace older stellar populations and the quenching process. \textbf{\SMtab{tab:lines}} in \SMsec{subsubsec:supplement_observed_spectral_features} lists the prominent rest-frame optical absorption and emission features in quiescent systems (adapted from \citealt{Hamadouche2026}), including their primary source (stellar, ISM, AGN), physical origin, and diagnostic use, noting both individual lines and blended bandpasses. These features can be analyzed individually (e.g., through Lick indices) or modeled in aggregate via full spectral or spectrophotometric fitting. Building on this empirical foundation, we next summarize the current understanding of stellar population properties, ordered by increasing complexity.  

\subsubsection{Stellar masses}
Total stellar mass is a generally well-constrained physical property of massive galaxies when rest-frame optical to NIR photometry is available (to $\lesssim0.3$ dex \citealt{Muzzin2009}), owing to the minimal M/L variation in the NIR \citep[e.g.,][]{Bell2001}. A major source of uncertainty is outshining: A small fraction of UV-bright stars can dominate a galaxy’s light, masking older, mass-dominant populations, particularly when only rest UV/optical data are available or at high redshift \citep[e.g.,][]{Conroy2013}. As a result, stellar masses can be biased in systems having rapidly varying SFHs, such as poststarburst galaxies, where the choice of SFH prior directly affects inferred M/L ratios \citep[e.g.,][]{Pacifici2012,Pacifici2016}. We provide a more detailed discussion of SFH priors, mock-recovery tests, and model sensitivities in \SMsec{subsubsec:supplement_stellar_mass}.

JWST spectroscopy now reduces these uncertainties by providing rest-frame optical and NIR constraints for quiescent galaxies at $z>3$ \citep[e.g.,][]{Carnall2023b,Valentino2023,Nanayakkara2024}. These data directly measure the Balmer/4000\AA\ break and the rest-NIR continuum dominated by evolved stars, whereas MIRI photometry extends this leverage to $z\gtrsim5$, breaking outshining-driven M/L degeneracies \citep{Papovich2023}.

In addition, stellar mass estimates depend on assumptions about chemical abundance patterns and the IMF \citep[e.g.,][see also \autoref{subsec:chemical_abundances}]{Forrest2022,Kriek2024}. Current SPS models do not yet self-consistently treat nonsolar abundance ratios, which is particularly important in early quiescent galaxies that are likely to be $\alpha$-enhanced. Initial work suggests continuum flux differences of up to $\sim$80\% relative to solar-scaled models \citep{Park2025}, potentially leading to overestimated stellar masses, though the impact of abundance variations is case-dependent and complex. Improved abundance treatments may help alleviate tensions between inferred SFHs and stellar masses of $z\sim3$--5 quiescent galaxies, which otherwise imply star-formation efficiencies approaching the baryonic limit \citep[e.g.,][]{Glazebrook2024,Carnall2023b,deGraaff2025}.

\subsubsection{Average stellar ages}
\label{subsec:stellarages}

The average stellar ages of quiescent galaxies at $z>1$ provide key constraints on their formation epochs. Although such measurements are challenging to obtain, the data are in some ways easier to interpret at high redshift where galaxies are younger and spectral age indicators evolve most rapidly, a point long appreciated by the community \citep[e.g.,][]{RenziniCimatti1999}. Three common definitions of the `average' age of a stellar population are used: the light-weighted age, which averages stellar ages by their contribution to the integrated light; the mass-weighted age, which averages over the total stellar mass formed; and a simple stellar population (SSP) equivalent age, which assumes a single burst. Light-weighted ages are typically biased younger by several hundred Myr to a few Gyr compared to mass-weighted ages due to the disproportionate luminosity of young stars \citep[e.g.,][]{Gallazzi2005,Choi2014,Leja2019}, especially in systems with extended or bursty SFHs \citep[e.g.,][]{Carnall2019,Park2025}. While distinguishing recently quenched from older systems is generally straightforward \citep[e.g.,][]{Whitaker2013,Belli2019}, the classic problem of outshining \citep[e.g.,][]{Papovich2001,Maraston2010,Suess2022} hampers reconstruction of their detailed SFHs, limiting our ability to recover the earliest star-formation episodes of recently quenched galaxies and the ages of the oldest quiescent systems at high redshift \citep[e.g.,][]{Glazebrook2017,Glazebrook2024}.

Age-sensitive spectral features such as the Balmer/4000$\mathrm{\AA}$ break and Balmer absorption lines provide strong diagnostics (\textbf{\autoref{fig:deepdive}}), but degeneracies with metallicity, dust, and $\alpha$-element abundances complicate interpretation \citep[e.g.,][]{Worthey1994,Thomas2005,Gallazzi2005,Conroy2013}. Classical Lick indices target these variations, with certain combinations of spectral features designed to break degeneracies in local quiescent galaxies, but their utility at $z>1$ is limited by blended features, coupled abundance effects, and reduced intrinsic variation in younger stellar populations.  We expand on these classical index diagnostics and their limitations in \SMsec{subsubsec:lick}.  

Because of these complexities, most recent studies now favor full spectral fitting, which models the continuum and absorption features simultaneously and can capture the coupled effects of age, metallicity, and elemental abundances. Constraining detailed chemical compositions remains particularly difficult for young quiescent galaxies that dominate the population at $z>1$, where A-type stars weaken metal-sensitive features and data quality is often limited. These challenges, along with uncertainties in the SPS models themselves \citep[e.g.,][]{Maraston2005,Conroy2010,Gallazzi2026}, mean that, while stellar ages can be constrained moderately well, detailed chemical abundances remain significantly more uncertain (see also \autoref{subsec:chemical_abundances}).

\begin{figure}[h]
\includegraphics[width=0.96\linewidth]{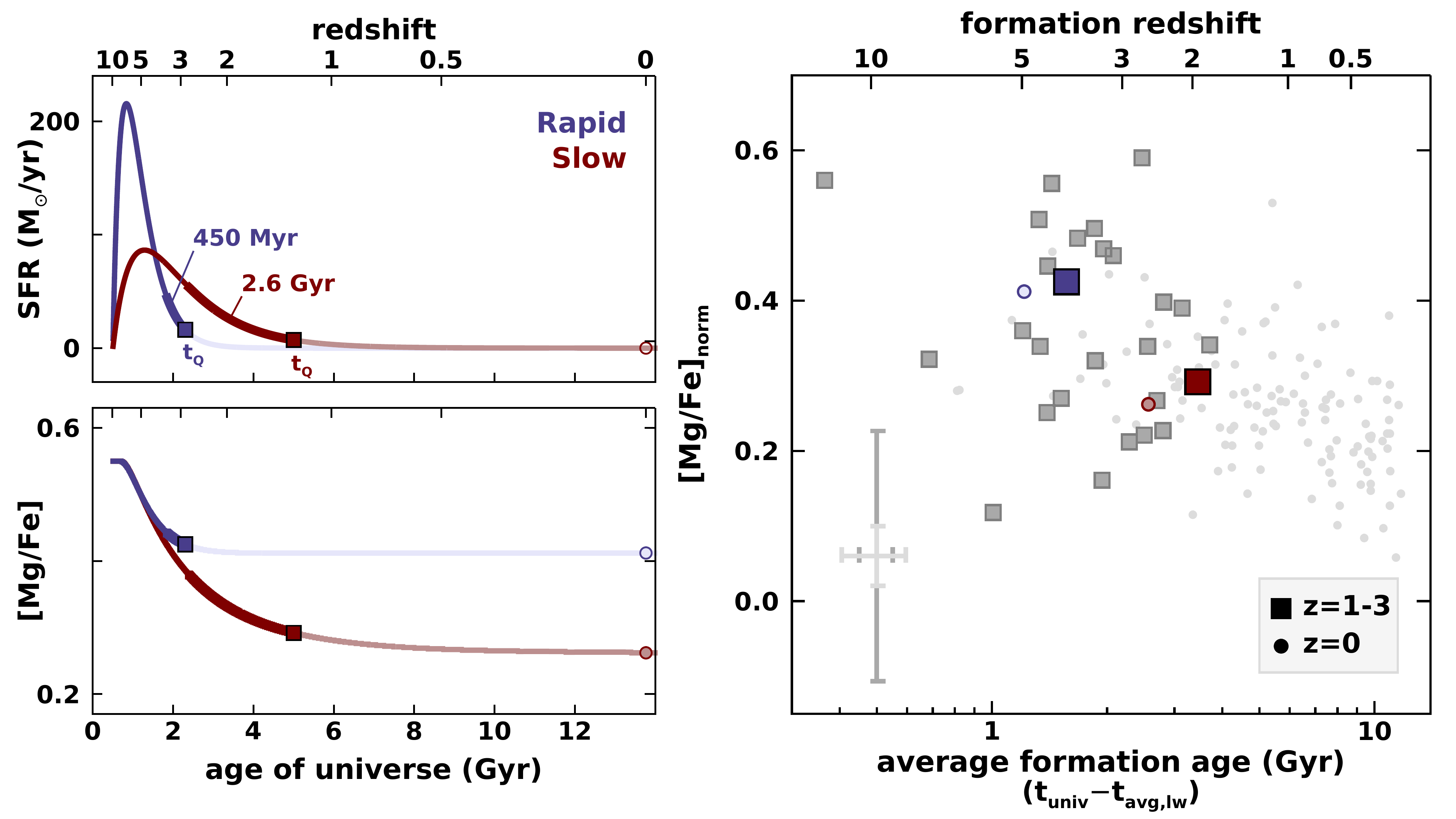}
\vspace{-0.2cm}
\caption{Example star formation (top left) and chemical evolution (bottom left) histories for a present-day massive galaxy ($10^{11}$ M$_\odot$) that quenches rapidly (\emph{blue}) versus one that quenches slowly (\emph{red}), following \citet{Gountanis2025}. In the rapid case, core-collapse SNe from short-lived massive stars enrich the ISM with $\alpha$-elements (e.g., Mg) before Type Ia SNe contribute additional Fe on longer timescales, leaving  [Mg/Fe] nearly unchanged from quenching  (3$\sigma$ below the SFMS; \emph{square}) to the present (\emph{circle}). Compared with the \citet{Beverage2025} compilation, $z=1$–3 galaxies show similar [Mg/Fe] values and average formation ages. The latter is defined as the difference between the age of the Universe at observation and the light-weighted stellar age. Average uncertainties are shown for context. Although some $z=0$ systems retain high $\alpha$-enhancements consistent with rapid early formation, most show lower [Mg/Fe] and older formation ages,likely reflecting post-quenching mergers and accretion that dilute both [Mg/Fe] and inferred timescales. Figure adapted from \citet{Beverage2025} (CC BY 4.0). Abbreviation: SNe, supernovae.} 
\label{fig:chemev}
\end{figure}

\subsubsection{Star-formation histories and quenching timescales}  

It is often useful to extend analyses of quiescent galaxies beyond average ages to their full SFHs. Several timescales help characterize formation and quenching histories. These include the $t_{\mathrm{50}}$ formation time or redshift \citep[when 50\% of the stellar mass formed, more details in \autoref{subsec:stellarages};][]{Pacifici2016,Carnall2018,Carnall2019}, the time since quenching \citep[interval between crossing a quenching threshold and the epoch of observation;][]{Carnall2019,Park2024}, and the quenching duration itself. The latter lacks a consistent definition, leading to systematic differences due to methodology. One approach defines the duration as the interval from $t_{\mathrm{50}}$ to quiescence \citep{Carnall2018,Akhshik2023}, analogous to the time between peak SFR and quiescence. By contrast, parameteric SFHs often use an exponential e-folding timescale $\tau_q$ \citep{Wild2016,Belli2019}, which generally yields shorter estimates because $\tau_q$ describes only the post-quenching rate of decline; the total time from peak SFR to quiescence is typically several $\tau_q$. Because $\tau_q$ depends strongly on the assumed SFH shape, it may not capture true quenching durations, especially in systems with late bursts. 

Motivated by the framework of \citet{Tacchella2016}, another approach tracks the time required for a galaxy to evolve from $1\sigma$ to $3\sigma$ below the SFMS (e.g., top left, \textbf{\autoref{fig:chemev}}), which is broadly consistent with both $\tau_q$ and the duration of galaxies with transitional colors \citep{Schawinski2014,Carnall2019}. Because the SFMS evolves with redshift, this definition is inherently tied to cosmic evolution and thus avoids the need for an explicit normalization by the age of the Universe, as done in other studies \citep[e.g.,][]{Pacifici2016,Carnall2018,Akhshik2023}. Overall, quenching timescales span a few hundred Myr to several Gyr, consistent with the wide diversity expected theoretically (\autoref{sec:quenching_mechanisms}) and observed in post-starburst galaxies and compact star-forming progenitors \citep[e.g.,][]{Barro2013,Barro2017}. We caution that the inferred shapes of SFHs and quenching durations can be significantly biased by modeling assumptions, including analytic forms or even ``non-parametric'' approaches \citep[e.g.,][]{Carnall2019a,Leja2019a,Kaushal2024}. 

With methods used to infer SFHs continuing to improve \citep[e.g.,][]{Iyer2026}, they hold significant potential for distinguishing quenching mechanisms. Well-constrained SFHs can reveal past episodes of extreme star formation (accelerated astration) and quenching timescales, separating rapid processes (e.g., AGN-driven gas removal) from slower ones (e.g., cosmological starvation, environmental processes, or morphological quenching). With sufficiently high-quality data, bursty SFHs may also be identified \citep[e.g.,][]{Wang2025}, offering clues to rejuvenation events \citep[e.g.,][]{Pandya2017,Iyer2019}. While such events appear rare observationally, simulations predict they should be common, especially in galaxy outskirts and at high redshift \citep[e.g.,][]{Tacchella2016,Nelson2018,Dolag2025}, although centrally-weighted observations may bias against detection. 

Since stellar mass closely tracks halo mass \citep[e.g.,][]{Behroozi2013,Moster2013}, SFHs also encode information about the connection between star formation suppression and halo growth, including the transition from cold- to hot-mode accretion (cosmological starvation) and the role of AGN feedback, which depend on SMBH energetics and the spatial distribution of cold gas. Thus, intrinsic SFHs provide not only a record of galaxy assembly but also key diagnostics of dominant quenching processes.

\subsubsection{Chemical Abundances} 
\label{subsec:chemical_abundances}

The stellar metallicities and abundance patterns of quiescent galaxies at $z>1$ provide key complementary constraints on their SFHs and enrichment timescales. Massive quenched galaxies at these epochs already exhibit near-solar or super-solar metallicities, implying rapid early enrichment \citep[e.g.,][and references therein]{Kriek2016,Onodera2015,Gallazzi2021}
Spectroscopic studies targeting absorption features such as Fe, Mg, and Ca show that the stellar metallicities of $z\sim1.5$–2 quiescent galaxies are broadly consistent with those of local early-types at fixed mass, suggesting that most 
chemical enrichment occurred prior to quenching \citep[e.g.,][]{Gallazzi2014, Kriek2019, Beverage2024}. 
Lower-mass quiescent galaxies, however, display substantial scatter and often sub-solar metallicities, consistent with more extended SFHs \citep[e.g.,][]{Lonoce2015,Saracco2019}.
While minor mergers and late accretion contribute to the observed diversity \citep[e.g.][]{Oser2010,vanDokkum2010}, these processes primarily affect galaxy outskirts rather than central chemical properties.  These secondary effects are discussed in more detail in \SMsec{subsubsec:supplement_chemical_evolution}.

The relative abundance of $\alpha$-elements, typically parameterized as [$\alpha$/Fe], are powerful tracers of enrichment timescales. 
Because $\alpha$-elements 
and Fe are released promptly by core-collapse supernovae (SNe) while Fe also accumulates in comparable amounts from delayed Type Ia SNe, elevated [$\alpha$/Fe] ratios signal rapid star formation followed by early quenching \citep[e.g.,][]{Thomas2005}. 
Measurements of [Mg/Fe] in massive quiescent galaxies at $z\sim2$ often fall in the range $\sim$0.2-0.5 \citep[e.g.,][and references therein]{Beverage2024,Beverage2025,Jafariyazani2025}, consistent with short ($<1$ Gyr) formation timescales \citep[\textbf{\autoref{fig:chemev}}
; compilation adapted from][]{Beverage2025}.   
These chemical timescales support a picture in which massive galaxies form and quench rapidly -- likely via ejective AGN feedback -- while intermediate-mass systems quench more slowly, potentially through radiative heating \citep[e.g.,][see \autoref{sec:framework}]{Hopkins2008,Choi2015}. 

Interpreting these abundance ratios through chemical evolution models, however, remains non-trivial. As \textbf{\autoref{fig:chemev}} illustrates, [$\alpha$/Fe] evolves jointly with the SFH and is highly sensitive to assumptions about the IMF and the baryon cycle, including gas inflows, outflows, and star formation efficiency \citep{Gountanis2025}. Rapid and slow quenching pathways can yield overlapping abundance signatures, complicating attempts to map [$\alpha$/Fe] values directly to quenching timescales \citep[e.g.,][]{Thomas2005,deLaRosa2011}. Thus, while elevated [$\alpha$/Fe] ratios imply short formation timescales, they do not uniquely determine the details of the quenching process.

Despite their importance, measuring abundance ratios at high redshift remains notoriously difficult. Classical absorption indices weaken substantially at young ages, and age–metallicity–abundance degeneracies are severe, especially at modest SNR (see \SMsec{subsubsec:lick}). A major limitation is that most SPS models adopt scaled-solar abundance patterns \citep[e.g., see][]{Kriek2016}, which can introduce systematic biases for $\alpha$-enhanced populations. Full spectral modeling that leverages the entire continuum and line spectrum is therefore essential \citep[e.g.,][]{Beverage2024,Gountanis2025}. Additional details on abundance-sensitive feature behavior, limitations of current SPS models, and spectral library calibration are presented in \SMsec{subsubsec:supplement_chemical_models}.

Finally, robust interpretation of [$\alpha$/Fe] is hindered by uncertainties in the IMF and its impact on $\alpha$-element yields. Variations in the IMF that change the proportion of massive stars \citep{vanDokkumConroy2024}, and thus the $\alpha$-element yields resulting from core-collapse SNe, directly influence the expected [$\alpha$/Fe]. The super-solar $\alpha$-enhancements observed in many massive quiescent galaxies at high redshift may therefore reflect a combination of extreme, short-timescale SFHs and/or IMF variations \citep[e.g.,][]{Forrest2022,Kriek2024}. 
Continued advances in SPS models and chemical evolution frameworks will be essential for interpreting high-redshift quiescent galaxy chemical abundances, particularly in systems with extreme or unusual enrichment patterns. 
Even high-quality spectroscopic observations of quiescent galaxies will remain difficult to interpret reliably without corresponding progress in chemical-evolution theory.

\subsubsection{Dust Attenuation}
\label{subsubsec:dust_attenuation}

Although massive quiescent galaxies at $z>1$ are generally considered dust-poor relative to star-forming systems, recent observations show that dust attenuation can still play a non-negligible role in shaping their observed SEDs. Their colors become increasingly reddened, from 0.1 mag at $z\sim0$ to 0.4-0.7 mag at $z\sim2$ \citep{Martis2016,Siegel2025}, a trend that is degenerate with age and metallicity in the rest-frame optical. To further complicate matters, the geometry of this residual dust may also  differ from that of the stars \cite[e.g.,][]{Setton2024,Ji2024_MIRI,Siegel2025}.  The global SFRs of massive quiescent galaxies at $z\sim2$ are $\sim0.5$ dex higher than at $z\sim0$ (\textbf{\autoref{fig:ssfr}}), which may naturally contribute to the corresponding evolution in $A_{V}$ for an otherwise passively evolving population.  Because dust attenuation tracks evolving SFRs, it depends on the quenching timescale, where more rapid quenching quickly halts new dust production from massive stars, whereas slowly quenched systems may retain dust from recent star formation. The quenching mechanism itself (e.g., radiatively-efficient AGN feedback) may also influence dust survival, either  destroying grains in energetic outflows or redistributing them into the CGM \citep[e.g.,][]{Donevski2023,Lorenzon2025}.

Most of far-IR emission in quiescent galaxies is from dust heated by stars older than 100 Myr \citep[e.g.,][]{Fumagalli2014,Hayward2014,Utomo2014}, including new dust formed in the circumstellar envelopes of evolved stars (\SMsec{subsubsection:TPAGB_dust}).
Although the same processes operate in star-forming galaxies, the heating from young stars ($<100$ Myr) far outweighs that from older populations. Even modest  attenuation (e.g., $A_V\sim0.3$ mag) can bias stellar ages and masses if not properly modeled (see Figure~2 in the review by \citealt{Salim2020} and references therein). This is especially problematic when fixed attenuation curves (often Calzetti-like; \citealt{Martis2019}), 
are adopted, despite evidence that quiescent galaxies may not follow such laws.

Broad UV–optical–IR coverage, ideally including Balmer decrements, can help constrain dust attenuation and break degeneracies with other SPS parameters, although significant uncertainties remain \citep[see also review by][]{Conroy2013}. First, the dust attenuation curve for quiescent galaxies is poorly constrained because two competing trends act simultaneously: Low sSFRs are associated with grayer, shallower curves, while very low dust columns ($A_V\lesssim0.3$) observed to exhibit steeper, more UV-sensitive curves \citep[e.g.,][]{Salim2018}. The balance between these trends varies within the quiescent population, contributing to the observed diversity in their inferred attenuation properties. Second, dust geometry -- whether mixed with stars, centrally concentrated, or patchy -- strongly affects the effective attenuation \citep[e.g.,][]{Chevallard2013}.  
Third, degeneracies with metallicity, IMF, and abundance patterns further complicate isolating dust effects at high redshift. Altogether, while quiescent galaxies contain little dust compared to star-forming systems, the modest amounts that remain can meaningfully bias inferred  stellar population properties, reinforcing the need for flexible attenuation models in SPS fitting.

\subsection{Interstellar Medium of Quiescent Galaxies}
\label{subsec:ism}

Massive galaxy halos in the early Universe are expected to host substantial gas reservoirs \citep[e.g.,][]{SilkRees1998}, confirmed observationally through scaling relations and molecular gas detections \citep[][and references therein]{Tacconi2020}. 
Given that this gas should cool efficiently, fueling at least one substantial early episode of star formation, and, in some cases, sustaining activity for extended periods \citep[e.g.,][]{Keres2005}, 
the very existence of quenched galaxies at early times was not anticipated. Understanding the presence, quantity, and spatial distribution of gas in quiescent galaxies is therefore critical for constraining how star formation shuts down and what physical processes are responsible \citep{ManBelli2018}. Such measurements not only reveal the current and future star-forming potential of these systems but also provide key insight into their long-term evolutionary pathways.

Distinguishing among different quenching mechanisms requires determining whether gas has been consumed, expelled, heated, or prevented from cooling (\autoref{sec:quenching_mechanisms}). Yet many gas phases are inherently difficult to detect directly. Cold molecular hydrogen (H$_2$) lacks a permanent dipole moment and radiates inefficiently at low temperatures, rendering it elusive.  Consequently, observations rely on indirect tracers such as carbon monoxide (CO) spectroscopy, which probes cold H$_2$ \citep{Bolatto2013}, and far-IR/sub-mm dust continuum emission \citep{Scoville2016}, which can provide estimates of the total gas mass under assumptions about dust-to-gas ratios \citep[e.g., see review by][]{Tacconi2020}. 
Complementary probes of ionized and hot gas components, through optical/UV line emission \citep{Tumlinson2017} and X-ray data \citep{Werner2014}, 
are essential for a complete multiphase view of the ISM and CGM in quiescent galaxies, with early evidence already hinting at CGM–quenching connections \citep{Tchernyshyov2023}. However, at the low–to–moderate spectral resolution of most high-redshift data, absorption features often indicate the presence of ionized or neutral gas but do not uniquely constrain its location or kinematic state. Thus, observationally identified ISM may include outflowing (or inflowing) material, and some gas classified as outflowing could instead reside within the galaxy.

\begin{figure}
    \centering
    \includegraphics[width=0.97\linewidth]{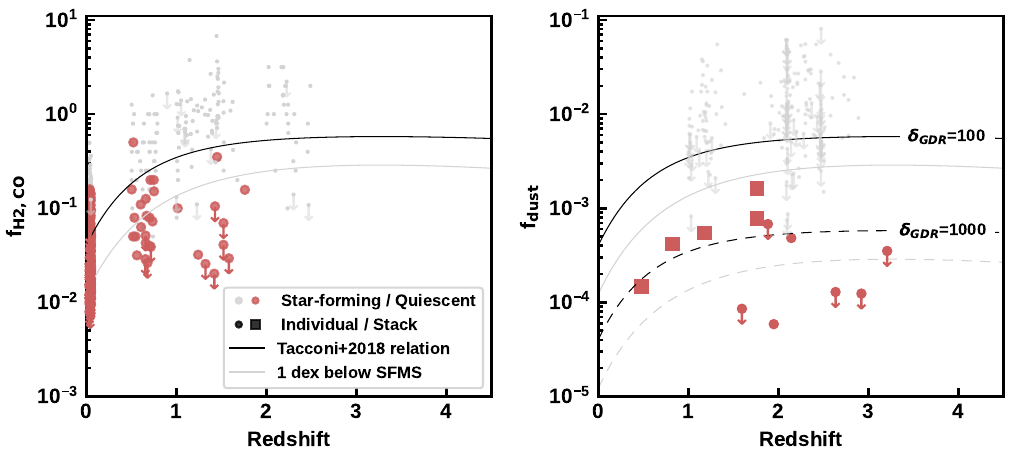}
    \vspace{-0.2cm}
    \caption{The molecular gas reservoirs of quiescent galaxies are deficient relative to well-studied scaling relations for star forming counterparts \citep[\emph{black line}, ][]{Tacconi2018}. Indirect studies of the molecular gas reservoirs of galaxies below the SFMS via CO lines (left panel) and dust continuum (right panel) are somewhat sparse (\emph{red symbols}) and each is sensitive to a different collection of systematic uncertainties. CO data from \cite[][and references therein]{Williams2021} and dust continuum from \cite[][and references therein]{Whitaker2021b}. Abbreviation: SFMS, star-formation main sequence.}
    \label{fig:gasfractions}
\end{figure}

\subsubsection{Molecular Gas}
\label{subsubsec:coldgas}

A robust census of cold molecular gas in quiescent galaxies lags significantly behind that of their stellar populations, largely due to the observational difficulty of constraining H$_2$ content in systems with low SFRs \citep[e.g.,][]{Sargent2015,Spilker2018}. This gap is critical, as most quenching mechanisms require cold molecular gas to be expelled, depleted, or heated, and therefore predict a deficit of H$_2$ in post-starburst or quiescent systems (\autoref{sec:quenching_mechanisms}). The only well-established scenario in which a galaxy may retain significant amounts of cold molecular gas without ongoing star formation is morphological quenching, in which a stabilizing stellar potential suppresses fragmentation and collapse of the gas \citep[][see \SMsec{subsubsec:gas_stabilization_galaxyscales}]{Martig2009}. 

\begin{marginnote}
\entry{Cold gas main sequence (CGMS)}{tight, redshift- dependent relation between stellar mass and inferred molecular gas mass for star-forming galaxies.}
\end{marginnote}

CO remains the most widely used tracer of cold H$_2$ \citep[e.g.,][]{Bolatto2013}, and deep ALMA/NOEMA observations targeting the lowest-energy CO lines, such as CO(2–1) and CO(3–2), in massive quiescent galaxies at $z>1$ generally find very low molecular gas fractions, $f_{\mathrm{H_2,CO}} \lesssim 10\%$, well below the equivalent SFMS in M$_{H_2}$ -- the cold gas main sequence (CGMS; left panel, \textbf{\autoref{fig:gasfractions}}). Some low- to intermediate-redshift post-starburst and recently quenched systems show higher gas fractions (up to $\sim$20–40\%), indicating that substantial molecular gas can persist for some time after quenching \citep[e.g.,][]{French2015,Socolovsky2019,Belli2021,Setton2025,Suess2025}. However, local studies using dense-gas tracers such as HCN and HCO$^{+}$ show that some CO-rich post-starbursts are severely deficient in the denser gas phases that directly fuel star formation \citep[e.g.,][]{French2018}. This deficiency could indicate that CO-based M$_{H_2}$ estimates are inflated because much of the gas is diffuse, or that a substantial H$_2$ reservoir is present but unable to condense into dense clumps. In the latter case, stabilization mechanisms such as morphological quenching or merger-driven turbulence likely prevent dense-gas and star formation even when molecular gas is present. In some systems, mergers may also resupply gas and trigger rejuvenation. Nevertheless, such gas-rich quenched galaxies appear to be the exception rather than the rule, as the bulk of high-redshift quiescent systems are H$_2$-poor.

When CO is detected in other quiescent systems, inferred depletion timescales are often $\lesssim1$~Gyr \citep[e.g.,][]{Spilker2018,Bezanson2019,Williams2021}, supporting a scenario in which star formation ceased rapidly and has not resumed due to a lack of significant molecular fuel. In galaxies with more moderate gas reservoirs, short depletion times of a few hundred Myr indicate that the remaining molecular gas cannot sustain long-term star formation without replenishment \citep[e.g.,][]{Woodrum2022}. These constraints are broadly consistent with rapid quenching mechanisms such as AGN-driven gas removal, accelerated astration during starbursts, or more rapid environmental processes, all of which predict both low gas fractions and short consumption timescales. Detailed discussions of the uncertainties in the CO-to-H$_2$ conversion factor, CO-faint molecular gas, and comparisons to alternative tracers are provided in \SMsec{subsec:supplement_ism}.

Fine-structure lines such as [CI] and [CII] offer promising complementary probes of cold gas in regimes where CO is faint or photodissociated. [CI] can trace more diffuse molecular gas, while [CII] 158~$\mu$m is one of the brightest ISM cooling lines and is readily accessible at high redshift with ALMA \citep[e.g.,][]{CarilliWalter2013,Zanella2018}. To date, however, [CI] and [CII] remain undetected in spectroscopically confirmed quiescent galaxies at $z>1$, with current observations yielding upper limits on their cold gas content \citep[e.g.,][]{DEugenio2023}. Both lines trace multiple phases and require assumptions about excitation and carbon abundance, so they are best interpreted in combination with CO and dust continuum; we expand on these caveats in \SMsec{subsubsec:CO}.

\subsubsection{Cold Dust}
\label{subsubsec:colddust}

Dust continuum observations offer an appealing alternative tracer of H$_2$, as millimeter wavelengths are sensitive to the large grains that dominate the total dust mass \citep[e.g.][]{Casey2014, Scoville2016}. Compared to relatively expensive CO spectroscopy, dust continuum enables more efficient measurements that can reach extremely low dust mass limits, making it well suited for statistical studies of the cold ISM in quiescent galaxies. Current efforts rely either on stacked dust analyses \citep[e.g.,][]{Gobat2018,Magdis2021} or on targeted dust-continuum observations of small, carefully selected samples \citep[e.g.,][]{Whitaker2021a}. Of the few direct detections that exist, $f_{\mathrm{H_2}}$ inferred from dust in $z>0$ quiescent galaxies spans more than a dex (\textbf{\autoref{fig:gasfractions}}), underscoring that the difficulty lies not only in detectability but also in genuine intrinsic variation across systems. 

The origin of the variation in dust mass for quiescent galaxies remains unclear.  Strongly lensed systems at $z\sim2$ exhibit extremely low dust-to-stellar mass fractions ($f_{\mathrm{dust}}\lesssim0.01\%$), implying gas fractions $f_{\mathrm{H_2}}\lesssim1\%$, for typical dust-to-gas ratios ($\delta_{\mathrm{GDR}}$), and placing them well below the star-forming locus at these redshifts (\textbf{\autoref{fig:gasfractions}}, right). By contrast, stacking of color-selected quiescent samples in low-resolution surveys infer higher dust fractions, $f_{\mathrm{dust}}\sim0.06$–0.08\%, corresponding to $f_{\mathrm{H_2}}\sim6$–8\% with similar $\delta_{\mathrm{GDR}}$ assumptions \citep[e.g.,][]{Gobat2018,Magdis2021,BlanquezSese2023}. These stacked results roughly follow extrapolations of the CGMS \citep{Tacconi2020}, yet they stand in tension with the very low dust and gas masses inferred from individual lensed systems.

This diversity underscores how dust-based M$_{\mathrm{H2}}$ estimates depend not only on assumptions about $\delta_{\mathrm{GDR}}$, but also dust temperature and emissivity, all of which may evolve rapidly as galaxies quench. As the (s)SFR declines, thermal sputtering, driven by collisions with energetic electrons in the hot, diffuse halos of quiescent galaxies, can erode grain surfaces and ultimately destroy dust \citep{Tsai1995,Li2019}, driving $\delta_{\mathrm{GDR}}$ as high as $10^4{-}10^5$ \citep{Whitaker2021b}. In this picture dust declines more rapidly than $H_2$, making dust-based M$_{\mathrm{H2}}$ estimates unreliable at low sSFR, consistent with recent observations \citep{Spilker2025}. At the same time, simulations show that the opposite trend is possible: gas can be consumed or expelled faster than dust, leaving behind significant dust reservoirs (or rapidly replenished ones) even after star formation has ceased \citep{Hayward2014,Lorenzon2025}. Simulations also show that stacking analyses are biased, preferentially recovering emission from the dustiest subset of the quenched population and thereby overestimating average dust masses \citep{Lorenzon2025}. Reconciling these contrasting theoretical predictions, as well as the wide spread in observed dust properties, will require direct, unbiased measurements of dust masses across quenching timescales. Progress will also depend on moving beyond traditional stacking approaches, or developing stacking methods that mitigate their bias toward the dustiest systems. We discuss these physical and observational systematics, as well as the role of late-stage dust production by TP-AGB stars, in \SMsec{subsubsec:dust}.

Overall, CO, dust, and fine-structure line measurements point to a consistent picture in which most high-redshift quiescent galaxies are H$_2$-poor, with short depletion times. A minority of post-starburst or recently quenched systems retain more substantial reservoirs, particularly at later cosmic times. The detailed interplay between gas content, dust physics, and quenching timescales remains an active area of investigation.

\subsubsection{Neutral and Ionized Gas}
\label{subsec:outflows}

Understanding the presence and role of neutral and ionized gas in quiescent galaxies at $z>1$, when combined with studies of cold molecular gas, is critical for constraining the physical processes that drive and maintain galaxy quenching in the early Universe. Despite being classified as ``dead'' in terms of star formation, many massive quiescent galaxies at high redshift are still embedded in rich gaseous environments \citep[e.g.,][]{Genzel2014},
raising questions about how star formation was halted and why it remains suppressed. Neutral gas, traced through rest-UV and optical absorption lines \citep[e.g.,][]{Concas2017,Valentino2020} or HI 21 cm in the local Universe \citep[e.g.,][]{Serra2012}, offers insight into the fuel reservoir that could in principle cool and reignite star formation. Ionized gas, on the other hand, indicates the presence of either residual star formation, AGN photoionization, or shock heating \citep[see review by][]{Kewley2019}.

\begin{figure}
    \centering
    \includegraphics[width=0.98\linewidth]{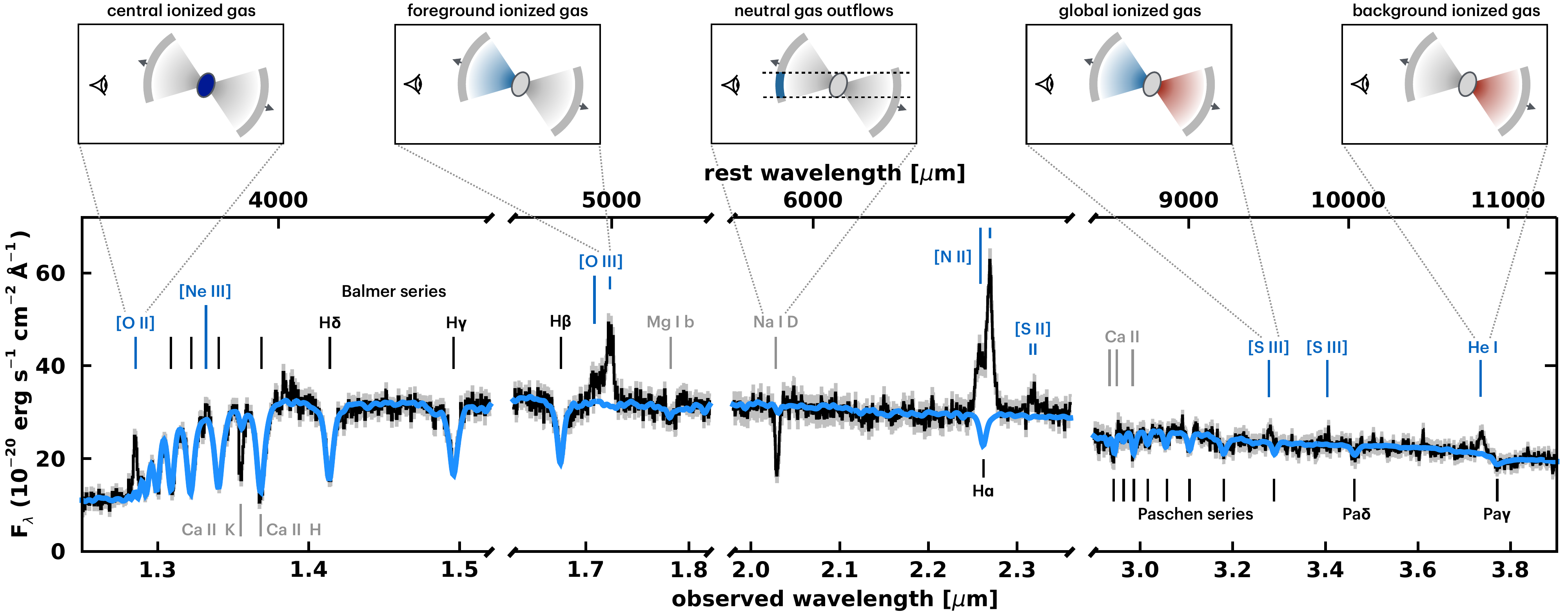}
    \caption{A massive quiescent galaxy at $z=2.5$, $\sim$1 dex below the SFMS, shows clear neutral outflows in its spectrum from the Near Infrared Spectrograph on the \emph{James Webb Space Telescope} traced by blueshifted NaI absorption and excess CaII K. This provides strong empirical evidence of a causal connection between the quenching of star formation and radiatively efficient AGN feedback. The mismatch between the model (\emph{blue}) and data (\emph{black}) further reveals ionized gas via [OII], [OIII], and [NII] emission. Top panels illustrate the spatial regions probed by each kinematic gas component. Absorption lines from hydrogen are marked in black, metals in gray, and ionized-gas emission features in blue, with more information found in \textbf{\SMtab{tab:lines}}. Figure adapted from \citet{Belli2024} (CC BY 4.0). Abbreviation: SFMS, star-formation main sequence.}
    \label{fig:outflows}
\end{figure}

Observations of massive quiescent galaxies at high redshift reveal a rich collection of spectral lines tracing the ISM, including both emission lines tracing ionized gas that is warm ($\sim10^4$ K) relative to typical ISM temperatures ([OII], [NeIII], H$\beta$, [OIII], H$\alpha$, [NII], [SII], [SIII], and HeI) and absorption lines that are sensitive to warm ($\sim10^{4}$ K) ionized gas (FeII, MgII, CaII K) and cold ($\sim100-500$ K) neutral gas (NaI D) \citep[e.g.,][]{Maltby2019, Man2021, Belli2024, Davies2024, DEugenio2024, Valentino2025}. These absorption lines are resonant transitions out of the atomic ground state and can originate in either stellar photospheres or the ISM (see \textbf{\SMtab{tab:lines}}). For example, Na D and CaII K absorption may arise from both stars and neutral ISM gas \citep[e.g.,][]{Belli2024}, whereas FeII and MgII absorption are purely interstellar in origin. Their kinematics then provide additional information: a blueshift relative to the systemic velocity signals outflowing gas \citep[e.g.,][]{Tremonti2007,Wu2025}. \textbf{\autoref{fig:outflows}} illustrates this in a massive quiescent galaxy at $z=2.5$, where radiatively efficient AGN feedback is revealed by emission-line ratio diagnostics together with blueshifted absorption features that trace both ionized and neutral outflows. 

Strong emission lines in quiescent galaxy spectra are common at intermediate \citep[e.g.,][]{Maseda2021} and high redshift \citep[e.g., \textbf{\autoref{fig:deepdive}} and  \textbf{\ref{fig:outflows}};][]{Belli2024,Ito2025}. Distinguishing the physical origin of this emission -- whether residual star formation, AGN photoionization, and/or shock heating -- requires additional data. Multi-wavelength observations (X-ray, IR, or radio) offer independent AGN indicators \citep[e.g.,][]{Stern2005,Ellison2021}, while emission line ratios \citep[][and references therein]{Kewley2019} 
help distinguish AGN from pure star formation. However, commonly used line-ratio diagnostics cannot uniquely differentiate AGN from shocks \citep[e.g.,][]{Alatalo2016}. Spatial information provides further constraints, as AGN photoionization is typically nuclear \citep[e.g.,][]{Whitaker2013}, whereas shocks may produce extended or kinematically broadened emission. Comparing observed line ratios and line widths to shock models offers a viable path forward \citep{Zhu2025}.
Local post-starburst galaxies show evidence for both shock heating \citep[e.g.,][]{Alatalo2016} and ejective AGN feedback \citep[e.g.,][]{Cheung2016}.
 
The unmistakable signature of blueshifted ISM-sensitive spectral features, in absorption or emission, points to galaxy-scale outflows. 
Although debate remains over which phase (neutral, ionized, or molecular) dominates mass loss rates, 
outflows are ubiquitous in massive star-forming galaxies at $z\sim2$ \citep{Genzel2014,ForsterSchreiber2019}. JWST now reveals that neutral outflows may play a decisive role in quiescent systems: \citet{Belli2024} demonstrate that, while ionized outflow rates are negligible, the neutral mass outflow rate exceeds the residual SFR, implying that neutral winds can strongly suppress star formation. Similar evidence is emerging from other recent JWST samples \citep{Davies2024,DEugenio2024,Wu2025,Valentino2025,Ito2025,Onoue2025}, while pre-JWST detections were already reported \citep{Maltby2019,Man2021}. For example, \citet{Davies2024} find that $\sim50\%$ of their $z\sim2$ quiescent sample shows excess Na D absorption, with roughly half of those also exhibiting the blueshifted hallmark of neutral outflows. JWST is now probing down to intermediate stellar mass regimes at $z>3$, revealing AGN in quenched systems \citep{Marchesini2023,Sato2024} -- such empirical constraints are especially important for testing the stellar mass threshold above which AGN feedback becomes important. Taken together, these results strongly support quenching via radiatively efficient, ejective AGN feedback, though improved calibrations of mass outflow rates are necessary \citep{Moretti2026} and X-ray studies offer conflicting evidence \citep[e.g.,][]{Almaini2025}.

Whether the observed neutral winds are able to escape the deep potential wells of massive halos or instead recycle back into the ISM remains an open question; the answer bears directly on how quiescence is maintained over long timescales. Neutral gas with modest outflow velocities ($\sim$200 km/s) is unlikely to escape entirely, consistent with simulations in which outflowing gas is primarily recycled rather than permanently expelled \citep[e.g.,][]{AnglesAlcazar2017}. 
Maintaining long-term quiescence therefore likely requires ongoing preventative processes, such as radiatively-inefficient AGN feedback or virial shock heating. Central [OIII]/H$\beta$ ratios in stacks of older quiescent galaxies at $z\sim2$ are consistent with low-level AGN or shocks \citep{Whitaker2013}, supporting this picture.

Finally, some quiescent systems retain moderate reservoirs of neutral gas \citep[e.g.,][]{Davies2024} and, in a subset, molecular gas \citep[e.g.,][see also \autoref{subsubsec:coldgas}]{Bezanson2022_psb}. In such galaxies, quenching may not stem from wholesale gas ejection but instead from processes that inhibit star formation, such as gas heating by AGN feedback (mechanical or photoionization), virial shock heating, or morphological quenching. Expanding multiphase gas studies to larger, more diverse samples will be essential to disentangle the interplay of feedback, accretion, and stabilization processes that govern long-term quiescence.

\section{QUIESCENT POPULATION DEMOGRAPHICS}
\label{sec:demographics}

This section reviews key galaxy scaling relations in the context of quiescent systems. The quenching of star formation correlates with several global structural properties, including stellar mass, surface (or central) mass density, and central velocity dispersion \citep[e.g.,][]{Franx2008,Barro2013,Cheung2012}.  
However, causal relationships are difficult to establish given the strong covariances among these parameters and the complex interplay of internal and external processes \citep[see reviews by][]{Somerville2015,ForsterSchreiber2020,Tacconi2020}. 
Despite this, scaling relations are powerful tools for tracing the broad redshift evolution of galaxy properties and for identifying the physical regimes most closely associated with quenching \citep[e.g.,][]{Whitaker2017,Bluck2023}. With demographic progress driven mainly by large photometric samples \citep[e.g.,][]{Whitaker2011,Muzzin2013,Skelton2014,Laigle2016,Weaver2023a},
future wide-area spectroscopic datasets are critically needed for precise measurements of kinematics, metallicities, and stellar ages across diverse quenching pathways.

\subsection{Redshift Evolution of Number Density and Stellar Mass Functions}
\label{subsec:numberdensity}

\begin{figure}
    \centering\includegraphics[width=0.98\linewidth]{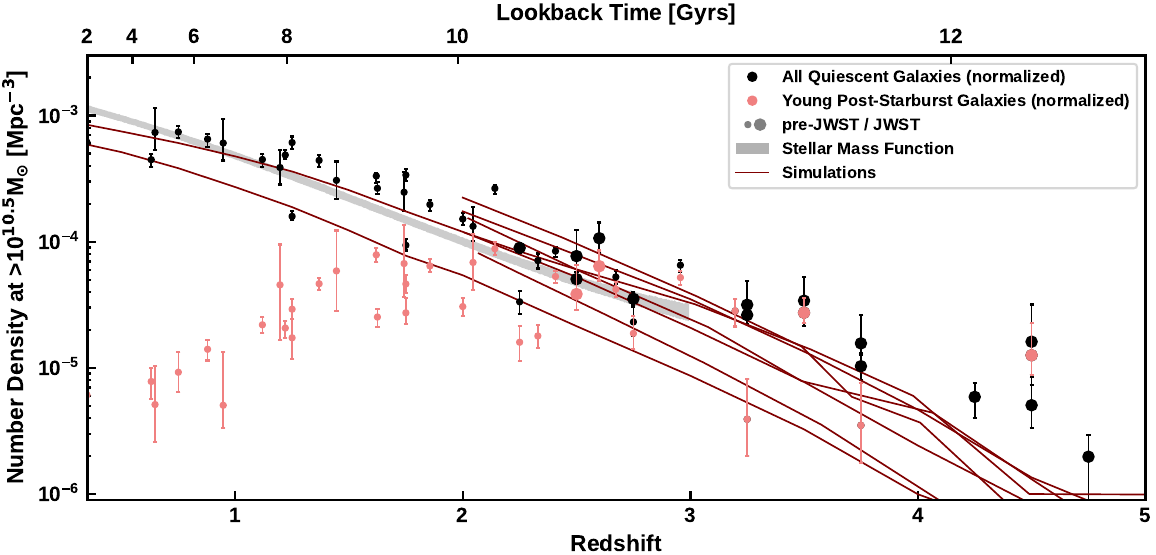}
    \caption{The number density of massive quiescent galaxies (corrected to M$_{\star}\geq$ 10$^{10.5}$ M$_{\odot}$) measured from a host of photometric and spectroscopic surveys \citep[\emph{filled symbols},][and references therein]{Zhang2026},     
    the quiescent stellar mass function \citep[\emph{gray band}]{Leja2020}, and predictions from simulations \citep[\emph{maroon lines};][and references therein]{Lagos2024}. Before $z\gtrsim$2.5, the population is dominated by younger or post-starburst galaxies (\emph{pink}), at which point the full population grows as galaxies age. At the earliest times ($z>3$), theoretical models underpredict the number of massive quiescent galaxies found by recent JWST studies. Abbreviation: JWST, \emph{James Webb Space Telescope}.}
    \label{fig:numberdensities}
\end{figure}

Stellar mass functions (SMF) and the number-density evolution of quiescent galaxies provide key insight into the build-up of this population and the physical mechanisms governing quenching \citep{Somerville2015}.  The simple exercise of counting galaxies in fixed stellar-mass bins places stringent constraints on the hierarchical growth of dark matter halos and the baryon-to-star conversion efficiency \citep{Behroozi2013}. Benchmark observational studies \citep[e.g.,][]{Fontana2006,Muzzin2013,Ilbert2013,Tomczak2014,Davidzon2017,Weaver2023a} 
show that while the universe contained only $\sim$1\% of its present-day stellar mass at $z\sim3.5$, massive quiescent galaxies already contributed significantly. Since then, their stellar mass density has risen by more than two orders of magnitude, accompanied by a tenfold increase in number density for $\log(M_{\star}/M_{\odot})\gtrsim10.5$ from $z\sim3$ to $z\sim1$ (\textbf{\autoref{fig:numberdensities}}). All data are normalized to a consistent mass threshold in \textbf{\autoref{fig:numberdensities}}, with more extensive discussion in \SMsec{sec:supplement_demographics}. This rapid early build-up suggests that quenching was already highly efficient; the primary era of massive galaxy quenching coincides with the peak of cosmic star formation.

The redshift evolution of the population of quiescent galaxies traces the combined effects of early star formation, rapid quenching, and subsequent structural growth (\autoref{sec:framework}). At the highest redshifts ($z{>}3$), quiescent galaxies are exceedingly uncommon, with number densities dropping sharply beyond $z{\sim}2$ \citep{Marchesini2010,Straatman2014,Schreiber18,Merlin19,Girelli2019,Valentino20,Valentino2023,Carnall2023b,Zhang2026}, 
indicating that quenching becomes increasingly rare and likely stochastic at early times. Despite their scarcity, JWST has extended the redshift frontier of spectroscopically confirmed massive, compact quiescent systems, with sizable samples now identified at $z{\sim}3{-}5$ \citep[e.g.,][and references therein]{Carnall2024,Nanayakkara2025,Ito2025}, and even returns clear detections at $z{\sim}5{-}7$ \citep{deGraaff2025,Weibel2025}, pointing to the existence of rapid and efficient early quenching channels \citep[see also][]{Chittenden2026}.  By $z{\sim}1$, the massive quiescent population ($M_{\star}>10^{10.5}$\,M$_{\odot}$) is effectively in place \citep[\textbf{\autoref{fig:numberdensities}}; e.g.,][]{Whitaker2012,Wild2016,Forrest2018,Belli2019,Kawinwanichakij2020,Clausen2024,Zhang2026}, 
with subsequent evolution dominated by structural growth of individual galaxies rather than a growing population (\autoref{subsec:mergers}). 
Morphological decompositions further show that spheroid-dominated systems govern the buildup of the quiescent SMF across time \citep{HuertasCompany2016}, underscoring a link between quenching and structure. Toward lower redshifts, the buildup of the quiescent population increasingly reflects the addition of galaxies at lower stellar masses \citep[e.g.,][]{Santini2022,Weaver2023a,Hamadouche2025}.
The observed mass- and redshift-dependent evolution of the quiescent SMF is consistent with a downsizing scenario \citep{Thomas2005,Thomas2010}, wherein the number densities of quiescent galaxies at $M_{\star}{\sim}10^{10}$\,M$_{\odot}$ plateau by $z{\sim}1$, while populations below $M_{\star}{\lesssim}10^{9}$ M$_{\odot}$ continue to assemble at later epochs. Self-consistent measurements of the SMF therefore provide strong constraints on galaxy formation models, particularly regarding the timing, efficiency, and mass dependence of quenching mechanisms \citep{DeLucia2024,Lagos2024}. 

Despite decades of progress on both fronts, there are two long-standing tensions within observations of massive quiescent galaxies and theoretical predictions.  First, there is a large spread ($\sim$1.5 dex) in the inferred number densities of massive quiescent galaxies across observational studies and simulations \citep[\textbf{\autoref{fig:numberdensities}}; e.g.,][]{Schreiber18, Girelli2019, Merlin19, Valentino20, Valentino2023, Lagos2024}.
This directly correlates with the long-recognized systematic uncertainties at the massive end of the SMF \citep[e.g.,][]{Marchesini2009}.
Observationally, much of the variance is attributed to the fact that massive quiescent galaxies are rare signposts of distant overdensities and protoclusters \citep[e.g.,][]{Strazzullo2013}, with increasing numbers of new JWST studies \citep[e.g.,][]{Ito2025,Ito2025a,Pan2025,McConachie2025},
and are therefore subject to significant cosmic variance \citep[e.g.,][]{Somerville2004,Moster2011,Jespersen2025}. Additional scatter arises from systematics, including sample selection and photometric data quality (\autoref{sec:selection}), with contamination becoming increasingly severe at $z>3$ \citep[e.g.,][]{Forrest2020, Forrest2024, Xie2024, Antwi-Danso2025}. Theoretically, the limited volumes of current large-scale simulations, typically $\sim$100–300 cMpc on a side, introduce comparable cosmic variance effects \citep[e.g.,][]{Genel2014, Nelson2018, Dave2019}. 
More fundamentally, feedback prescriptions in simulations remain wide-ranging and degenerate \citep[e.g.,][]{Weinberger2018}, yielding divergent predictions for the abundance of massive quenched systems \citep[\autoref{sec:quenching_mechanisms}; see also][and references therein]{Lagos2024}. 

The second tension is that observations suggest massive galaxies assembled faster and earlier than predicted by most simulations \citep{Kriek2016,Glazebrook2017}. This issue resurfaced and has become a central concern with the advent of early JWST spectroscopy, which has revealed a population of spectroscopically confirmed quiescent galaxies at $z>3$–5 that are both massive and compact \citep[e.g.,][]{Carnall2024,Glazebrook2024,deGraaff2025}. \textbf{\autoref{fig:numberdensities}} illustrates that the number densities of these bonafide quiescent systems exceed theoretical predictions by factors of several \citep{Hartley2023,Lagos2024}, 
underscoring the difficulty current simulations face in producing galaxies that quench so early.   While zoom-in simulations targeting overdense regions show promise \citep{Lovell2023}, this tension highlights the need to re-evaluate feedback prescriptions in cosmological models. In particular, reassessing the efficiency, timing, and coupling of stellar and AGN feedback \citep[e.g.,][]{Terrazas2020, Bluck2023} in a data-driven manner that leverages new empirical constraints from JWST is warranted.  

\begin{figure}
    \centering
    \includegraphics[width=0.98\linewidth]{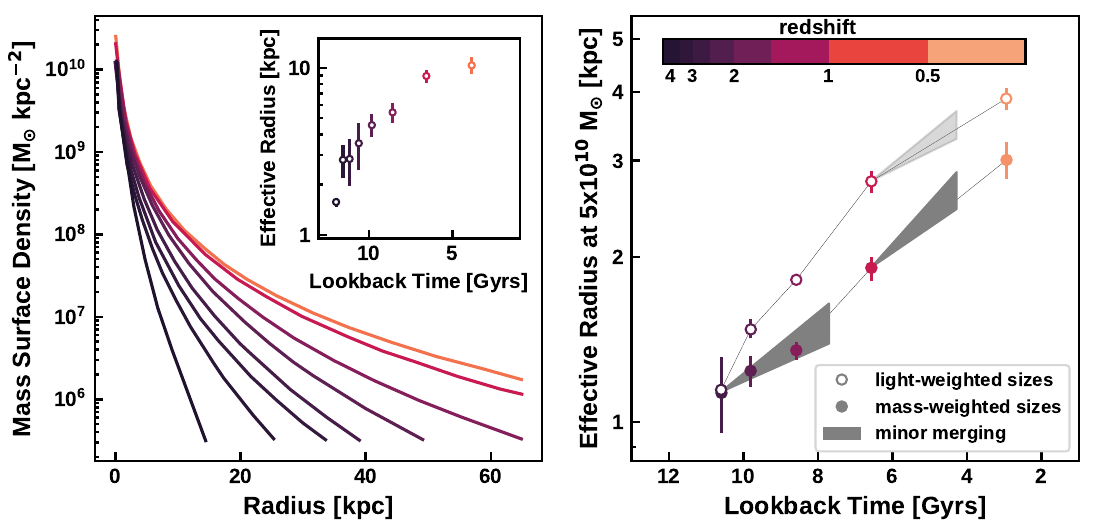}
    \caption{The (light-weighted) stellar mass profile shapes of quiescent galaxies evolve strongly in the outskirts (left), efficiently growing the effective radii (left, inset). This evolution is less dramatic mass-weighted sizes (\emph{filled symbols} in right panel, light-weighted size growth in open symbols). Minor merging models \cite[\emph{gray bands},][]{Newman2012} can account for the factor of $\sim3$ size growth since $z\sim2$. Left panel adapted from \citet{Hill2017} with permission of the AAS. Right panel adapted from \citet{Suess2019} with permission of the AAS.}
    \label{fig:hillsuess}
\end{figure}

\subsection{Morphological Evolution}
\label{subsec:morphology}

Locally, quiescent elliptical galaxies are rounder and more compact (at fixed mass) than their star-forming counterparts, exhibiting a distinct and steeper size–mass relation \citep[e.g.,][]{Shen2003,Lange2015}.
They also differ systematically in their structural profiles: spiral star-forming galaxies typically show exponential light distributions ($n\lesssim2$), whereas quiescent ellipticals are characterized by higher S\'{e}rsic indices, with denser cores and extended wings \citep[see review by][]{Kormendy2004}. 
This structural bimodality has often been interpreted as evidence that quenching and morphological transformation occur together at earlier epochs (e.g., \citealt{Kormendy2004}), although simulations demonstrate they can proceed on different timescales \citep{Park2022}.

Against this local benchmark, one of the most surprising discoveries from early HST imaging was the identification of massive quiescent galaxies at $z>2$ with extraordinarily compact half-light radii, in some cases as small as 1 kpc \citep[e.g.,][]{Daddi2005,vanDokkum2008}. Initially viewed as extreme relative to present-day ellipticals, these systems raised questions about their formation and growth. Subsequent studies confirmed that the stellar mass-size relation extends across cosmic time 
\citep[e.g.,][]{vanderWel2014,Mowla2019,Cutler2022,Nedkova2021,Wright2024,Martorano2023,Kawinwanichakij2026},
typically quantified as half-light radii measured from 2D S\'{e}rsic profile fitting, with this relation already in place as early as $z\sim6$ \citep[e.g.,][]{miller2025}. These morphological samples are generally $\sim1$ dex shallower than nominal survey completeness limits \citep[e.g.,][]{vanderWel2014}, as robust structural measurements require higher SNR than simple detection, and may still miss the intrinsically reddest and/or lowest-surface brightness galaxies. Although such biases likely preferentially exclude dusty star-forming galaxies and lower mass systems, the extent to which quiescent galaxies are also omitted -- and how this impacts quenching demographics -- remains uncertain. 

\begin{figure}
    \centering
    \includegraphics[width=0.98\linewidth]{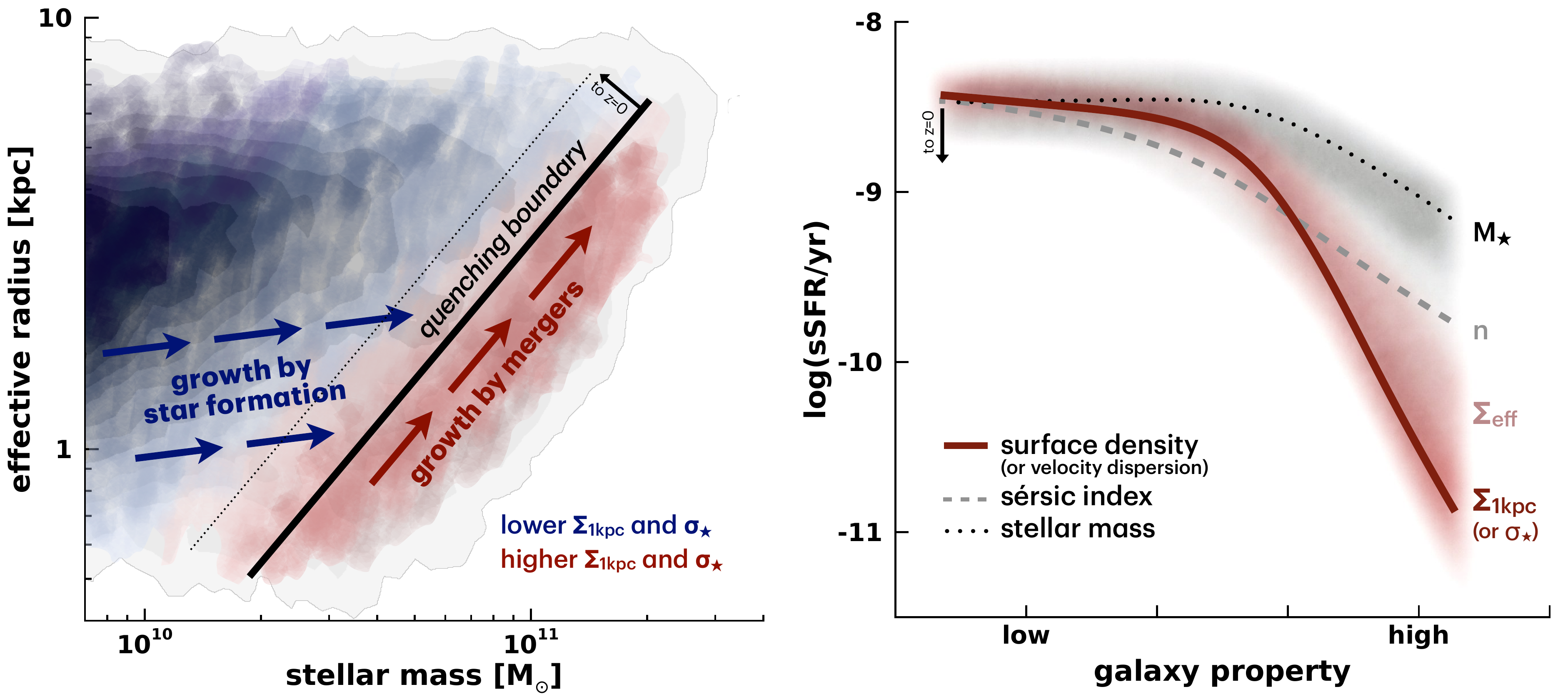}
    \caption{Schematic illustration of galaxy growth trajectories in the size–mass plane, with grayscale contours derived from the \citet{vanderWel2014} catalog, showing the observed distribution of galaxies (left). The right panel highlights structural parameters in log scale most strongly linked to quiescence \citep[based on][]{Whitaker2017,Chen2020}. Star-forming galaxies grow primarily in mass with modest increases in size until crossing a redshift-dependent threshold in stellar density or velocity dispersion (\emph{black line}, left), beyond which quenching ensues. The central SMBH may at this point be sufficiently massive to drive feedback, though other density-regulated processes may contribute (\autoref{sec:quenching_mechanisms}). Post-quenching evolution proceeds mainly through dry mergers or minor accretion, driving steep size growth.  
    While stellar mass and S\'ersic index broadly correlate with sSFR, only high surface density or velocity dispersion uniquely predict quiescence, especially within the inner kiloparsec. Left panel adapted from \citet{vanDokkum2015} with permission of the AAS. Right panel adapted from \citet{Whitaker2017}. Abbreviation: sSFR, specific star-formation rate.}
    \label{fig:size-mass}
\end{figure}
The stellar mass-size relations of star-forming and quiescent galaxies evolve with time, with the latter population exhibiting far more dramatic size growth than expected from passive structural evolution alone \citep[e.g.,][]{Trujillo2007,vanderWel2014}. 
Some of this apparent growth reflects progenitor bias, whereby larger galaxies quenching later inflate the average size of the quiescent population without requiring dramatic structural evolution of earlier quenched systems \citep[e.g.,][]{vanderWel2009,Carollo2013,Fagioli2016,Belli2015}. 
At the same time, the central densities of quiescent galaxies decline only slightly with time \citep[e.g.,][]{Bezanson2009,Saracco2012,Tacchella2015,Whitaker2017,Mosleh2017}, 
indicating that most size growth occurs in the outskirts \citep[\textbf{\autoref{fig:hillsuess}a}; e.g.,][]{vanDokkum2010,Hill2017}. Building on this picture, while the evolution of quiescent light profiles is largely attributed to ex situ growth at $z<1$ \citep{vanDokkum2015,Matharu2019}, 
it has been suggested that the frequency of minor merging may not be sufficient to fully explain the observed size growth at higher redshift \citep[e.g.,][]{Newman2012,Matharu2020}. This tension has been alleviated by the fact that the mass-weighted sizes are somewhat smaller than the light-weighted properties \citep[\textbf{\autoref{fig:hillsuess}b};][]{Suess2019,Suess2020,Suess2022,Martorano2023}, although progenitor bias will still play a role as long as quenching does not perfectly coincide with structural transformation \citep{Ji2026_Size,Clausen2025}.

An intuitive way to synthesize these trends is to view galaxy evolution in the stellar mass-size plane, where galaxies follow distinct growth trajectories before and after quenching \citep[\textbf{\autoref{fig:size-mass}a}; adapted from][]{vanDokkum2015}. Star-forming galaxies primarily grow in mass with relatively modest changes in their structure, gradually building stellar density over time \citep[e.g.,][]{Barro2017}. Quenching typically occurs once galaxies cross a threshold in central density \citep[e.g.,][]{Voit2015}, which has been empirically established as one of the strongest predictors of low sSFR and quiescence \citep[\textbf{\autoref{fig:size-mass}b}; e.g.,][]{Franx2008,Williams2010,Fang2013,Mosleh2017,Chen2020}. 

The central density, quantified as a projected 3D stellar surface density within a kpc \citep[$\rho_1$ or $\Sigma_1$; e.g.,][]{Bezanson2009}, shows the tightest correlation with sSFR  (\textbf{\autoref{fig:size-mass}b}), comparable to stellar velocity dispersion \citep[$\sigma_{\star}$; e.g.,][]{Wake2012}.  This points to a direct or indirect (e.g., SMBH-mediated) causal link between quiescence and central density \citep{Chen2020}. This critical boundary is redshift-dependent \citep{vanDokkum2015,Voit2015} and may mark the stage at which the central SMBH has grown sufficiently to drive powerful outflows \citep[e.g.,][]{Terrazas2016,HeckmanBest2014}, although other mechanisms linked to stellar density could play a role (\autoref{sec:quenching_mechanisms}). After quenching, subsequent evolution is dominated by ex situ growth, either through major gas-poor (``dry'') mergers or minor merging and accretion in the outskirts. Depending on the mass-ratio \citep{Bezanson2009}, mergers generally increase galaxy size more than mass and steadily increase S\'ersic indices. While global stellar mass and light-profile shapes (S\'ersic index $n$)  also correlate with reduced sSFRs, the scatter is substantial at low sSFRs \citep{Whitaker2017}.  We emphasize that the observed evolution in S\'ersic index implies it should not be used as a reliable separator of late- and early-type galaxies at high redshift. Structural parameters themselves evolve, and classifications based solely on profile shape risk conflating genuine morphological transformation with demographic or observational effects. Projected axis-ratio distributions offer a more robust statistical probe of 3D shapes, although they can only be interpreted at the population level \citep[e.g.,][]{vanderWel2014b}.

\subsection{Fundamental Plane and Dynamical Evolution}
\label{subsec:dynamics}

Quiescent galaxies in the local Universe are generally dispersion-supported, though a clear kinematic dichotomy between fast and slow rotators has been established by integral-field surveys \citep[e.g.,][]{Cappellari2011,Emsellem2011} and comprehensively reviewed in \citet{Cappellari2016}.
Spatially integrated stellar dynamics, quantified by the stellar velocity dispersion) of these quiescent systems are often studied in conjunction with their structures (half-light radii and surface brightnesses) in the context of the fundamental plane \citep{Dressler1987,Djorgovski1987}, with the \citet{Faber1976} scaling as a conceptual precursor. 
The fundamental plane evolves strongly with redshift since $z\sim2$  \citep[e.g.,][]{Bezanson2013, vandeSande2014},  driven in part by M/L evolution due to passive aging, with additional contributions from changing inner dark matter fractions and late time structural evolution \citep[e.g.,][]{deGraaff2021,Stockmann2021}. 

However, the dynamical state of galaxies immediately after quenching, and whether the process destroys the ordered disk rotation of progenitors, has yet to be fully established. The main limitation stems from the difficulty of measuring spatially resolved stellar absorption-line kinematics in representative samples before ex situ mergers and accretion dilute existing ordered orbits. At intermediate redshifts, massive quiescent galaxies exhibit substantially more rotation than similarly massive ellipticals today \citep{Bezanson2018, vanHoudt2021}, 
but these systems are already billions of years post-quenching. Whereas gravitational lensing has revealed a handful of striking, magnified quiescent galaxies at $z\sim2$ with clear rotational gradients \citep[e.g.,][]{Toft2017, Newman2018}, JWST/NIRSpec sensitivity and resolution now enable larger samples that robustly establish the rotational support of distant massive quiescent galaxies \citep[e.g.,][]{Pascalau2026, Slob2025}. 

Together, these findings indicate that quenching does not universally destroy rotation, disfavoring highly disruptive merger-driven scenarios \citep[e.g.,][]{Robertson2006} or pure monolithic collapse. Instead, the emerging picture is that most high-redshift quiescent galaxies retain significant rotation at quenching, although rare early slow rotators \citep[e.g.,][]{Forrest2025} show that more disruptive pathways also occur. Collectively, this supports a framework in which local pressure-supported ellipticals arise through extended assembly histories that gradually transform rotating progenitors into slow rotators \citep{Naab2014}.

\section{QUENCHING IN AN EVOLUTIONARY FRAMEWORK}
\label{sec:framework}

This final section synthesizes results from the literature to outline testable predictions for distinguishing quenching pathways. As key observations remain incomplete, these predictions rely on simulations and should be treated as provisional until confirmed empirically.

\begin{figure}
    \centering
    \includegraphics[width=\textwidth]{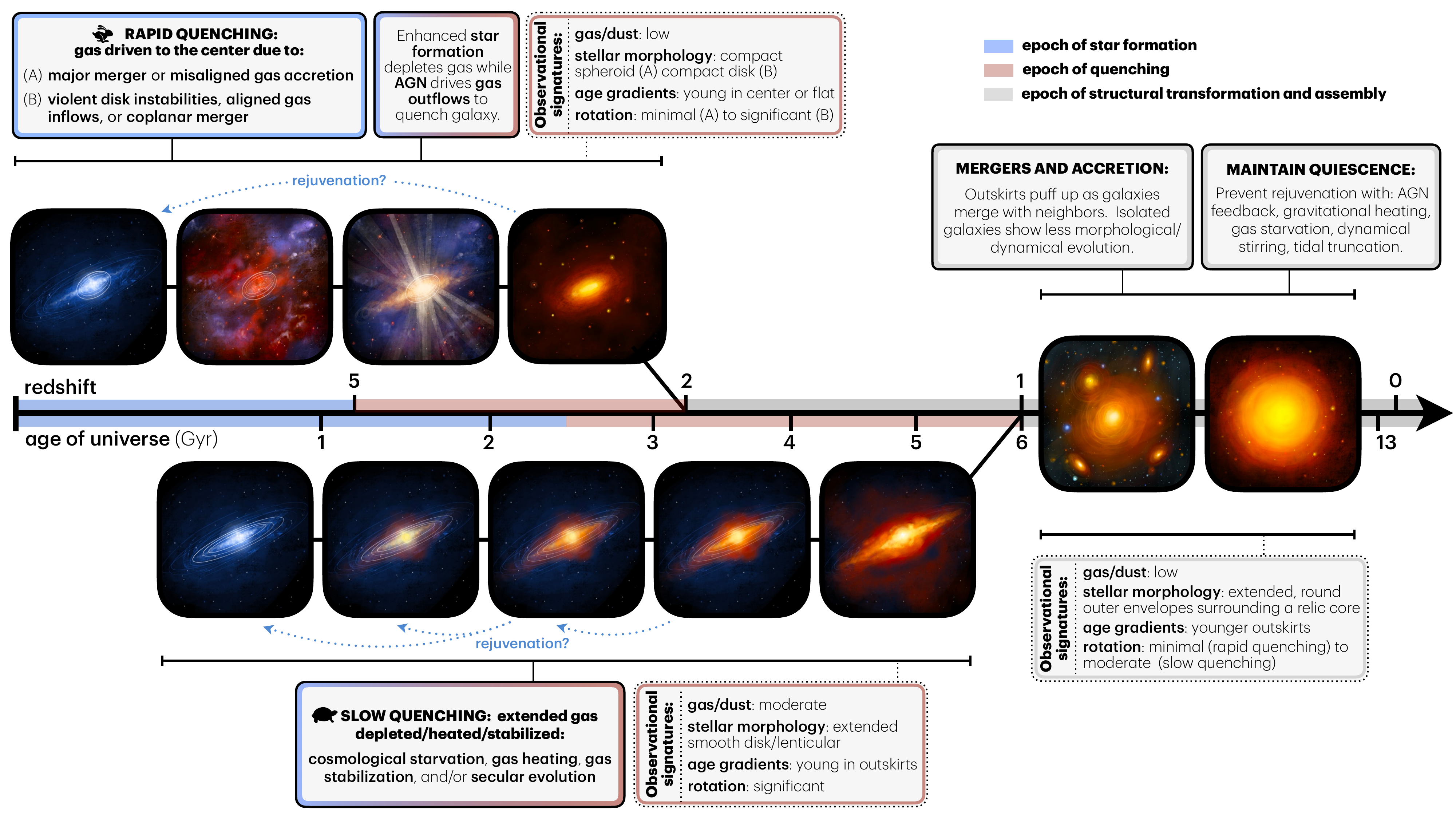}
    \caption{Evolutionary framework for quenching in galaxies with $M_\star \gtrsim 10^{10}$ M$_\odot$, contrasting rapid ($<$1 Gyr; \emph{top}) and slow ($>$1 Gyr; \emph{bottom}) modes. Pathways are divided into three main phases: star formation (\emph{blue}), quenching (\emph{maroon}), and structural assembly (\emph{grey}). Observational signatures for key phases are indicated, with cold gas shown as white contours. Massive galaxies typically form and quench rapidly at high redshift via central gas inflows expelled by AGN feedback, whereas slower quenching arises when gas remains extended in disks of intermediate-mass systems, where secular processes and gradual heating dominate. After quenching, galaxies evolve structurally through minor mergers and accretion. Illustrations by Carrington Bryan (University of Pittsburgh, 2025 Physics \& Astronomy Artist in Residence),  available at: \href{https://zenodo.org/records/17123339}{doi:10.5281/zenodo.17123339}. Abbreviation: AGN, active galactic nucleus. }
    \label{fig:evolutionary-framework}
\end{figure}

The wide diversity of quenching channels \citep[e.g.,][]{Park2022} can be distilled into two broad pathways (\textbf{\autoref{fig:evolutionary-framework}}): (1) rapid quenching whose onset is triggered by ejective, radiatively-efficient AGN feedback (\autoref{subsec:rapid}); and (2) slow quenching in which gas is gradually heated, depleted, and/or stabilized (\autoref{subsec:slow}). Regardless of the progenitor morphology (compact spheroid or a disk growing in central density), radiative/quasar-mode AGN can act rapidly when gas is centrally concentrated. Both theoretical models and observations now demonstrate that radiation pressure on dust and energy-conserving winds can couple strongly to dense central gas, driving multiphase outflows with mass-loading factors comparable to or exceeding SFRs \citep[see review by][]{Veilleux2020}.

Conversely, when the cold gas reservoir is more extended, the same radiatively-efficient AGN tends to heat and ionize the CGM rather than expelling all central fuel, regulating star formation more gently and promoting slow quenching \citep[e.g.,][]{Fabian2012}. 
In practice, slow pathways often coincide with secular evolution in which gas consumption, morphological stabilization, halo heating, and cosmological or environmental starvation gradually suppress inflows and cooling \citep[e.g.,][]{DekelBirnboim2006,Martig2009,Peng2015}.

Another key parameter influencing the evolutionary pathway is the halo mass, which regulates gas supply and thus indirectly governs SMBH fueling and energetics \citep{Iyer2025}. Via the stellar–halo connection, galaxies in the most massive halos tend to host higher stellar masses and more massive SMBHs \citep{Behroozi2013,KormendyHo2013}, 
and are overrepresented in overdensities \citep[e.g.,][]{Overzier2016}. Such systems more readily funnel gas to the nucleus, drive high-power AGN, and thus likely follow the rapid pathway (\autoref{subsec:rapid}), crossing a quenching boundary (\autoref{subsec:morphology})
earlier and more abruptly \citep[e.g.,][]{Fang2013,Chen2020}. Galaxies in intermediate-mass halos (i.e., more modest SMBH accretion rates with extended cold gas) more often evolve along slow, preventative tracks (\autoref{subsec:slow}). Although we focus on observables herein, these trends likely reflect underlying halo-scale physics (e.g., virial shock stability, CGM entropy, and AGN coupling efficiency).

\subsection{Early Rapid Quenching and Compact Core Formation}
\label{subsec:rapid}

At the onset of quenching, galaxies with compact gas configurations -- and therefore stronger coupling between gas and the black hole accretion rate (BHAR) -- are predicted to unbind their central gas reservoirs, quenching the system rapidly through radiatively efficient AGN feedback that drives strong outflows. The basic energetic requirement is that the coupled luminosity of the AGN exceeds the binding energy of the central gas, in which case the ejection can occur on short ($\sim10^{7-8}$ yr) timescales \citep[e.g.,][]{Fabian2012}.  
This centrally concentrated gas can either originate from a past major-merger event \citep[e.g.,][and references therein]{Somerville2015}, 
which typically disrupts stellar disks and transforms galaxies into compact spheroids, or from smooth gas accretion \citep[e.g.,][]{Dekel2009}. In rare cases with high gas fractions, disks may regrow rapidly after a merger \citep[e.g.,][]{Robertson2006}, potentially blurring this morphological distinction. Alternatively, violent disk instabilities can drive gas toward the center of a disky galaxy without necessarily destroying its morphology \citep[e.g.,][]{Dekel2009_instabilities}, 
offering a second pathway for building the dense central configurations required for rapid feedback-driven quenching. 

\begin{marginnote}
    \entry{Black hole accretion rate (BHAR)}{the rate at which mass is accreted onto a  SMBH.}
\end{marginnote}

Rapid quenching likely dominates at the highest redshifts, slow-quenched systems are more common later \citep[e.g.,][]{Peng2015}. This reflects basic timescale arguments and compact early morphologies that favor efficient quenching (\autoref{subsec:morphology}). However, current methods are biased to more easily identify rapidly quenched galaxies with post-starburst features (e.g., strong Balmer absorption, weak nebular emission), 
whereas slow quenching leaves subtler color and spectral changes that blend with dusty star-forming populations \citep[e.g.,][see also number density arguments in \SMsec{sec:supplement_demographics}]{Carnall2019}.  

Whether galaxies follow the canonical pathway with morphological transformation or preserve their disks, the resulting accumulation of central cold gas is the same. This triggers an extreme dust-enshrouded central starburst, during which the system may appear as a bright submillimeter galaxy \citep[e.g.,][]{Blain2002,Toft2014, Toft2017, Nelson2014}, and may mark the onset of central bulge growth \citep{Benton2024,Bodansky2026}. 
This compact gas reservoir is critical for enabling radiative-mode AGN feedback and rapid quenching, even in the absence of a classical bulge.  
Although simulations allow for a $\sim$Gyr delay between the dusty starburst and quenching \citep{Park2022}, this timescale is uncertain, particularly given that empirical studies suggest most massive galaxies quench before $z\sim1.5$ (\autoref{subsec:numberdensity}).  Regardless of this delay, once centrally concentrated cold gas has accumulated, it can readily fuel radiatively-efficient AGN activity, driving high BHARs capable of launching powerful neutral and ionized outflows \citep[e.g.,][]{DiMatteo2005}.

Recent JWST observations demonstrate that neutral outflows in massive quiescent galaxies may be sufficiently powerful to quench star formation \citep[e.g.,][]{Davies2024}, driving rapid quenching (\autoref{subsec:outflows}).  This efficiency reflects not only the large mass of gas expelled, but also the already depleted cold-gas fractions at the onset of quenching due to rapid consumption during the preceding compact starburst phase \citep[e.g.,][]{Zolotov2015,Tacchella2016}. 
A key testable prediction is therefore that compact quiescent remnants at cosmic noon should be both gas- and dust-poor. Their degree of rotational support then depends on how gas is funneled to the center: major mergers or misaligned accretion typically build dispersion-supported spheroids \citep[e.g.,][]{Hopkins2009_cores}, 
whereas violent disk instabilities, aligned gas accretion, or coplanar mergers can preserve angular momentum and produce rotating compact disks \citep[e.g.,][]{DekelBurkert2014}.
A critical next step is to link these predictions directly: galaxies showing evidence for recent compact starbursts and depleted gas/dust reservoirs should also display distinct stellar kinematics, with dispersion-dominated systems tracing merger-driven quenching and rotating systems pointing to disk-instability or aligned-accretion origins.

Prior to quenching, massive galaxies may exhibit constant, bursty, or even single-burst SFHs, reflecting sustained, episodic, or rapid one-time star formation phases \citep{Tacchella2016}. While rejuvenation events are possible, once galaxies are quenched via a strong ejective event (quasar-mode AGN), they almost never rejuvenate in a significant way \citep[e.g.,][]{Nelson2018}. 
Thus, if quenching is permanent, sustaining many Gyrs of passive evolution requires ongoing maintenance energy input, often attributed to radiatively inefficient AGN feedback \citep[see reviews:][]{Fabian2012,HeckmanBest2014}, with additional support from gravitational heating in hot halos. Such heating may also drive irreversible dust loss through thermal sputtering, producing extreme $\delta_{\mathrm{GDR}}$ values (\autoref{subsubsec:colddust}). This framework yields testable predictions, in which permanent and rapidly quenched galaxies (compact, high $\Sigma_1$ systems) should display high $\delta_{GDR}$ ratios that rise with stellar age and low-level radio/X-ray AGN signatures indicative of maintenance-mode heating. 

\subsection{Slow Quenching}
\label{subsec:slow}

Contrary to the scenarios presented in \autoref{subsec:rapid}, traditional models link radiative-mode (`quasar') AGN feedback to a slow regulation of star-formation \citep{Fabian2012}, acting over Gyr timescales in extended gas disks.
In this scenario, intermediate-mass, rotationally supported star-forming galaxies gradually exhaust their gas supply via secular processes \citep{Peng2015}.  At the onset of quenching, such systems are predicted to retain higher gas fractions overall, with gas distributions that more extended than starlight \citep{Park2022}.  The longer timescales may also permit dust regrowth \citep{Lorenzon2025}, raising the question of whether slow-quenched galaxies show more typical $\delta_{\mathrm{GDR}}$ values and detectable dust reservoirs.  Their quiescent descendants should preserve some ordered rotation, though reduced by disk heating and minor mergers or accretion (see \autoref{subsec:mergers}). 

Observationally, direct evidence for slow quenching remains sparse compared to the rapidly quenched population, in part because current spectroscopic and photometric selection techniques are biased toward systems with strong Balmer absorption and compact morphologies \citep[e.g.,][]{Carnall2019}. Nonetheless, several studies have identified candidate slow-quenching galaxies with extended disks, moderate gas reservoirs, and residual dust attenuation at $z\sim1$–2 \citep[e.g.,][]{Belli2019, Carnall2020}.
These systems often have lower central densities and more extended morphologies than compact post-starbursts, consistent with gradual gas exhaustion and heating rather than abrupt removal. 

A natural consequence of this secular pathway is that rejuvenation episodes are more likely at intermediate stellar masses ($M_\star \sim 10^{10}$–$10^{11} M_\odot$), where galaxies retain extended morphologies and can re-accrete gas \citep[e.g.,][]{Nelson2018}.  
Simulations and spectral reconstructions suggest such events are relatively common, although typically modest and occurring in the outskirts at 2–4$\times$ the effective radius \citep[e.g.,][]{Iyer2019,RemusKimmig2025}. Current selection techniques likely miss this low-level rejuvenation (\autoref{sec:selection}), even if it is common in reality. Observationally, rejuvenation may manifest as UV-bright outer disks, age gradients, or secondary Balmer features in integrated spectra. By contrast, the most massive, compact quiescent galaxies with low gas fractions are unlikely to rejuvenate, providing a clear discriminant between rapid and slow quenching channels.

While not the focus of this review, environmental processes may play an outsized role at intermediate to low stellar masses, where galaxies are more vulnerable to external influences. In dense environments such as groups and clusters, satellites quench at higher rates than centrals of the same stellar mass, both locally and out to $z\sim1$ \citep[e.g.,][]{YPeng2010,Wetzel2013}. Galaxies in X-ray–bright clusters also show severe depletion of CGM HI \citep{Burchett2018}, likely driven by stripping, heating, and ionization by the hot, diffuse intracluster medium (ICM). Multiple mechanisms contribute (\autoref{sec:quenching_mechanisms}), typically acting on Gyr timescales, with quenching efficiencies depending sensitively on both galaxy and host halo mass \citep{Donnari2021,ScholzDiaz2024}. Centrals, by contrast, quench mainly through internal processes, though their growth and suppression can be accelerated in dense environments and protoclusters \citep[e.g.,][]{Papovich2012,Strazzullo2013}.

\begin{marginnote}
    \entry{Intracluster medium (ICM)}{the hot, diffuse gas permeating galaxy clusters.}
\end{marginnote}

Despite the confirmed existence of low-mass quiescent galaxies at $z\sim1-3$ \citep[e.g.,][]{Santini2022,Cutler2024},
current evidence suggests that environmental processes may not dominate at $z\gtrsim1$ \citep[e.g.,][]{Balogh2016,Werner2022,Pan2025}, and many such systems are only temporarily quenched amid bursty SFHs \citep{Mintz2026}.  Recent discoveries extend this picture to a new class of low-mass 'napping' galaxies at $z\sim5$–8 \citep[e.g.,][]{Looser2023,Trussler2025}. While our focus has been on massive systems, where quenching is rarely reversible, these results highlight the need to move beyond a simple fast/slow framework for low-mass populations, accommodating extremely short ($<40$ Myr) quiescent phases now revealed by JWST.
More broadly, wide-field surveys increasingly indicate that environment often accelerates quenching that would have occurred anyway, leaving open key questions about the relative roles of nature and nurture in shaping the quenched population across cosmic time\footnote{A more detailed discussion of environmental processes is beyond the scope of this review; nonetheless, wide-area spectroscopic surveys pushing to higher redshifts [e.g., PFS (Prime Focus Spectrograph on Subaru), MOONS, (Multi Object Optical and Near-infrared Spectrograph on the Very Large Telescope), Euclid, the Nancy Grace Roman Space Telescope] should soon provide a more unified picture of their role in galaxy quenching.}.

\subsection{Maintaining Permanent Quiescence}

Rejuvenation appears vanishingly rare in the most massive quenched galaxies \citep[e.g.,][]{Chauke2019,Akhshik2021}, but both observations \citep[e.g.,][]{Peng2015,Belli2019} and simulations \citep[e.g.,][]{Tacchella2016,RemusKimmig2025} suggest that it becomes increasingly plausible at intermediate masses, typically manifesting as mild, low-level episodes in the outskirts.
Consequently, quiescence has long been considered effectively permanent in the most massive galaxies, maintained by mechanisms that continually heat the gas and prevent collapse.  That said, simulations paint a more nuanced picture for the recently discovered quenched populations at $z>3$, arguably the first generation of massive quiescent galaxies. Tracing these systems forward in the Magneticum simulation, \citet{RemusKimmig2025} find that roughly half remain quenched or are accreted into larger structures by $z=2$, while the rest rejuvenate -- mostly weakly in their outskirts, but with a notable minority experiencing substantial episodes that persist to $z\sim2$ \citep[see also][]{Wellons2016}.

Regardless of whether quenching is rapid or slow, hot halo gas can sustain low-BHAR, radiatively inefficient (`maintenance mode') AGN feedback over long timescales \citep[e.g.,][]{Croton2006,HeckmanBest2014}.
As discussed in \autoref{sec:quenching_mechanisms}, episodic mechanical energy injection from AGN-driven jets and outflows, together with 
gravitational processes such as virial shocks and dynamical friction, continually heats the CGM and suppresses cooling. These long-timescale mechanisms likely act in concert to keep the halo hot and render quiescence effectively permanent in most massive galaxies.

\subsection{Late-stage Growth through Mergers and Accretion}
\label{subsec:mergers}

Massive quiescent galaxies, while largely inactive in forming new stars, continue to grow significantly in stellar mass and size after quenching \citep[e.g.,][]{vanderWel2014}, primarily through the accretion of satellite galaxies via minor mergers \citep[e.g.][]{Newman2012,Suess2019,Suess2023}. Observational studies demonstrate that while the dense stellar cores of present-day ellipticals were largely in place by $z\sim2$ \citep{Bezanson2009}, these systems can double in size from $z\sim2$ to the present day without a significant corresponding increase in central stellar density \citep[e.g.,][]{vanDokkum2010,Hill2017}. This scenario is consistent with ex situ growth through the deposition of stars in the galaxy outskirts \citep[e.g.,][]{vanDokkum2015}, producing extended, low-surface-brightness envelopes around compact cores. This ``two-phase’’ formation picture \citep[e.g.,][]{Oser2010} reconciles the early, rapid in situ star formation of massive galaxies with their later structural assembly; \citet{Thomas2005} established that the bulk of stars in massive ellipticals form remarkably early, but the galaxies themselves only reach their present-day sizes and structures through prolonged accretion-driven evolution.

Complementary kinematic studies of quiescent galaxies at high redshift \citep{Toft2017,Newman2018,Bezanson2018} compared with local samples \citep[e.g.,][]{Emsellem2011,Cappellari2016} support the same conclusion. Minor mergers are predicted to leave clear dynamical signatures (heating and puffing up of stellar orbits, reduced net rotation, and building more complex velocity structures in the outskirts), while largely preserving the compact, dispersion-supported cores \citep[e.g.,][]{Naab2014}. Thus, while many high-redshift quiescent galaxies show significant rotational support, billions of years of minor mergers and accretion events naturally explain their transformation into the predominantly dispersion-dominated systems seen today (i.e., where the two branches merge on the right side of \textbf{\autoref{fig:evolutionary-framework}}; see \autoref{subsec:rapid} and \autoref{subsec:slow}).

\section{SUMMARY AND OUTLOOK}
\label{sec:summary} 

This article reviews the formation and evolution of massive quiescent galaxies, emphasizing how stellar populations, gas and dust content, structures, and dynamics constrain the physical processes that drive and maintain quenching.  The emphasis is primarily empirical, focusing on insights gained from the selection and inferred physical properties of massive quiescent galaxies at $z>1$, although we also draw on complementary work at lower redshift and theoretical modeling. Taken together, the following summary points synthesize the current state of knowledge, while also outlining open questions and speculative ideas that will require qualitatively new data to confirm.

Individual strongly lensed systems provide a fertile playground for discovery, offering a glimpse of what may become routine with larger samples in the future. A prime example is MRG-M0138 \citep{Newman2018}, which continues to yield groundbreaking results. This massive quiescent galaxy at $z=2$ has hosted two multiply-imaged, long-time-delay Type Ia SNe \citep[e.g.,][]{Peirel24},
providing valuable cosmological constraints and insights into galaxy evolution. It provided some of the earliest dynamical evidence that high redshift quiescent galaxies can be rotation-dominated \citep[\autoref{subsec:dynamics};][]{Newman2018}, the first resolved metallicity and age gradients consistent with extended, slow SFHs \citep[\autoref{subsec:slow};][and earlier work therein]{Akhshik2023,Jafariyazani2025}, and the first direct detection of compact cold dust in a quenched system outside the nearby universe \citep{Whitaker2021a}. Most recently, \citet{Newman2025} weighed its SMBH, a measurement reminiscent of those in nearby galaxies like M87 but now billions of years earlier, implying that SMBHs were already established at cosmic noon. Together, these discoveries show how a single strongly lensed system can transform our view of quiescent galaxies, and highlight a clear future direction: expanding such detailed, multi-faceted experiments beyond rare lensed laboratories to the general quiescent population with next-generation facilities.

\begin{summary}[SUMMARY POINTS]
\begin{enumerate}
\item Robust identification of quiescent galaxies requires not only careful diagnostics but also high-quality multi-wavelength photometry and accurate redshift estimates, since contamination often arises from photometric scatter, dusty star-forming interlopers, and template mismatches. Rest-frame color methods (e.g., UVJ) have been invaluable for building statistical samples across cosmic time, but they are ultimately proxies and remain vulnerable to contamination, especially at $z>3$. With advances in Bayesian SED modeling, redshift-dependent sSFR–based selections (e.g., $>2$–$3\sigma$ below the star-forming main sequence) now provide a more physically motivated and flexible approach to defining quiescence. Thus, while quiescent galaxies can be identified with empirical color selections, an evolving sSFR threshold may better control contamination as SPS modeling continues to mature. Ultimately, photometric techniques must be complemented by spectroscopy to robustly constrain stellar ages, metallicities, and kinematics -- an effort that will be transformed by upcoming wide-area spectroscopic surveys.

\item High-redshift quiescent galaxies exhibit rapid SFHs, reaching near-solar or even super-solar metallicities within a few hundred Myr and displaying elevated [$\alpha$/Fe] ratios consistent with quenching timescales $<$1 Gyr. These chemical signatures indicate intense, brief starbursts followed by early truncation, particularly in the most massive systems. By contrast, intermediate-mass quiescent galaxies show more extended SFHs, lower metallicities, and greater diversity in [$\alpha$/Fe], reflecting downsizing and the influence of slower quenching channels. Despite these broad trends, significant scatter remains, shaped by merger histories, minor accretion, and environmental context. Robust interpretation is hampered by current limitations of SPS models, which often assume solar-scaled abundance ratios, and by uncertainties in the stellar IMF, both of which directly affect derived ages, metallicities, and enrichment timescales. Overcoming these systematics will be critical for turning abundance patterns into precise constraints on quenching pathways.

\item Quiescent galaxies exhibit diverse multiphase gas and dust properties. While many show low cold-gas fractions and dust-poor conditions, others, particularly post-starburst systems, retain significant molecular reservoirs and dust masses. Theory predicts that gas-to-dust ratios span a wide range in quiescent systems, with early empirical support; if correct, the standard calibrations used for star-forming galaxies break down, becoming sensitive to quenching timescale and pathway. Neutral outflows provide strong evidence for ejective AGN feedback driving rapid quenching, while it remains unclear whether slow-quenching systems preserve moderate gas and dust reservoirs with reduced star formation efficiencies and occasional rejuvenation, particularly at intermediate masses. Disentangling the roles of outflows, heating, and residual accretion across quenching pathways is a key priority for future work.  

\item Quiescent galaxies follow distinct time-evolving scaling relations, linking stellar mass, sSFR, size, surface density, and velocity dispersion; central density emerges as the strongest predictor of quiescence. Their sharp rise in number density at $z\sim2$–3 and extreme early compactness point to rapid, efficient quenching. Subsequent size growth is largely ex situ, consistent with two-phase formation, while progenitor bias and evolving population demographics further shape their evolution.   

\item A unifying framework distinguishes two primary quenching channels: (1) fast quenching, where compact, gas-rich systems undergo AGN-driven ejective outflows that halt star formation within $\lesssim$1 Gyr; and (2) slow quenching, where extended gaseous disks evolve through secular gas exhaustion, virial shock heating, or episodic AGN heating of the CGM over Gyr timescales. These pathways bookend a continuum of possibilities, yielding clear predictions for ISM content, gas-to-dust ratios, and dynamics, which in turn provide testable observational signatures.
\end{enumerate}
\end{summary}

\begin{issues}[FUTURE ISSUES]
\label{subsec:future}
\noindent \textbf{Improving the Models:} 
\vspace{-0.1cm}
\begin{enumerate}
\item SPS modeling: Break degeneracies between age, metallicity, and dust by incorporating non-solar abundance patterns, more accurate stellar evolution phases (e.g., TP-AGB, HB), and improved priors for flexible nonparametric SFHs.  
\item IMF and chemical abundance variations: Establish whether IMF variations and/or $\alpha$-enhancement biases drive current tensions in stellar mass and SFH inferences.
\item Simulations: Refine treatment of AGN feedback (radiative and mechanical) and dust physics, and better constrain rejuvenation frequencies, merger-driven disk survival, and the balance between fast and slow quenching channels. 
\end{enumerate}

\noindent\textbf{Expanding the Scope:}
\vspace{-0.1cm}
\begin{enumerate}
\item High-redshift regime ($z>3$): Systematically constrain the earliest quiescent galaxies with JWST spectroscopy to more robustly determine quenching timescales and the outflow energetics related to gas removal.
\item Gas and dust: Expand studies to better understand the origin of exotic gas-to-dust ratios by leveraging larger samples across cosmic time with better data quality, also leveraging PAH ratios to constrain the grain size distributions and spatial morphology, and confirm tentative correlations with quenching timescales. 
\item SFHs of intermediate-mass galaxies: Extend high redshift studies beyond the most massive systems to capture the diversity of SFHs, rejuvenation likelihoods, and slow-quenching pathways.
\item Environment and halos: Directly link quenching outcomes to halo mass and environment through joint galaxy–CGM/ICM surveys, probing virial shock heating, starvation, and environmental quenching across mass scales.
\item Resolved studies: Map gradients in stellar populations, dust, and gas at sub-kpc scales with JWST, ALMA, and future generation large aperture telescopes to distinguish between compaction-driven vs. secularly evolving pathways and search for residual star formation in outskirts as a sign of rejuvenation.  
\end{enumerate}

\noindent\textbf{New Directions}
\vspace{-0.1cm}
\begin{enumerate}
\item Multi-phase ISM tracers: Synthesize (expanded) CO studies with neutral and ionized gas diagnostics, testing ejective vs. preventative quenching signatures. 
\item Star formation diagnostics: Future studies of rest-NUV, PAHs, and Pa-$\alpha$ at high redshift are critical to rule out  hidden optically thick star formation, especially in the cores \citep[e.g., analog studies to][]{Smercina2018} and establish when UV upturns appear \citep[e.g.,][]{LaCras2016}. 
\item Dynamical tests: Build statistical quiescent samples with stellar kinematics for a wide range in age to assess whether quenching preserves or destroys rotation.
\item Globular cluster formation: Leverage cluster populations in quiescent galaxies $z>1$ to understand the formation channels, disentangling in situ formation from ex situ accretion events, and their unique chemical signatures  \citep[e.g.,][]{Whitaker2026}.
\item Constrain black hole masses: Improved telescope sensitivity and spectroscopic capabilities coupled with enhancements from strong gravitational lensing afford unique opportunities to weigh SMBHs at $z\sim2$ \citep[e.g.,][]{Newman2025}.
\end{enumerate} 
\end{issues}

\section*{DISCLOSURE STATEMENT}
The authors are not aware of any affiliations, memberships, funding, or financial holdings that might be perceived as affecting the objectivity of this review. 

\section*{ACKNOWLEDGMENTS}
We foremost acknowledge that many impactful contributions from the community could not be cited due to space constraints; omissions are unintentional, difficult choices were necessary in distilling decades of progress alongside rapid recent developments.
We gratefully acknowledge many colleagues for their conversations and thoughtful feedback that enriched this review: J. Antwi-Danso, S. Belli, A. Beverage, R. Feldmann, M. Hamadouche, M. Kriek, E. Lambrides, J. Leja, A. Long, D. Narayanan, D. Setton, J. Spilker, S. Toft, and T. Tripp.  
We are indebted to S. Belli, A. Beverage, K. Ito, C. Lovell, I. McConachie, D. Narayanan, and Y. Zhang for generously sharing materials and inspiration for figures, and R. Chandar, J. Greene, L. Kewley, and A. Pope for their mentorship. K.E.W. is grateful for discussions with D. Donevski, K. Iyer, S. Faber, A. Newman, and C. Maraston, which shaped key parts of this review. We thank our science editor J. Wu for their impactful contributions, and R. Lowe-Webb for her support through the process. We gratefully acknowledge funding and resources from UMass Amherst, supporting K.E.W.’s efforts during her sabbatical. This work builds on research supported by federal funding agencies, whose sustained investment in fundamental science has enabled the remarkable progress synthesized herein. Data products presented here were retrieved from the Dawn JWST Archive (DJA), an initiative of the Cosmic Dawn Center, which is funded by the Danish National Research Foundation under grant No. 140. AI tools were used during manuscript preparation: ChatGPT-5 (2025) for language refinement and Claude Opus 4.6 (2026) for editorial review and reference checks; all substantive content was written and verified by the authors. Finally, writing this review has been its own kind of era -- with both clarity and complications -- as we tried to hold on to golden threads of insight. We hope that this synthesis reflects a balance of kindness and cleverness. We are profoundly grateful to our families for their patience and support, helping us through long deadlines and a cruel summer with steady encouragement.

%





\end{bibunit}



\renewcommand{\thesection}{S\arabic{section}}
\renewcommand{\theHsection}{S\arabic{section}}
\renewcommand{\theHsubsection}{S\arabic{section}.\arabic{subsection}}
\renewcommand{\theHsubsubsection}{S\arabic{section}.\arabic{subsection}.\arabic{subsubsection}}
\setcounter{section}{0}
\renewcommand{\thefigure}{S\arabic{figure}}
\renewcommand{\theHfigure}{S\arabic{figure}}
\setcounter{figure}{0}
\renewcommand{\thetable}{S\arabic{table}}
\renewcommand{\theHtable}{S\arabic{table}}
\setcounter{table}{0}
\renewcommand{\figureautorefname}{Supplemental Figure}
\renewcommand{\tableautorefname}{Supplemental Table}
\clearpage
\begin{bibunit}
\includepdf[pages=1]{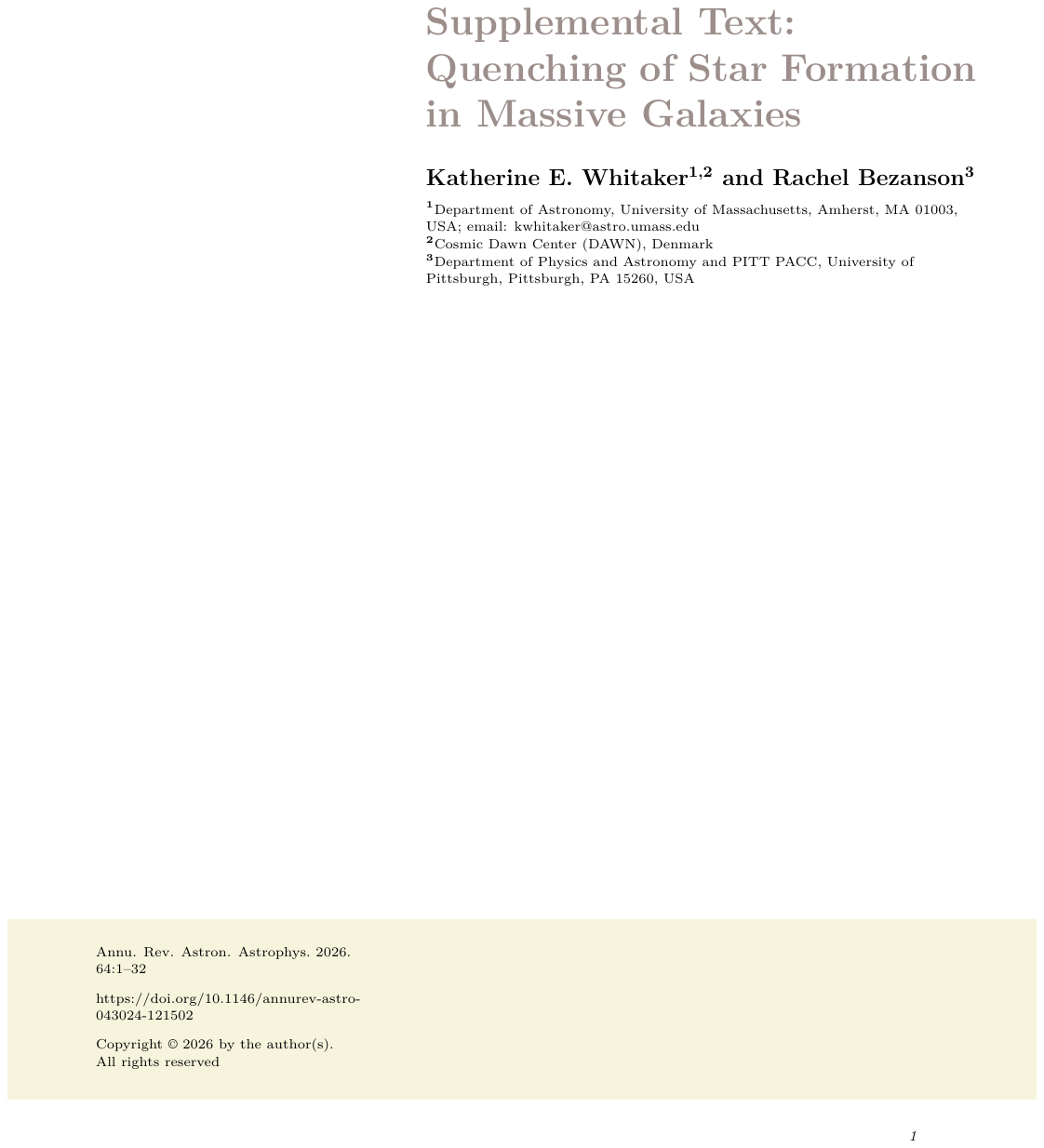}
%
%
%


\def\aap{A\&A}
\def\apj{ApJ}
\def\aapr{A\&A Rev.}
\def\apjl{ApJ}
\def\mnras{MNRAS}
\def\araa{ARA\&A}
\def\aj{AJ}
\def\qjras{QJRAS}
\def\physrep{Phys. Rep.}
\def\prd{Phys. Rev. D}
\def\nat{Nat}
\def\aaps{A\&A Supp.}
\def\apss{Ap\&SS}
\def\apjs{ApJS}
\def\pasj{PASJ}
\def\pasp{PASP}

\setcounter{secnumdepth}{4}

  







\section{OVERVIEW OF QUENCHING MECHANISMS}
\label{sec:supplement_quenching_mechanisms}

The quenching of star formation ultimately depends on the fate of the cold gas supply within and around a galaxy.  In this supplemental section, we provide extended descriptions of the individual physical processes that can halt or maintain the suppression of star formation, expanding upon the overview presented in \textbf{\mainfig{fig:quenching_mechanisms}} of the main article.

Following the framework of \citet{ManBelli2018}, we distinguish between mechanisms that initiate quenching (e.g., by removing or heating cold gas) and those that maintain quiescence over cosmological timescales.  For clarity, we group quenching mechanisms by the dominant physical scale on which they operate: cosmological ($>$1 Mpc), circumgalactic ($\sim$100–500 kpc), and galactic ($<$100 kpc). Each subsection summarizes the underlying physics, characteristic timescales, and notes supporting observational evidence, outlining how galaxies exhaust, heat, or lose their star-forming fuel (cold molecular gas) across scales. Throughout this supplemental section, the quenching timescale (i.e., the third column in \textbf{\mainfig{fig:quenching_mechanisms}} of the main article) 
refers to the characteristic time over which star formation declines once a quenching mechanism operates, typically corresponding to depletion or removal of the galaxy’s cold gas reservoir rather than the triggering event itself.

\subsection{Quenching Mechanisms on Cosmological Scales of $>$1 Mpc}
\label{subsec:quenching_cosmoscales}
\subsubsection{Prevention of gas accretion}
\label{subsubsec:prevention_cosmoscales}

When the accretion of pristine gas from the surrounding large-scale environment onto a dark matter halo is cut off or greatly reduced, star formation in the galaxy will decline over time. 
This effect, known as \textbf{cosmological starvation}, occurs when the supply of fresh gas from the cosmic web diminishes as halos transition from phases of rapid growth to slower accretion \citep[e.g.,][]{Feldmann2015,Feldmann2016}.
The relevant scales for this process are typically of order a few virial radii ($\gtrsim$500 
\begin{marginnote}[]
\entry{Cosmological starvation}{Reduced cosmological gas inflow onto halos, limiting the supply of fresh star-forming fuel.}
\end{marginnote}
kpc), where gas accretion onto the halo is regulated by the surrounding large-scale structure. As inflows weaken, the galaxy must increasingly rely on its existing cold-gas reservoir. The resulting decline in star formation typically begins over roughly a halo dynamical time (several hundred million years depending on redshift). The full quenching of the galaxy generally unfolds over several dynamical times as the cold-gas reservoir is gradually exhausted. Stellar mass loss from evolved stars can recycle gas into the interstellar medium (ISM) and temporarily sustain star formation, but this internal supply is generally insufficient to maintain star formation indefinitely once external accretion is suppressed. Consequently, star formation gradually declines as the available fuel is consumed. Cosmological starvation therefore suppresses star formation by limiting the supply of fresh gas, although additional heating or feedback processes are likely required to fully quench massive galaxies.

In dense environments such as clusters, a similar process of \textbf{environmental starvation}\footnote{also commonly referred to by the more violent term ``strangulation''} can occur \citep[e.g.][]{Larson1980,Balogh2000,Peng2015}.  
In this scenario, a galaxy is cut off from its external gas supply after infall into a more massive halo, preventing further accretion of star-forming fuel. In some cases, galaxies may already 
\begin{marginnote}[]
\entry{Environmental starvation}{Reduced gas inflow in dense environments due to interaction with the surrounding hot intracluster medium.}
\end{marginnote}
begin to experience reduced gas accretion or environmental preprocessing (see \SMsup{subsubsec:gas_removal_cgmscales}) within smaller groups prior to cluster infall \citep[e.g.,][]{McGee2009}. At $z<1$, environmental starvation is often invoked to explain the gradual aging of satellite galaxies in dense environments, inferred from colors and spectroscopic age diagnostics (e.g., D$_n$4000). The resulting evolution can proceed over a wide range of timescales, from several hundred million years \citep{Feldmann2011} to several billion years \citep{vanDenBosch2008,Wetzel2013,Peng2015,Maier2016}, depending on halo mass, orbital history, and the efficiency of gas removal processes. This slow evolution 
at late times is consistent with a steady exhaustion of their remaining gas once fresh inflows are suppressed \citep{Baxter2022,McNab2021}, leading to a gradual and often extended decline in star formation.
Massive group and cluster centrals can host faint extended X-ray halos \citep[e.g.,][]{Gobat2011}, indicating that hot intragroup or intracluster gas is already in place in at least some systems by $z\sim2$ and may further inhibit cooling and accretion.  Together, these signatures point to a slow-acting environment-driven quenching channel that can lead to a gradual decline in star formation once fresh inflows are suppressed. Large-scale quenching thus reflects a gradual but protracted ramp-down of star formation.

\subsubsection{Gas heating}
\label{subsubsec:gas_heating_cosmoscales}

Infalling gas that makes its way further into a galaxy can be shock-heated to high temperatures, preventing it from cooling and condensing onto the galaxy \citep{WhiteRees1978}.
Known as \textbf{gravitational heating}, this broad term encompasses heating from accretion, compression, and shocks operating across scales from galactic and circumgalactic environments to the collapse of large-scale structure. 
At cosmological scales, 
\begin{marginnote}[]
\entry{Gravitational heating}{Heating of gas through shocks and compression as gravitational energy is released during structure formation.}
\end{marginnote}
gas is heated along filaments by shocks that arise during structure formation, one of the most energetic manifestations of gravitational heating in the Universe \citep{CenOstriker1999,Ryu2003}. 
This gravitational energy conversion continues down to smaller scales, from the heating of the CGM ($<$500 kpc) by infalling satellites  \citep{KhochfarOstriker2008} to the accretion of cold gas clumps within the inner $\sim$10 kpc \citep[\SMsup{subsubsec:gas_heating_galaxyscale};][]{DekelBirnboim2008}. Heating from large-scale collapse unfolds over roughly a halo dynamical time (of order 1 Gyr, depending on redshift), gradually establishing the hot gas environment that suppresses further cold inflows \citep[e.g.,][]{DekelBirnboim2006,FaucherGiguere2011,vandeVoort2011,Nelson2019}.  The principal role of gravitational heating is to maintain the hot gas reservoir in the CGM, intracluster medium (ICM), and beyond. A well-known example on cluster scales (a few Mpc) is the shock-heating of infalling gas through large-scale accretion shocks \citep{Voit2005}.  The gas can be heated up to 10$^{8}$ K, forming a pressure-supported halo in which cooling times are long compared to the halo dynamical time, thereby suppressing the condensation of gas onto the galaxy. 
At $z>1$, emerging hot (10$^{7}$–10$^{8}$ K) X-ray halos and Sunyaev–Zel’dovich (SZ) detections suggest the onset of hot ICM assembly around massive quiescent galaxies \citep[e.g.,][]{Gobat2011,Andreon2016,Chiu2018}. However, the hot ICM is not fully developed or in equilibrium until later times ($z\lesssim1.5$), at which point environmental quenching mechanisms become increasingly effective \citep{Sarazin1986} (\SMsup{subsubsec:gas_removal_cgmscales}).

\subsection{Quenching Mechanisms on Circumgalactic Scales of 100-500 kpc}
\label{subsec:quenching_cgmscales}

\subsubsection{Gas heating}
\label{subsubsec:gas_heating_cgmscales}

Gas heating can both initiate quenching and help maintain quiescence, depending on the relevant spatial and temporal scales. A key mechanism is \textbf{virial shock heating}, in which infalling gas is shock-heated near the virial radius to the halo’s virial temperature ($\sim$100–300 kpc for massive halos), preventing it from cooling and condensing onto the galaxy \citep{WhiteRees1978,DekelBirnboim2006}. 
Stable virial shocks develop 
\begin{marginnote}[]
\entry{Virial shock heating}{Shock heating of infalling gas near the virial radius to the halo’s virial temperature.}
\end{marginnote}
when the cooling time of infalling gas exceeds its compression time, producing hot gaseous halos that inhibit efficient cooling and suppress cold-mode accretion \citep{BirnboimDekel2003,Keres2005}.
In simulations, this transition occurs once halos surpass a critical mass of $\gtrsim$10$^{12}$ M$_\odot$, marking the transition to a regime where hot-mode accretion becomes dominant \citep{DekelBirnboim2006,Keres2005}, although cold filamentary streams may still penetrate the halo at high redshift \citep{Dekel2009}.
Gas accreting along cosmic filaments can be shock-heated to ${\sim} 10^{6}–10^{7}$ K upon encountering the virial shock near the halo boundary, at which point the cooling time becomes long compared to the halo dynamical time, suppressing the supply of cold gas to the galaxy. In the hot-mode regime, gas first enters a quasi-static halo and cools only slowly. The transition from the onset of hot-mode accretion to full quenching typically occurs of order 1–2 billion years, after which residual star formation fades as the hot halo suppresses further cold accretion onto the galaxy.

Indirect observational support for halo mass-dependent quenching comes from the sharp increase in the quenched fraction above log(M/M$_{\odot}$)$>$10.5 at $z\lesssim2$ \citep[e.g.,][]{Muzzin2013}, consistent with expectations for virial shock heating.  More direct evidence is provided by diffuse X-ray detections of extended hot halos around massive quiescent galaxies \citep[e.g.,][]{Gobat2011,Andreon2016}. However, diffuse X-ray and thermal SZ detections of hot halos are consistent with several additional quenching processes that maintain virialized gas, including gravitational heating (\SMsup{subsubsec:gas_heating_cosmoscales}) and the suppression of cold accretion at cosmological scales (\SMsup{subsubsec:prevention_cosmoscales}). 

As the majority of galaxies above the $\gtrsim10^{12} M_\odot$ dark matter halo mass threshold are quenched \citep{Behroozi2019}, virial shock heating is often cited as an important trigger of quenching in massive systems. More broadly, this falls under ‘mass quenching’ \citep{YPeng2010}, in which massive halos maintain a quasi-static, virialized hot gas reservoir with long radiative cooling times that strongly reduce cold gas inflow. 
We emphasize, however, that this umbrella includes a variety of processes that also scale with galaxy or halo mass, including AGN feedback (\SMsup{subsubsec:gas_heating_galaxyscale} and \ref{subsubsec:gas_removal_galaxyscales}), morphological quenching (\SMsup{subsubsec:gas_stabilization_galaxyscales}), and cosmological starvation (\SMsup{subsubsec:prevention_cosmoscales}). Once gas passes a virial shock it joins a hot, quasi-static halo and can in principle cool and accrete onto the galaxy if its cooling time becomes sufficiently short.  Virial shock heating therefore contributes to the onset of quenching but does not by itself guarantee its long-term maintenance, which likely requires additional heating processes.

The temperature gradient between the hot ICM and cooler CGM or ISM gas drives \textbf{thermal conduction}, transferring heat via particle collisions \citep{Cowie1977}. Conduction alone does not expel gas, but over gigayear timescales it can gradually raise the temperature of the surrounding medium and alter its thermodynamic state. If sufficiently efficient, conduction can lead to \textbf{thermal evaporation}, slowly absorbing cooler ISM gas into the hot halo \citep{CowieMcKee1977}. Observationally, smooth X-ray temperature and entropy profiles, together with a relative lack of multiphase gas in non-cool-core cluster cores \citep[e.g.,][]{Cavagnolo2008,Hudson2010}, point to maintenance heating via conduction. These processes typically act on long timescales, often comparable to or exceeding the Hubble time in lower-density halos, and are therefore most effective in helping maintain quiescence in already hot, pressure-supported systems, typically in concert with other heating processes \citep[e.g.,][]{Voit2005}.

\begin{marginnote}[]
\entry{Thermal conduction and evaporation}{Heat transfer across temperature gradients via particle collisions, which can evaporate cooler gas into a hot medium.}
\end{marginnote}

\subsubsection{Gas removal and redistribution}
\label{subsubsec:gas_removal_cgmscales}

Within denser environments, galaxies can be quenched through mechanisms that remove or strip gas, broadly grouped under the phenomenological term environmental quenching, in which star formation is suppressed by external factors. Although often associated with rich clusters, many such processes also act in groups and less virialized environments. A modern reinterpretation of the term first introduced by \citet{Moore1996}, \textbf{dynamical stirring}\footnote{Historically referred to as ``galaxy harassment''; \href{https://folklife.si.edu/magazine/intergalactic-pachamama-kichwa-cosmology-vs-western-astrophysics}{see here for alternative language suggestions}.} describes the high-velocity encounters and tidal interactions that perturb galaxy structures and gas reservoirs, from rapid close passages to cumulative forces from the cluster or group potential \citep[e.g.,][]{Moore1999}. Stronger tidal forces can also cause \textbf{tidal truncation}, stripping gas and stars under the influence of a massive neighbor or the surrounding halo potential \citep{Merritt1983}. Low-mass satellites are most vulnerable, with outcomes ranging from morphological transformation to gas redistribution and the disruption of star-forming conditions \citep[e.g.,][]{Toloba2014,Cutler2025}. Although tidal effects generally act over gigayear timescales, pericentric passages can induce stripping or distortion within a single orbit ($\sim$100–300 Myr). 

In addition to tidal interactions, galaxies in clusters (and to a lesser extent in groups) can also lose gas via \textbf{ram pressure stripping}, where gas is removed as the galaxy moves through the hot, dense ICM \citep{GunnGott1972,BoselliGavazzi2006}.  This process can act rapidly when galaxies move at high velocities through dense hot gas, clearing gas within $\sim$100–500 Myr, or more gradually over $\sim$1–2 Gyr in less extreme conditions \citep{RoedigerHensler2005}. At $z<1$, observational evidence includes HI-deficient disks \citep{Denes2016}, asymmetric gas tails tracing ongoing stripping \citep[e.g.,][]{Crowl2008}, and UV–optical color gradients indicative of outside-in quenching in dense environments \citep{Finn2018}.  Ram pressure stripping remains one of the most clearly established environmental mechanisms capable of driving rapid quenching, whereas most others operate more slowly (except under special circumstances such as strong tidal interactions). Regardless of timescale, environmental quenching mechanisms generally lead to long-lasting suppression of star formation because the galaxy's supply of fresh gas is reduced or removed.
\begin{marginnote}
\entry{Dynamical stirring}{Rapid close encounters that perturb morphology and redistribute gas within or between galaxies.}
\entry{Tidal truncation}{Loss of stars and gas through gravitational stripping by a more massive neighboring system.}
\entry{Ram-pressure stripping}{Removal of galactic gas due to pressure from motion through a dense intracluster medium.}
\end{marginnote}

\subsection{Quenching Mechanisms on Galaxy Scales of $<$100 kpc}
\label{subsec:quenching_galaxyscales}

\subsubsection{Gas heating} \label{subsubsec:gas_heating_galaxyscale}  
While kpc-scale heating can contribute to quenching, particularly during mergers, compaction, or by triggering AGN \citep[see review by][]{Somerville2015}, sustained quenching typically requires larger-scale (halo or cluster) processes (\SMsup{subsec:quenching_cosmoscales}) or continuous energy input, such as from AGN. 
In addition, evolved stellar populations heat (and mechanically redistribute, see \SMsup{subsubsec:gas_removal_galaxyscales}) gas through \textbf{stellar feedback (thermal)}, arising from both Type Ia supernovae (SNe) and evolved stars. 
Late-stage thermally-pulsating AGB (TP-AGB) stars, especially prominent in poststarburst galaxies, drive strong winds and experience heavy mass loss, producing dust and injecting mechanical energy into the ISM while also contributing to gas heating through dust-mediated photoelectric processes \citep[see review by][]{HoefnerOlofsson2018}. 
\begin{marginnote}[]
\entry{Stellar feedback (thermal)}{Heating of gas by evolved stars and supernovae, including AGB mass loss and supernova energy injection.}
\end{marginnote}
The TP-AGB contribution peaks $\sim$0.2–1.5 Gyr after a burst, set by the main sequence lifetimes of the $\sim$2–5 M$_\odot$ stars that dominate this phase. At later epochs ($\gtrsim$1 Gyr), post-AGB stars provide a long-lived, low-level ionizing field that photoionizes diffuse gas, producing emission-line ratios characteristic of low-ionization nuclear emission-line regions (LINERs) with narrow [OIII] profiles (dispersions of $\sigma$$\sim$100 km s$^{-1}$) typical of older quiescent systems \citep[e.g.,][]{Sarzi2010, Belfiore2016}. 
Type Ia SNe act on similar long timescales, injecting mechanical energy and shock heating into the hot, diffuse ISM \citep[e.g., review by][]{MathewsBrighenti2003}. For stellar feedback, the quoted timescale reflects the onset and persistence of heating rather than a discrete quenching timescale, as these processes can operate continuously over gigayear timescales.

The other widely invoked feedback mechanisms are the two AGN modes, which heat gas through distinct energy release channels \citep[see reviews by][]{Hickox2018,Harrison2024}. 
For \textbf{radiatively-inefficient AGN feedback}, or maintenance-mode, most accretion energy is retained in a hot accretion flow and released mechanically via jets and bubbles \citep[e.g.,][]{Fabian2012}. Observationally, extended radio jets and X-ray cavities or bubbles are well-established hallmarks of mechanical heating in massive early-type galaxies and cluster cores \citep[e.g.,][]{McNamaraNulsen2007,Fabian2012,HeckmanBest2014}. 
The giant elliptical M87 exemplifies this process: its large-scale jet and radio bubbles show how repeated mechanical energy injection maintains quiescence by reheating halo gas and suppressing cooling flows \citep{Forman2005}. The associated low Eddington ratios and LINER-like optical line ratios reflect radiatively inefficient accretion flows that dominate at low accretion rates \citep[e.g.,][]{BestHeckman2012}.  This maintenance phase can operate periodically over gigayear timescales, potentially explaining the continued suppression of residual cooling and star formation once the system is already quiescent.  More broadly, AGN activity should be viewed cumulatively, as individual episodes are brief compared to longer-lived quenching signatures they sustain \citep{MartinNavarro2018}. 

\begin{marginnote}[]
\entry{Radiatively-inefficient AGN feedback}{Mechanical heating by jets and bubbles from low-accretion active galactic nuclei (maintenance-mode).}
\entry{Radiatively-efficient AGN feedback (preventative)}{Radiative heating from luminous active galactic nuclei that ionizes gas and raises thermal pressure (quasar mode).}
\end{marginnote}

In contrast, \textbf{radiatively-efficient AGN feedback (preventative)} (quasar-mode) heats gas through photoionization, Compton heating, and radiation-driven winds, raising the thermal and turbulent pressure of the surrounding medium. While this process can stabilize gas against collapse, it does not always expel it entirely; instead, it may drive slower quenching by maintaining elevated gas turbulence and pressure while star formation continues. Under extreme conditions, such as when gas is centrally concentrated and dense, quasar-mode feedback can rapidly expel large fractions of the ISM, thus quenching star formation on short timescales \citep[\SMsup{subsubsec:gas_removal_galaxyscales};][]{DiMatteo2005}. Observationally, kinematically broadened emission lines, blueshifted [OIII] velocity offsets exceeding $\sim$200 km s$^{-1}$, and optical line ratios consistent with high ionization are characteristic of such radiatively-efficient feedback in recently quenched galaxies at high redshift \citep[e.g.,][]{Belli2024}. This quiescence need not be final; rejuvenation can occur due to partial cooling or re-accretion of residual gas once the AGN fades. The cumulative effects of quasar-mode feedback may influence galaxies over $\sim$1–3~Gyr timescales, extending beyond the initial outflow event. During this period the gas reservoir can gradually re-equilibrate, with occasional rejuvenation episodes that can temporarily reignite star formation, particularly in intermediate-mass systems \citep{Harrison2024}.

\subsubsection{Gas removal and redistribution}  
\label{subsubsec:gas_removal_galaxyscales}
High accretion rates onto SMBHs can produce \textbf{radiatively-efficient AGN feedback (ejective)}, releasing radiation pressure and winds that drive large-scale outflows under favorable conditions. This ejective feedback, triggered when a substantial inflow of gas reaches the nucleus, often coincides with accelerated astration (\SMsup{subsubsec:gas_consumption_galaxyscales}). Although brief, quasar-mode feedback can quench star 
\begin{marginnote}[]
\entry{Radiatively-efficient AGN feedback (ejective)}{Luminous active galactic nuclei driving winds that expel or redistribute gas.}
\end{marginnote}
formation if the central engine expels a large fraction of the ISM \citep{DiMatteo2005}. The active ejective phase is typically short-lived ($\sim$10$^{7}$–10$^{8}$ yr) while its cumulative impact on the host galaxy can extend over $\sim$0.1–1 Gyr, with observational signatures including fast AGN-driven ionized and/or molecular outflows ($v \gtrsim 500$~km~s$^{-1}$)
and high-ionization optical line ratios tracing radiatively-efficient accretion \citep[e.g.,][]{Belli2024,Harrison2024}, leaving compact molecular gas remnants. 
Although ionized gas outflows are widespread at high redshift \citep[e.g.,][]{ForsterSchreiber2019}, they typically expel only a small fraction of the ISM and are thus not always sufficient to drive quenching. By contrast, recent JWST observations show that early massive quiescent galaxies may drive substantial neutral gas outflows \citep[e.g.,][]{Belli2024,Davies2024}. 

While gas and metals can be slowly redistributed mechanically via \textbf{stellar feedback (mechanical)} from evolved populations, this mechanism is not strong enough to drive wholesale removal. Stellar feedback from Type~Ia~SNe energy injection and AGB mass loss acts over $\sim$1–3~Gyr, gently stirring and partially thermalizing the ISM (\SMsup{subsubsec:gas_heating_galaxyscale}) while enriching it with metals and dust. Observationally, this process is traced by diffuse, soft X-ray emission that follows the stellar light distribution in massive quiescent galaxies, consistent with this picture \citep[see review by][]{MathewsBrighenti2003}. Although such feedback cannot expel gas, it sustains low-level turbulence that can inhibit new star formation.

\subsubsection{Rapid gas consumption}
\label{subsubsec:gas_consumption_galaxyscales}
A direct consequence of extreme bursts of star formation is the rapid consumption of molecular gas, most aptly described as \textbf{accelerated astration}. If a cold gas reservoir is not replenished on timescales shorter than the gas depletion time (the inverse of the star formation efficiency, i.e., SFR per unit gas mass), the galaxy will exhaust its fuel and quench. Such starbursts can be triggered by major mergers, violent disk instabilities, or even AGN activity that stimulates star formation without expelling gas \citep[see reviews by][]{Veilleux2020,Harrison2024}. In all cases, angular momentum loss funnels gas inward, where it is shock-compressed, reaches high densities, and ignites intense star formation. The characteristic depletion timescales of these bursts (typically $\sim$0.3–1 Gyr) are consistent with the lifetimes of compact star-forming galaxies observed at $z$$\sim$2, which are thought to evolve into compact, red quiescent systems \citep[e.g.,][]{Barro2017}.  Observationally, this rapid consumption phase is supported by short depletion times \citep[e.g.,][]{Williams2021} and elevated [$\alpha$/Fe] ratios seen in massive quiescent galaxies, indicative of brief, intense star formation \citep[e.g.,][]{Beverage2024}.  Although gas consumption can efficiently suppress star formation in the short term, sustaining quiescence requires additional processes that prevent fresh gas accretion or re-cooling.	

\begin{marginnote}[]
\entry{Stellar feedback (mechanical)}{Gas redistribution through SN-driven outflows and mass loss from evolved stars.}
\entry{Accelerated astration}{Rapid gas consumption through star formation on timescales shorter than replenishment.}
\entry{Morphological quenching}{Suppression of star formation as a stellar bulge stabilizes gas against gravitational collapse.}
\entry{Stellar bar quenching}{Suppression of star formation as bar-driven dynamics stabilize gas against collapse.}
\entry{Magnetic field quenching}{Magnetic pressure and tension stabilize cold gas, inhibiting gravitational collapse and star formation.}
\end{marginnote}
\subsubsection{Stabilization of gas} 
\label{subsubsec:gas_stabilization_galaxyscales}
Even when cold gas remains bound within a galaxy’s potential well, turbulence can suppress star formation \citep[e.g.,][]{KrumholzMcKee2005,MacLowKlessen2004}. 
Gas that cannot efficiently dissipate turbulent energy through radiative cooling remains warm and inefficient at forming stars, while turbulent or magnetic energy injection lengthens cooling times and stabilizes gas against fragmentation. Turbulence may be sustained by a massive stellar bulge \citep[\textbf{morphological quenching};][]{Martig2009}, \textbf{stellar bar quenching} \citep[][]{Athanassoula1992}, \textbf{magnetic field quenching} \citep[][]{Krumholz2019}, or periodic, low-level AGN feedback that injects mechanical energy (\SMsup{subsubsec:gas_heating_galaxyscale}). These processes stabilize cold gas by increasing turbulent, magnetic, or dynamical pressure support, thereby reducing star formation efficiency and producing gradual quenching on gigayear timescales without immediate gas removal. However, we caution that both the magnitude and timescales associated with magnetic stabilization remain poorly constrained, both theoretically and observationally, and lack clear, direct, and unambiguous observational tracers.

Observationally, gas stabilization is seen in gas-rich galaxies with suppressed star formation, including compact systems with high central densities and velocity dispersions \citep[e.g.,][]{Suess2017,Bezanson2022_psb}, or barred galaxies with centrally suppressed star formation and LINER-like central emission, consistent with bar-driven gas stabilization \citep[e.g.,][]{Cheung2013,FraserMcKelvie2021}. Magnetic or AGN-driven turbulence is harder to isolate, but may manifest as enhanced gas velocity dispersions and warm molecular-gas emission in otherwise passive systems. Such mechanisms primarily maintain quiescence at low redshift, when cold-gas reservoirs are limited and replenishment is inefficient \citep[e.g.,][]{Tacconi2018}. 

\section{SELECTING QUIESCENT GALAXIES}
\label{sec:supplement_selection}

The main review highlights the continued utility of rest-frame color selection methods for identifying quiescent galaxies, even as the field increasingly moves toward SED-based sSFR selections. In this supplemental section, we expand on key physical considerations that merit additional discussion: physically motivated modifications to the UVJ diagram (\SMsup{subsection:uvj_modifications}), known physical trends encoded in rest-frame colors (\SMsup{subsection:uvjtrends}), and sources of contamination that become especially important at high redshift (\SMsup{subsec:uvj_highredshift}).

\subsection{Modifications to the UVJ Diagram}
\label{subsection:uvj_modifications}
Modifications to the UVJ diagram (\textbf{\autoref{fig:ssfr}a}) 
-- still adopted inconsistently -- include omitting the upper boundary \citep[e.g.,][]{vanderWel2014} or the lower boundary \citep[e.g.,][]{Belli2019}. The upper regime of the diagram remains poorly explored: these red sources are either genuinely low-sSFR but dust-rich and pushed beyond the standard UVJ quiescent selection box \citep[e.g.,][]{Alberts2023,Setton2024}, dusty star-forming contaminants \citep[e.g.,][]{Patel2012,Spitler2014,Schreiber2018,Lustig2023}, or a combination thereof. The lower boundary, by contrast, can exclude recently quenched galaxies with ages less than a few hundred megayears, an issue that becomes increasingly severe at higher redshift \citep{Antwi-Danso2023}. Opening this parameter space can also admit bursty star-forming galaxies caught in temporary lulls, though such systems appear to be rare among the massive population. 

To mitigate these shortcomings, alternative diagnostics have been proposed, such as combining submillimeter or radio diagnostics with optical colors to reduce misclassification of dusty star-forming galaxies \citep[e.g.,][]{Eales2018}.  In parallel, several studies have calibrated UVJ selections against independent sSFR thresholds and developed probabilistic or mixture-model classifications in both sSFR space \citep{Carnall2018,Leja2019,Gould2023,Antwi-Danso2023} and color space \citep{Davidzon2017}, often leveraging large-area surveys \citep[e.g.,][]{Muzzin2013,Laigle2016} to better quantify contamination and completeness.

\subsection{Physical Drivers of Rest-frame Color Trends}
\label{subsection:uvjtrends}
The UVJ quiescent region exhibits several key trends. The most fundamental is the increase in average stellar age along the quiescent sequence (from blue to red colors), with both colors reddening in agreement with SPS model predictions at $z>1$ \citep{Whitaker2012,Whitaker2013,Belli2019}.  The dust vector runs roughly parallel to the aging vector, introducing degeneracies that will require deeper spectroscopy, improved modeling, and broader wavelength coverage to resolve. \citet{Cheng25} show that metallicity varies roughly perpendicular to the age trend, with higher metallicities at redder $U-V$ and bluer $V-J$, indicating that age and metallicity leave distinct imprints within the UVJ plane. 

In general, sSFR declines along a trend similar to metallicity, but \citet{Leja2019} demonstrate that UVJ colors saturate below log(sSFR/year$^{-1}$)$<$-10.5, such that further decreases no longer shift the colors significantly. To probe this regime, they show that rest-frame far-UV ($\sim$1500\,\AA) and/or MIR ($\sim$12 $\mu$m) fluxes remain correlated with sSFR to much lower levels, and thus color–color diagrams incorporating these bands can efficiently target galaxies with extremely low ongoing star formation. Current facilities can access the rest-frame FUV (e.g., HST at $z>0.8$ and JWST at $z>3.5$), but data remain  too shallow for robust selection. Rest-frame MIR coverage at these redshifts awaits next-generation FIR instruments, so the landscape of color-based selection is unlikely to change substantially for at least the next decade.

\subsection{Contamination and Photometric Challenges at High Redshift}
\label{subsec:uvj_highredshift}

Classical rest-frame color selections lose utility at $z>3$ for several reasons. First, recently quenched galaxies appear blue in their $U-V$ and NUV–$r$ colors (without dust), especially if they quenched only a few hundred megayears ago \citep[e.g.,][]{Belli2019}. Indeed, \citet{Lustig2023} and \citet{Baker2025a} find that $\sim$30\% of young quiescent galaxies at $z\sim3$ fall outside the classical UVJ-selected quiescent region. This limitation is well recognized \citep[e.g.,][]{Park2023,Baker2025a} and also affects identification of low-mass quiescent populations at cosmic noon \citep{Cutler2024}. A practical solution is to extend the UVJ boundaries to bluer colors and apply sSFR constraints to exclude interlopers (see \autoref{subsec:sSFRs} of the main article). 

More generally, UVJ selections at $z>3$ face additional challenges. As rest-frame optical bands shift into the observed NIR and MIR, where data are shallower and filter coverage limited, photometry becomes noisier and usually requires extrapolation, increasing color uncertainties and systematic biases \citep[e.g.,][]{Gould2023}.
Emission line contamination also becomes significant, especially in the rest-frame $V$-band, leading to misclassifications; while present at lower redshift, it is more severe at $z>3$. To mitigate this, \citet{Antwi-Danso2023} propose synthetic $u_s$,$g_s$,$i_s$ filters that avoid strong emission lines and minimize extrapolation out to $z\sim6$, reducing contamination compared to UVJ, particularly at $z>3$ where the rest-frame $J$ band redshifts beyond JWST/NIRCam’s $4.5~\mu$m limit. Avoiding emission line contamination by star-forming galaxies with moderate dust attenuation improves the fidelity of color selection, but comes at the expense of reduced sensitivity to extreme dust obscuration ($A_V>1$).

The reddest UVJ region, which is typically populated by older quiescent galaxies, remains prone to significant contamination by dusty star-forming galaxies even at $z\sim3$ \citep[up to 60\%,][]{Lustig2023}. The limited wavelength baseline at the reddest wavelengths makes it difficult to distinguish these populations, and this problem only worsens at lower redshift. Thus, while the synthetic filter approach of \citet{Antwi-Danso2023} is promising at $z>3$ if rest-frame coverage disappears, its broader utility may be limited. An alternative is to use observed-frame color selections (using redshift-dependent observed color-cuts to identify relatively red breaks). As in early extragalactic searches, these avoid rest-frame extrapolation and are less model-dependent, potentially yielding more complete samples at $z>3$ \citep{Long2024}, albeit likely at the cost of purity as dusty galaxies and red AGN are included.

\section{ANATOMY OF A QUIESCENT GALAXY}
\label{sec:supplement_anatomy}

\begin{figure}
    \centering
    \includegraphics[width=0.98\linewidth]{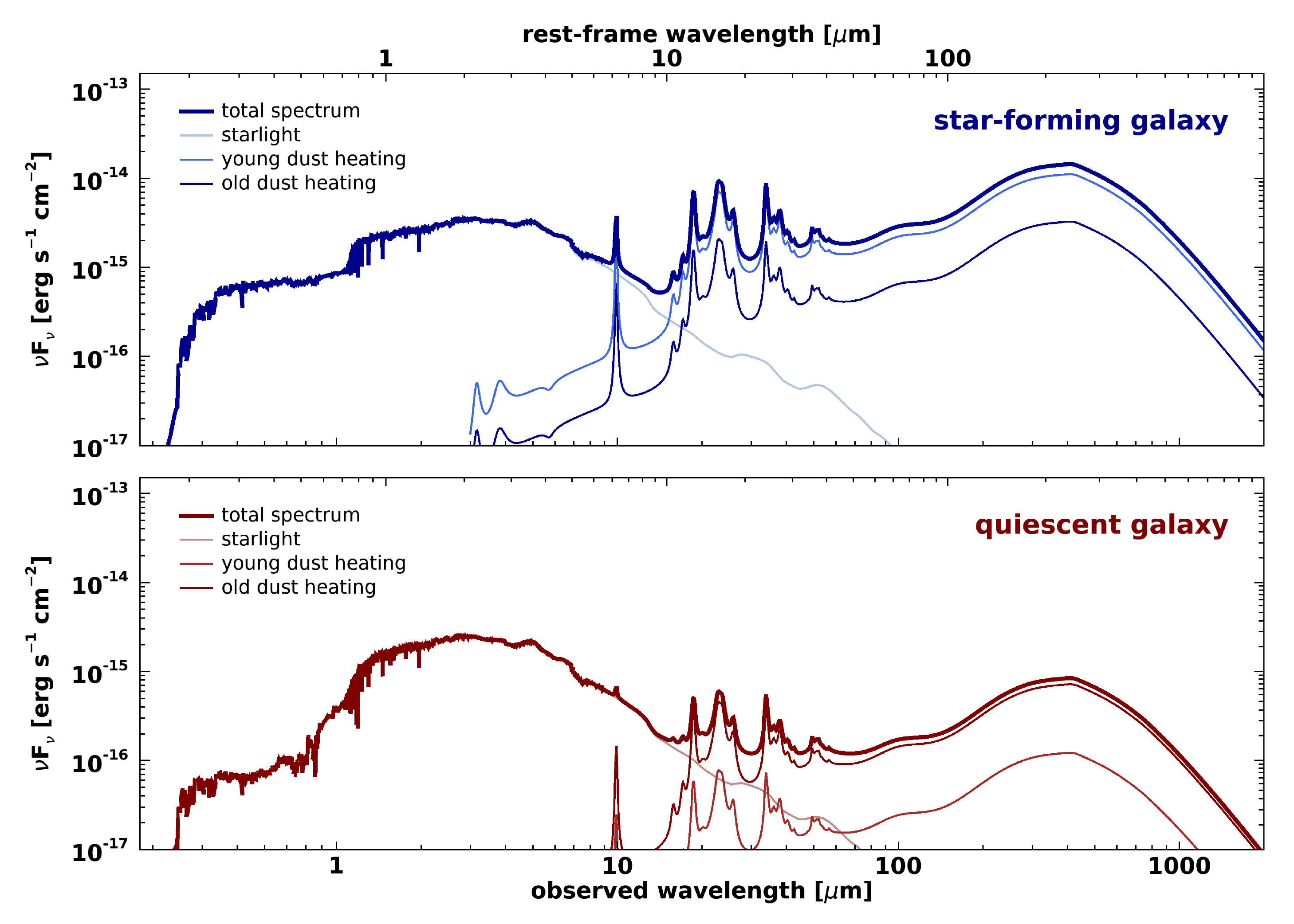}
    \caption{Example spectral energy distribution of a typical star-forming (log(sSFR/year$^{-1}$)=-9.2, $\tau_{\mathrm{V}}$=1.2; \emph{blue}, top) and quiescent (log(sSFR/year$^{-1}$)=-11.0, $\tau_{\mathrm{V}}$=0.45; \emph{maroon}, bottom) galaxy, based on Flexible Stellar Population Synthesis models that adopt the average properties inferred in \citet{Leja2019b} for log(M$_{\star}$/M$_{\odot}$)=11 at $z=2$.  Each total spectrum (\emph{thick lines}) is split into three components: starlight, old dust heating (from stars $>$100 Myr), and young dust heating (from stars $<$100 Myr).  Whereas the FIR luminosity of the star-forming galaxy is dominated by young dust heating, accurately tracing new star formation, heating for the quiescent galaxy is instead dominated by old stars, complicating SFR inferences. Abbreviations: FIR, far-infrared; SFR, star formation rate. }
    \label{fig:sed}
\end{figure}

Understanding the stellar populations and ISM of quiescent galaxies begins with a clear picture of what fundamentally distinguishes their SEDs from those of star-forming systems. To set the stage, \textbf{\autoref{fig:sed}} presents Flexible Stellar Population Synthesis models \citep{Conroy2010} of a typical massive quiescent galaxy at $z\sim2$ alongside a similarly massive star-forming galaxy, decomposing the emission into the stellar continuum and dust heated by stars younger and older than 100 Myr. This comparison highlights two essential points. First, the SEDs of quiescent galaxies are markedly different from those of star-forming systems, with steeply declining UV fluxes, prominent Balmer and 4000$\mathrm{\AA}$ breaks, and substantially weaker MIR and FIR emission. Second, and crucial for interpreting dust measurements, the FIR luminosity of quiescent galaxies is dominated by dust heated by older stars rather than by ongoing star formation. As a result, dust emission in these systems cannot be straightforwardly converted into SFRs or molecular gas masses without accounting for the underlying stellar population and dust-heating physics.

These insights motivate the more detailed material contained in this supplemental section. Here we expand on the physical ingredients that shape the SEDs and spectral signatures of quiescent galaxies, the limitations of commonly used stellar-population and ISM tracers, and the modeling uncertainties that become especially important at intermediate and high redshift. The goal is to provide a deeper, more technical foundation for the interpretations presented in the main text and to clarify the assumptions, caveats, and open questions that underlie current studies of quiescent galaxy formation and evolution.

\subsection{Stellar Populations of Quiescent Galaxies}
\subsubsection{Observed Spectral Features}
\label{subsubsec:supplement_observed_spectral_features}
\textbf{\autoref{tab:lines}} provides a detailed compilation of the rest-frame optical spectral features that are most relevant for diagnosing the stellar populations and ionized gas conditions of quiescent galaxies, based on a more extensive list in \citet{Hamadouche2026}. For each absorption and emission feature, the table lists the rest-frame vacuum wavelength, the broad source of the line (stellar photospheres, neutral or ionized ISM gas, or AGN), and its primary diagnostic use. The absorption features span the major age- and metallicity-sensitive indices used in stellar population studies, from Balmer lines tracing intermediate-age A-F stars and recent quenching to metal lines (e.g., CaII H\&K, Mgb, Fe and TiO complexes) probing enrichment and $\alpha$-element ratios in evolved G-K giants. Several bandpasses correspond to classical Lick spectral indices, while others represent blended molecular or atomic features that encode abundance ratios or surface gravity.

The emission features in \textbf{\autoref{tab:lines}} play a complementary role. Although quiescent galaxies lack strong nebular emission from star formation, weak lines such as [NII], [OII], or [OIII] may appear due to shocks, low-level AGN activity, or ionization by post-AGB stars (hot evolved stellar cores of low- and intermediate-mass stars $\lesssim$8 M$_{\odot}$ in the brief UV-bright phase after the asymptotic giant branch). The table distinguishes emission excited by such evolved stellar populations from that driven by AGN. This distinction is important when interpreting the weak emission lines commonly detected in high-redshift quiescent spectra with JWST.

Together, the spectral features of quiescent galaxies represent the core diagnostics used throughout the literature to extract their stellar ages, metallicities, abundance patterns, and ionization conditions. \textbf{\autoref{tab:lines}} is intended to guide interpretation of both individual absorption features and integrated spectral indices and to support the more comprehensive discussions of stellar population modeling and quenching physics that follow in the main review.  Additional details and references are presented in \citet{Hamadouche2026}.

\begin{sidewaystable}
    \caption{Rest-frame optical spectral features prominent in quiescent galaxies.}
    \begin{center}
    \begin{tabular}{@{}lllll@{}}
    \hline
    \hline
    Element & $\lambda_{\mathrm{rest,vacuum}}$ & Source & Line Origin & Diagnostic  \\
    \hline
    \hline
    \emph{Absorption} \\
    \hline
    H$\infty$ & 3645.98 & stellar & photospheres (late B-A-F stars) & age (blended) \\
    H10 & 3798.98 & stellar & photospheres (late B-A-F stars) & age (blended)\\
    H9 & 3836.47  & stellar & photospheres (late B-A-F stars) & age (blended) \\
    H$\theta$ & 3889.06 & stellar & photospheres (late B-A-F stars) & age (blended) \\
    CaII K & 3934.78 & stellar, ISM & photospheres (F–K), neutral ISM & age, ISM \\
    CaII H & 3969.59 & stellar, ISM & photospheres (F–K), neutral ISM$^{\rm a}$ & -- \\ 
    H$\epsilon$ & 3971.20 & stellar & photospheres (late B-A-F stars) & age (blended) \\
    MnI & 4018.10 & stellar & photospheres (G-K stars) & metallicity$^{\rm b}$  \\
    H${\delta}_{A,F}$$^{\rm e}$ & 4102.89$^{\rm c}$ & stellar & photospheres (late B-A-F stars) & age \\ 
     CNI$^{\rm e}$ & 4161.15$^{\rm c}$ & stellar & photospheres (G-K giants) & metallicity$^{\rm j}$ \\ 
    SrII & 4216.00 & stellar & photospheres (F–K giants) & s-process (AGB) enrichment \\
    Ca4227$^{\rm e}$  & 4228.5$^{\rm c}$ & stellar & photospheres (G-K giants) & metallicity, [$\alpha$/Fe] \\  
    CH (G4300)$^{\rm d}$ & 4305.61$^{\rm c}$ & stellar & photospheres (G-K giants) & age, metallicity \\
    H$\gamma$ & 4341.68 & stellar & photospheres (late B-A-F stars) & age  \\
    Fe4383$^{\rm e}$ & 4383.0$^{\rm c}$ & stellar & photospheres (G-K stars) & metallicity  \\
    Ca4455$^{\rm e}$ & 4464.6$^{\rm c}$ & stellar & photospheres (G-K stars) & metallicity  \\
    Fe4531$^{\rm e}$ & 4536.8$^{\rm c}$ & stellar & photospheres (G-K stars) & metallicity  \\
    C$_{2}$4668$^{\rm e}$ & 4677.1$^{\rm c}$ & stellar & photospheres (G-K giants) & metallicity  \\
    H$\beta$$^{\rm e}$ & 4862.68$^{\rm c}$ & stellar & photospheres (late B-A-F stars) & age, metallicity \\
    Fe5015$^{\rm e}$ & 5016.43$^{\rm c}$ & stellar & photospheres (G-K stars) & metallicity  \\ 
    Mgb$^{\rm f}$ & 5176.70 & stellar & photospheres (G-K giants) & [$\alpha$/Fe], gravity   \\
    Fe5270$^{\rm e}$ & 5265.65$^{\rm c}$ & stellar & photospheres (G-K stars) & metallicity, gravity  \\
    Fe5335$^{\rm e}$ & 5332.13$^{\rm c}$ & stellar & photospheres (G-K stars) & metallicity  \\
    Fe5406$^{\rm e}$ & 5401.25$^{\rm c}$ & stellar & photospheres (G-K stars) & metallicity (weak)  \\
    Fe5709$^{\rm e}$ & 5710.25$^{\rm c}$ & stellar & photospheres (G-K stars) & metallicity (weak)  \\
    Fe5782$^{\rm e}$ & 5788.38$^{\rm c}$ & stellar & photospheres (G-K stars) & metallicity (weak)  \\
    Na D$^{\rm g}$ & 5895.6  & stellar, ISM & photospheres (G–M stars), neutral ISM & ISM, $N_{\mathrm{H I}}$  \\
    TiO$_{1}$, TiO$_{2}^{\rm e}$ &  5965.4, 6230.9$^{\rm c}$  & stellar & photospheres (M stars) & IMF-sensitive  \\
    H$\alpha$ &  6564.61 & stellar & photospheres (late B-A-F stars) & age (weak)  \\ 
        \hline
    \emph{Emission} \\
    \hline
    {[NeV]}$^{\rm h}$ &  3346.75, 3426.85 & AGN & high-ionization NLR gas & AGN ionization \\
    {[OII]}$^{\rm i}$ & 3727.1, 3729.86 & ISM, AGN & ionized gas (post-AGB, shocks, AGN; SF in non-QGs) & ionization; SFR \\
    {[NeIII]}$^{\rm h}$ & 3869.86 & ISM, AGN & ionized gas (AGN, shocks, post-AGB) & ionization, outflows \\
    {[OIII]}$^{\rm i}$ & 4960.29, 5008.24 & ISM, AGN & ionized gas (AGN, shocks, post-AGB; SF in non-QGs) & ionization; AGN, outflows \\
    H$\alpha$ & 6564.61 & ISM & ionized gas (H II, AGN, post-AGB) & SFR, AGN/post-AGB ionization\\
    \hline
    \hline
    \end{tabular}
\end{center}
\begin{tabnote}
$^{\rm a}${Ca K also arises in diffuse neutral/warm ionized ISM gas. By contrast, Ca H is less useful as an ISM diagnostic as it coincides with H$\epsilon$, leaving little continuum for additional absorption even when present in the ISM.}; $^{\rm b}${Produced in the outer layers of massive stars from incomplete Si-burning processes. More Mn is produced in SNe Ia than Fe, some also produced via SNe II.}; $^{\rm c}${Taken as the central wavelength of the bandpass; this table mixes Lick Indices (noted) with individual elements.}; $^{\rm d}${ Fraunhofer G4300 index is a blend of the CH, Fe and Ca absorption lines.}; $^{\rm e}${Lick indices measuring the abundance of the main absorber: most dominated by the named atom, containing lesser contributions from other elements (except Fe5782, dominated by Cu, Cr and Mg absorption, with smaller Fe contribution).}; $^{\rm f}${Mgb index measures strength of three individual MgI unresolved lines.}; $^{\rm g}${NaI is a doublet from fine-structure splitting of NaI and excited states (5891.6{\AA}, 5897.6{\AA}).}; $^{\rm h}${[NeV], [NeIII] produced in high-ionization regions, mostly by AGN narrow-line region (NLR) but also intense star-formation.}; $^{\rm i}${[OII] and [OIII] are collisionally excited forbidden emission lines that are typically powered by post-AGB stars, shocks, or AGN in quiescent galaxies, rather than star formation}.; $^{\rm j}${Sensitive to N abundance when measured concurrently with G-band feature.}
\end{tabnote}
    \label{tab:lines}
\end{sidewaystable}

\subsubsection{Impact of SFH Priors on Stellar Mass Estimates}
\label{subsubsec:supplement_stellar_mass}

While the main text provides an overview of the current state of stellar mass uncertainties, here we expand on two areas that particularly benefit from a more technical discussion: the role of SFH priors in shaping inferred stellar masses, and insights from mock-recovery tests that assess the reliability of different SPS modeling frameworks for quiescent galaxies. Together, these considerations highlight how modeling choices, rather than data quality alone, can drive systematic differences in recovered stellar masses.

Stellar mass estimates can be highly sensitive to the choice of prior on the form of the SFH, as shown in early demonstrations of prior-driven biases in SPS modeling \citep[e.g.,][]{Pacifici2012,Pacifici2016}. Traditionally, many analyses adopt parametric SFHs (e.g., exponentially declining, delayed-$\tau$, double power-law or log-normal forms), which impose strong assumptions about the shape and smoothness of a galaxy’s past star formation. These forms can bias inferred stellar masses and mass-to-light ratios in systems with complex or rapidly varying SFHs, particularly poststarburst galaxies, by forcing the model toward overly smooth or monotonic histories \citep[e.g.,][]{Carnall2019a,Leja2019a}. 

The growing use of non-parametric SFHs relaxes these constraints by allowing the SFR to vary across independent time bins. However, non-parametric approaches still rely on priors that regulate the allowed temporal behavior of the SFH. For example, continuity priors suppress rapid fluctuations in the SFR \citep[e.g.,][]{Leja2019}, though such an assumption generally recovers reliable masses even for bursty systems \citep[e.g.,][]{Leja2019b,Lower2020}. That said, alternative priors or binning strategies can yield systematically different mass estimates \citep{Leja2019a}.  The flexible late-time refinement introduced by \citet{Suess2022}, which assigns increased resolution to recent epochs, tends to attribute more light to recent star formation, thereby often yielding lower inferred stellar masses when recent star formation is present, potentially reflecting a more accurate recovery of late-time SFH. 

Despite the growing popularity of non-parametric SFHs, it is important to note that the SFH prior itself can introduce significant biases in both the recovered SFH shape and the inferred stellar mass. Mock recovery tests show that non-parametric SFH models recover first-order quantities such as stellar mass, SFR, and mass-weighted age to $\sim$0.1–0.2 dex in stellar mass and $\sim$0.2–0.4 dex in SFR and stellar age \citep[e.g.,][]{Leja2019a,Lower2020}. Higher-order features such as burst fraction or quenching time require SFH priors that permit rapid temporal variability. However, stress tests against simulation-based ground truths suggest that stellar mass uncertainties can still reach an order of magnitude \citep{Narayanan2024,Wang2025}. The choice of prior, whether enforcing smoothness, continuity, or burstiness, effectively encodes assumptions about galaxy evolution, and these assumptions can dominate the inference, especially when observational constraints are limited. 

\subsubsection{The Utility and Limitations of Spectral Indices}
\label{subsubsec:lick}

\begin{figure}[h]
\includegraphics[width=\linewidth]{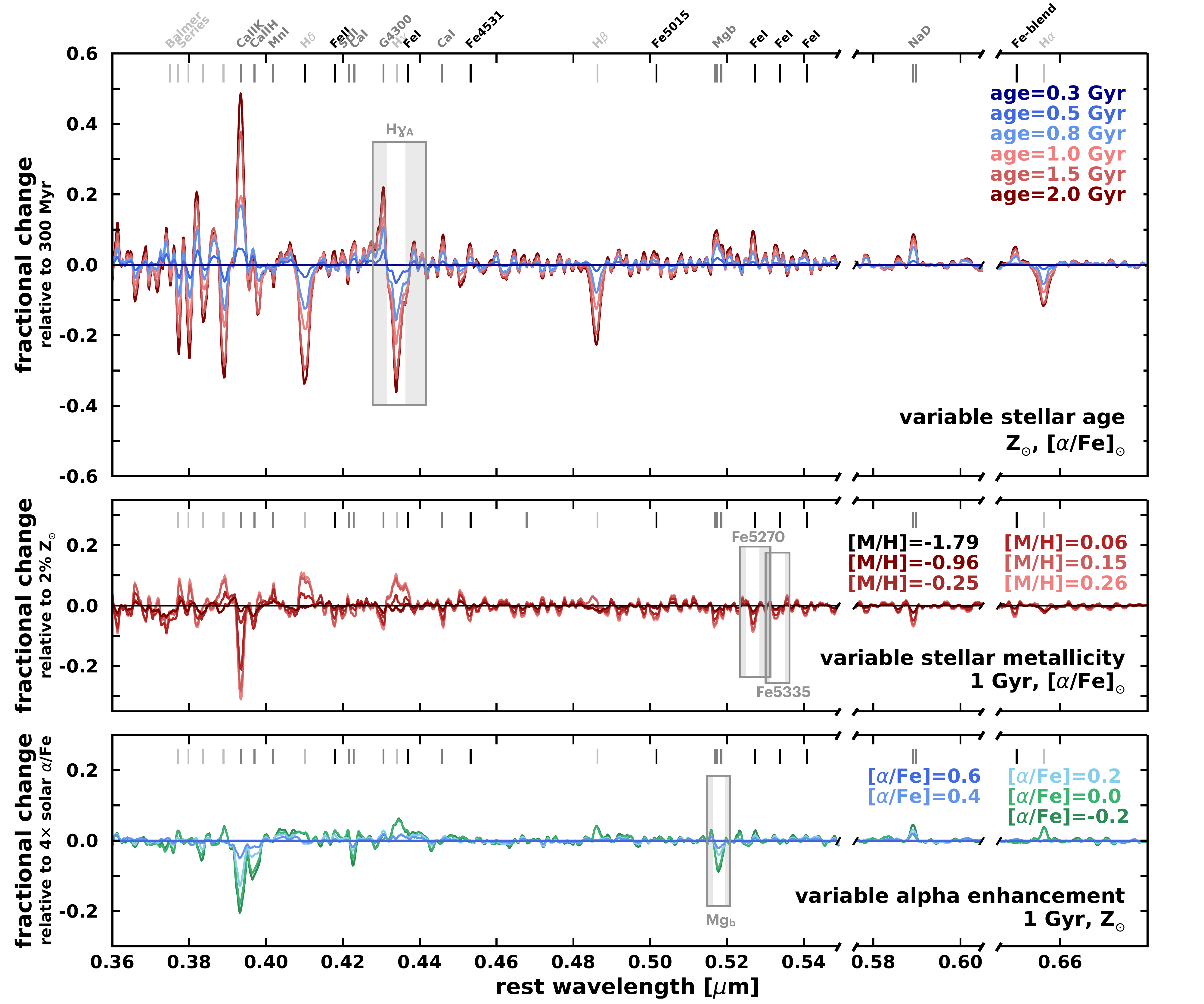}
\caption{The fractional change in stellar SSP spectra from the sMILES library \citep{Knowles2023}, normalized to remove the continuum, illustrates how rest–frame optical absorption features vary with age (top), metallicity (middle), and $\alpha$-enhancement (bottom). Features weaken toward negative values and strengthen toward positive values relative to a baseline: a 300 Myr population (top), $\sim$2\% Z$_\odot$ (middle), and 4$\times$ solar $\alpha$-enhancement ([$\alpha$/Fe]=0.6; bottom). Metallicities are parameterized as [M/H], which is equivalent to [Z/H] (or $\log(Z/Z_\odot)$) for solar abundance ratios ([$\alpha$/Fe]$\sim$0). Spectral lines are grouped as hydrogen (\emph{light gray}), $\alpha$-elements (\emph{gray}), and Fe (\emph{black}). In the top panel, strong Balmer lines fade as the population ages beyond 300 Myr. The middle and bottom panels fix the age to 1 Gyr, highlighting how metallicity (fixed to [$\alpha$/Fe]$_{\odot}$) and $\alpha$-enhancement (fixed to Z$_{\odot}$) produce subtler differences. While the age-sensitive Lick index H$\gamma_\mathrm{A}$ varies strongly, the metallicity-sensitive Fe5270/Fe5335 and $\alpha$-sensitive Mgb indices change far less. Thus, SSP-equivalent ages can generally be inferred robustly at $z>1$, but detailed chemical enrichment remains difficult to constrain. Abbreviations: SSP, simple stellar populations. }
\label{fig:smiles}
\end{figure}

\textbf{\autoref{fig:smiles}} illustrates how key spectral features vary in continuum-normalized model spectra when changing stellar age (top; at fixed solar metallicity and [$\alpha$/Fe]), metallicity (middle; at fixed age of 1 Gyr and solar [$\alpha$/Fe]), and $\alpha$-enhancement (bottom; at fixed age of 1 Gyr and solar metallicity). These comparisons highlight how individual absorption features respond to different stellar population parameters and why it is difficult to disentangle age, metallicity, and chemical abundance effects.

The classical Lick spectral index system was developed to isolate specific absorption features sensitive to stellar age (H$\gamma_{\mathrm{A}}$), total metallicity (Fe5270/Fe5335), and $\alpha$-enhancement (Mgb) 
\citep{Burstein1984,Worthey1994}. Defined as equivalent widths using feature and side-band regions, these indices are high-resolution analogs of rest-frame colors targeting key absorption features. However, several indices suffer from blending and coupled abundance effects; for example, Mg acts as an electron donor, altering the strength of Ca lines in ways that are not trivially separable \citep{Conroy2014}. As a result, interpreting isolated indices can be ambiguous when abundance patterns vary or multiple elements influence a given feature. 

At higher redshifts, these limitations become even more pronounced because metal- and abundance-sensitive features are intrinsically weaker in the younger quiescent populations that dominate $z>1$ samples. Strong metal lines and classical $\alpha$-enhancement indicators, such as the Mgb triplet, CaII H\&K, and FeI absorption lines around 5270–5335$\mathrm{\AA}$, are prominent in older quenched galaxies locally, but become far weaker in intermediate-age ($\sim$1 Gyr) populations that are dominated by A- and early F-type stars (\textbf{\autoref{fig:smiles}}). Broad TiO absorption features in the rest-frame NIR, which arise from cool giant stars and become prominent in older stellar populations, are likewise diminished at these intermediate ages.  The reduced sensitivity of metal and $\alpha$-element indices at early epochs therefore limits the diagnostic power of classical index measurements. These challenges are compounded by limitations in SPS models, uncertainties in the treatment of stellar evolutionary phases, and observational constraints at high redshift, especially low SNR and limited spectral resolution in the first deep spectra of quiescent galaxies \citep[e.g.,][]{Kriek2006}. 

Given the complexity and coupled behavior of spectral features, most modern studies now rely on full spectral fitting to model the entire spectrum and capture such coupled effects \citep[e.g.,][]{Johnson2021,Cappellari2023,Conroy2023,Carnall2018}, although some work still analyzes isolated indices along with photometric colors \citep[e.g.,][]{Gallazzi2024}. These full-spectrum approaches leverage the continuum shape and all absorption features simultaneously, enabling more robust constraints on stellar age, metallicity, and abundance patterns while naturally incorporating correlated uncertainties. Overall, classical index diagnostics provide valuable physical intuition but have limited utility for young quiescent galaxies at $z>1$. Full-spectrum modeling offers a more reliable pathway for deriving stellar population properties in these systems.

\begin{figure}[h]
\includegraphics[width=\linewidth]{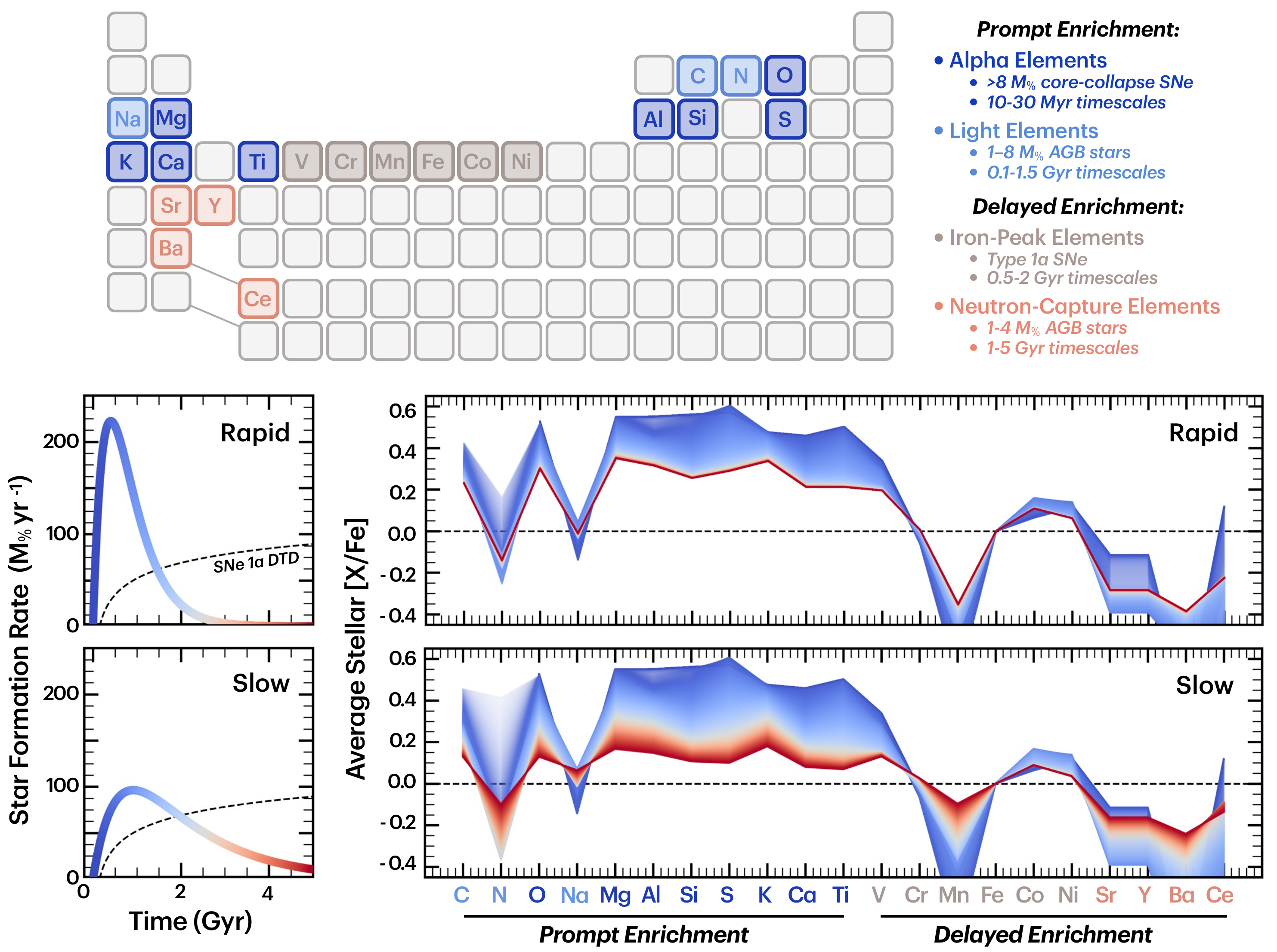}
\caption{Nucleosynthetic origins for major elemental groups that shape chemical evolution in massive galaxies. Elements are color-coded by the characteristic timescale of the dominant production site: $\alpha$-elements (\emph{blue}; produced promptly in core-collapse SNe of $>$8 M$_{\odot}$ stars within 10-30 Myr); light elements (\emph{light blue}; synthesized by 1–8 M$_{\odot}$ AGB stars, $\sim$0.1-1.5 Gyr); Fe-peak elements (\emph{gray}; released mostly by delayed Type Ia SNe following the delay–time distribution, or DTD); and s-process neutron-capture elements (\emph{salmon}; produced in 1–4 M$_\odot$ AGB stars over $\sim$1–5 Gyr).  
Although grouped by dominant channels, this chemical evolution model \citep[adapted from models by][]{Gountanis2025} includes only two enrichment pathways: a prompt channel tied to core-collapse SNe and a delayed channel tied to the SNe Ia DTD, with AGB contributions folded into these. Different timescale SFHs (rapid, \emph{top}; extended, \emph{bottom}) produce distinct mean stellar abundance ratio patterns relative to Fe: early, short bursts yield $\alpha$-enhancement, whereas extended histories allow delayed Fe-peak and neutron-capture enrichment to lower [X/Fe] over time, where X represents each of the individual chemical elements. Figure created by Aliza Beverage \citep{Beverage2026}. Abbreviations: SNe, supernovae; AGB, asymptotic giant branch; DTD, delay-time distribution; SFH, star formation history.}
\label{fig:periodictable}
\end{figure}

\subsubsection{Astrophysical Drivers of Chemical Abundance Patterns}
\label{subsubsec:supplement_chemical_evolution}
Chemical abundance patterns provide a complementary window into quenching beyond the SFH-based timescales discussed in the main text. Because elemental ratios trace the timing and duration of star formation, the efficiency of metal retention, and the relative contribution of different nucleosynthetic channels, they offer a powerful, though often underutilized, diagnostic of the physical processes that shape quiescent galaxies. A growing body of high-redshift spectroscopy has established that massive quenched systems at $z\gtrsim1$ frequently reach near-solar to super-solar metallicities \citep[e.g.,][]{Kriek2016,Onodera2015,Gallazzi2021,EstradaCarpenter2019, Morishita2019,Carnall2019,Esdaile2021}, indicating rapid early enrichment and efficient metal retention prior to quenching. In this supplemental section, we highlight several factors that complicate the interpretation of these metallicity and abundance measurements.

Minor mergers and extended accretion after quenching can introduce lower-metallicity stars and gas from satellites into the outskirts of galaxies, diluting global metallicities and contributing to the observed diversity in quiescent systems \citep[e.g.,][]{Oser2010,vanDokkum2010,Hirschmann2015,Beverage2024}. Such late-time assembly pathways naturally produce a broad range of metallicity profiles and global abundance patterns, consistent with the diversity seen among low- and intermediate-mass quiescent galaxies \citep[e.g.,][]{Lonoce2015,Saracco2019}. Observational evidence likewise shows that while the central regions of massive quiescent galaxies have passively evolved since $z\gtrsim1$, their global properties become increasingly inhomogeneous at lower redshift due to varied assembly histories \citep{Choi2014}.  Importantly, different assembly scenarios also imprint distinct abundance patterns: simulations by \citet{Choi2014} demonstrate how variations in ex-situ accretion, dissipational mergers, and centrally concentrated star formation episodes lead to different evolutionary tracks in metallicity and [$\alpha$/Fe] over time. However, extended accretion alone cannot easily explain the enhanced central [$\alpha$/Fe] ratios commonly observed in massive systems, which likely require rapid, centrally concentrated star-formation episodes or major-merger–driven starbursts \citep[e.g.,][]{Spolaor2010,Beverage2024}. These centrally peaked abundance patterns are fully consistent with the early, rapid-quenching pathways predicted for massive galaxies in our evolutionary framework (see \SMsup{sec:sm_evolutionary_framework}), in which the inner regions form quickly and remain largely unchanged while the outskirts continue to grow through later accretion.

\subsubsection{Modeling Limitations in Chemical Abundance Inference}
\label{subsubsec:supplement_chemical_models}
Most SPS models still adopt scaled-solar abundance patterns \citep[e.g.,][]{Onodera2015,Kriek2016}, potentially leading to systematic biases when applied to $\alpha$-enhanced populations. Work by \citet{Park2025} shows that $\alpha$-enhancement affects not only individual absorption features but also the full continuum shape, altering mass-to-light ratios (particularly in the optical) and reddening broadband colors. These issues are especially acute for high-redshift quiescent galaxies, where data quality is limited and galaxies are young by cosmic standards \citep[e.g.,][]{Wilkinson2017}.  For example, the sMILES stellar library \citep{Knowles2023}, a synthetic extension of the empirical MILES database \citep{SanchezBlazquez2006,Vazdekis2010}, reveals that although metallicity- and $\alpha$-sensitive features vary by $\sim$10–20\% for stellar populations older than 3 Gyr, this drops to $<$5\% at 1 Gyr relative to the baseline models shown in \textbf{\autoref{fig:smiles}} -- precisely the ages most relevant for high-redshift quiescent galaxies. This behavior underscores why degeneracies between age, metallicity, and [$\alpha$/Fe] become extremely difficult to break with classical Lick indices alone (see also \SMsup{subsubsec:lick}). Robust constraints therefore require spectro-photometric fitting that analyzes the full continuum and absorption-line spectrum \citep[e.g.,][]{Beverage2024,Gountanis2025}.

A major bottleneck in interpreting [$\alpha$/Fe] even when performing a full spectrophotometric analysis remains the limited availability of SPS models with flexible, non-solar chemical abundance patterns. Although individual stellar templates with variable element ratios exist, model grids with abundance variations and time-evolving SFHs have only recently begun to emerge. Until recently, \texttt{ALF} \citep[the Absorption Line Fitter;][]{Conroy2023} was the only tool capable of varying individual abundances, but it did so by applying response functions to SSP templates with fixed ages and metallicities. 

High-quality spectral studies have consistently shown that massive quiescent galaxies exhibit distinctly non-solar abundance patterns, with departures from solar that closely follow their SFHs \citep[e.g.,][]{Choi2014,Beverage2025}. These trends are illustrated in \textbf{\autoref{fig:periodictable}}, created by Aliza Beverage, which shows predictions from a simple leaky-box chemical-evolution model \citep{Beverage2026}. In this framework, Mg and Fe enrichment is first computed for a given SFH, and the corresponding core-collapse and Type Ia SNe rates at each timestep are combined with Milky Way–calibrated yields \citep{Weinberg2022} to generate time-dependent elemental abundances. Although \textbf{\autoref{fig:periodictable}} groups elements by their dominant nucleosynthetic channels, the model itself includes only two enrichment pathways: `prompt' core-collapse SNe and `delayed' Type Ia SNe. Any contribution from AGB stars is absorbed into these channels rather than treated independently. Elements such as C and N appear to track Fe-peak behavior, and metallicity-sensitive species like Na or Ca to reflect mixed production sources. The figure highlights how strongly [X/Fe] ratios, where X represents each chemical element, respond to the quenching timescale during the first few gigayears. Slow quenching SFHs produce a much broader diversity in [X/Fe] ratios than rapid quenching, whereas post-quenching ratios evolve very little once star formation ceases.

Altogether, the outlined challenges underscore why interpreting the chemical abundances of high-redshift quiescent galaxies remains fundamentally limited by the pace of theoretical model development. Continued advancement of models will not only improve constraints on quenching timescales across cosmic time but may also clarify the origin of systems with unusual abundance patterns, including ancient globular clusters \citep[e.g.,][]{Gratton2012}.

\subsection{Interstellar Medium}
\label{subsec:supplement_ism}

In this section, we expand on the discussion of cold ISM tracers in the main text by providing additional technical details on CO-based molecular gas measurements, dust continuum emission as a gas proxy,  alternative tracers such as [CI] and [CII], and late-stage dust production by TP-AGB stars. The goal is to highlight modeling assumptions and systematic uncertainties that underlie the conclusions presented in the main article.

\subsubsection{Tracing Molecular Hydrogen with Carbon Monoxide}
\label{subsubsec:CO}

CO is the most widely used tracer of cold H$_2$, as its emission correlates well with total H$_2$ mass in the local Universe despite typically being optically thick \citep[e.g.,][]{Bolatto2013}. The main uncertainty in interpreting CO observations lies in identifying the appropriate CO-to-H$_2$ conversion factor, $\alpha_\mathrm{CO}$, which is known to vary with physical conditions such as metallicity, gas density, and dynamical state \citep[e.g.,][]{Narayanan2012, Tacconi2020}.  A further limitation is the sensitivity required to detect faint CO emission lines, particularly in gas-poor systems. It is notoriously challenging to quantify the cold gas reservoirs of quiescent galaxies via CO: $\sim70\%$ of measurements at low sSFRs are upper limits even at $z=0$ \citep[e.g.,][]{Young2011,Saintonge2017},
and observations at high redshift remain sparse \citep[e.g.,][]{DEugenio2023}.  

Despite these challenges, mounting evidence suggests that massive quiescent systems contain very low molecular gas fractions, consistent with their suppressed SFRs. Deep ALMA and NOEMA observations targeting the lowest energy transitions (typically CO(2–1) or CO(3–2)) have yielded non-detections or marginal detections, implying molecular gas fractions $f_{\mathrm{H_2}} < 10\%$ in most cases at $z>1$ \citep[e.g.,][]{Williams2021,Zanella2023,Setton2025}. An important exception occurs in a subset of quiescent galaxies at $z\sim0 - 1$ which can host more substantial reservoirs up to 20–40\% \citep[e.g.,][]{French2015,Socolovsky2019,Belli2021,Zanella2023,Suess2025,Setton2025}.

These higher gas fractions are most common in the youngest poststarburst and recently quenched galaxies, showing that substantial molecular gas can persist after quenching, at least at $z\lesssim1$ \citep[e.g.,][]{French2015,French2018b, Alatalo2016,Bezanson2022_psb,Setton2025}. This supports scenarios in which morphological quenching, rather than depletion or expulsion, suppresses star formation at later times. \citet{Suess2025} further argue that mergers may play a dual role, simultaneously driving turbulence that suppresses star formation efficiency while also delivering fresh gas capable of triggering rejuvenation.

Nonetheless, such gas-rich cases appear to be the exception: available CO measurements in other quiescent systems typically imply depletion timescales shorter than 1 Gyr \citep[e.g.,][]{Spilker2018,Bezanson2019,Williams2021}, consistent with little remaining fuel to sustain star formation. This suggests that long-lived, morphologically quenched systems are likely uncommon at early times -- consistent with number-density arguments showing that slowly quenched populations primarily emerge later (\SMsup{subsec:SMnumberdensity}). Even so, morphological stabilization may still play an important role as a maintenance mechanism, helping to sustain quiescence once rapid quenching has occurred.

Altogether, interpreting CO measurements across cosmic time remains challenging. The inferred molecular content depends sensitively on $\alpha_\mathrm{CO}$, which may be elevated or suppressed in evolved, turbulent, or metal-rich systems \citep[e.g.,][]{Bolatto2013,Narayanan2012}, and CO itself may fail to trace diffuse, CO-faint molecular gas \citep[e.g.,][]{Popping2012,Tomassetti2014}.
Nonetheless, the existing detections and upper limits collectively suggest that genuinely H$_2$-poor massive quiescent galaxies appear to dominate at high redshift, even if their exact molecular content remains uncertain.

\subsubsection{Tracing Molecular Gas with Dust Continuum ($\lambda_{\mathrm{rest}}>250$ $\mu$m)}
\label{subsubsec:dust}

Interpreting dust emission in quiescent galaxies is challenging due to the complex and poorly constrained physics of dust evolution following quenching. Key uncertainties include the assumed molecular gas-to-dust mass ratio ($\delta_{\mathrm{GDR}}$; \citealt{Whitaker2021b,Lorenzon2025}), dust temperature \citep[e.g.,][]{daCunha2013,Magdis2012b,Schreiber2018}, and emissivity \citep{Draine2007,Casey2012}. 
Both empirical studies \citep[e.g.,][]{Scoville2016} and theoretical models \citep{Privon2018} suggest that long-wavelength dust continuum emission in the Rayleigh–Jeans regime ($
\gtrsim250\mu$m) provides a reasonable proxy for molecular gas mass for star-forming galaxies, but this calibration may break down in quiescent systems where dust properties and heating sources differ. Observational and theoretical estimates typically assume $\delta_{\mathrm{GDR}}\sim100$ and cold dust temperatures of $\sim$20–30 K \citep[e.g.,][]{Magdis2021}, similar to those observed in high-redshift star-forming galaxies \citep{Scoville2016} and local ellipticals \citep{Smith2012}. 

Direct constraints on the dust content of quiescent galaxies are incongruent, as emphasized in the main text. In strongly lensed quiescent galaxies at $z\sim2$, cold dust has been directly detected or constrained via stringent upper limits, consistent with other ALMA non-detections in unlensed systems \citep[e.g.,][]{Sargent2015,Zavala2019,Caliendo2021,Whitaker2021a,Morishita2022}.
While these lensed galaxies exhibit exceptionally low dust masses, more than two orders of magnitude below typical star-forming galaxies at the same epoch \citep{Tacconi2020}, the values are similar to constraints in lower redshift quiescent galaxies \citep{Donevski2023}. Such measurements are, however, in stark contrast with growing samples of dusty quiescent galaxies \citep[e.g.,][]{Donevski2020,Donevski2023,Lee2024, Setton2024},
although residual or obscured star formation is hard to rule out. 

Importantly, \citet{Lorenzon2025} show that stacking preferentially recovers emission from the dustiest $\sim$20\% of the quiescent population, biasing inferred average dust masses. This selection effect may help explain the discrepancy between stacked dust continuum detections and the non-detections in deep, lensed follow-up studies of individual galaxies. At the same time, lensed searches may still miss diffuse or sputtered dust components \citep[e.g.,][]{Gobat22}. Reconciling these approaches requires accounting for both the intrinsic diversity in dust properties and the observational biases in current surveys, motivating larger samples of individually detected quiescent galaxies using multiple ISM tracers \citep[e.g.,][]{Spilker2025,Lorenzon2025b}.

The diversity in dust content underscores how inferred dust masses depend on assumptions about dust physics, sample selection, and quenching stage. The inferred exotic $\delta_{\mathrm{GDR}}$ values \citep{Spilker2025,Lorenzon2025b} may not reflect a true absence of cold dust but instead arise from grain processing. As large grains are thermally sputtered, the remaining population becomes warmer and biased toward smaller sizes, naturally shifting the emission toward the MIR. Early results at $z\sim0.7$ show that quiescent galaxies have faint but diverse MIR dust and PAH emission \citep{BlanquezSese2023}; future work leveraging PAH ratios may help distinguish between complete dust destruction, as predicted in some cases, and a shift in grain properties toward warmer temperatures or smaller sizes.

Theoretical models may help resolve the tension between stacked analyses of large samples and individual detections of quiescent galaxies. Simulations predict that the observed diversity in $\delta_{\mathrm{GDR}}$ can arise from differences in quenching timescales, dust destruction efficiency (e.g., thermal sputtering), dust accretion through mergers, residual dust production from late stellar phases, and grain regrowth via metal accretion in the surrounding ISM \citep{Lorenzon2025}. These processes naturally generate a wide range of $\delta_{\mathrm{GDR}}$ ratios across quenched systems.

\subsubsection{Tracing Molecular Gas with [CI] and [CII]}
\label{subsubsec:finestructure}

Given the uncertainties in both CO and dust continuum studies of quiescent galaxies, there is a pressing need for more individual measurements, cross-comparisons of tracers in the same sources, and pilot studies of alternative approaches. Fine-structure lines such as [CI] and [CII] offer promising avenues for tracing cold H$_2$ in regimes where CO and/or dust are faint or undetectable \citep[e.g.,][]{Zanella2018}. The [CI] lines at 370 and 609 $\mu$m are thought to probe regions similar to those traced by CO but can remain visible in more diffuse or metal-poor gas where CO is photodissociated \citep[e.g.,][]{Papadopoulos2004,Bothwell2017}.
The [CII] 158 $\mu$m fine-structure line, one of the brightest ISM cooling transitions, traces both atomic and molecular gas and is readily accessible at high redshift with ALMA \citep[see review by][]{CarilliWalter2013}.

To date, [CI] and [CII] remain undetected in spectroscopically confirmed quiescent galaxies at $z>1$; the most notable attempt is \citet{DEugenio2023}, who conducted [CII] observations of a sample of massive quiescent galaxies at $z>2.8$ and derived upper limits on their cold gas content. Despite the lack of detections, [CI] and [CII] offer viable alternatives for probing the cold ISM in quiescent systems, potentially tracing molecular gas (in the case of [CI]) as well as the surrounding neutral medium (in the case of [CII]) when traditional tracers like CO and dust continuum fall short, especially as instrumentation and observing strategies improve.

However, interpreting these lines requires care.  [CII] can be emitted in multiple phases of the ISM. Although it is one of the brightest cooling lines, and thus an attractive target, it can arise in ionized, neutral atomic, and molecular gas, complicating its interpretation without supporting diagnostics \citep[e.g.,][]{CarilliWalter2013,Zanella2018}. For this reason, [CII] alone cannot be taken as a clean tracer of cold gas, but in combination with CO, [CI], or ionized-gas lines, it can provide valuable constraints on the multiphase ISM structure.  [CI], though more directly tied to molecular gas, is typically fainter than CO and requires long integrations in quiescent systems, even if it can be observed simultaneously with higher-J CO transitions such as CO(4-3) or CO(7-6) using ALMA. Both tracers depend on assumptions about excitation and carbon abundance \citep[e.g.,][]{Papadopoulos2004,Zanella2018,Valentino2020}, which may differ substantially in poststarburst galaxies and/or in metal-poor or $\alpha$-enhanced environments. Future studies that combine [CI], [CII], CO, and dust continuum in the same quenched systems will be essential for establishing robust cross-calibrations and disentangling the multiphase ISM. 

\subsubsection{Late-Stage Stellar Dust Production} 
\label{subsubsection:TPAGB_dust}

A final caveat in interpreting dust emission from quiescent galaxies is the uncertain physics of late-stage dust production, particularly from TP-AGB stars. Dust production in the circumstellar envelopes of TP-AGB stars peaks in intermediate-age ($\sim$0.6–2 Gyr) stellar populations \citep[e.g.,][]{Frogel1990,Maraston1998,Maraston2005}, making it especially relevant for recently quenched systems. This phase of stellar evolution is notoriously uncertain and TP-AGB dust production is often simplified or omitted in simulations, which typically rely on generic evolved-star yield prescriptions and dust-destruction prescriptions that remain poorly validated \citep[e.g.,][]{Conroy2010}. Pioneering theoretical work \citep{FerrarottiGail2006} established the basic microphysics of dust formation in TP-AGB winds, while subsequent models have incorporated increasingly sophisticated treatments of grain growth, chemistry, and pulsation-driven outflows \citep[e.g.,][]{Nanni2013,Nanni2014,Nanni2018}. 

Because TP-AGB dust production peaks on $\sim$1 Gyr timescales, it may transiently link dust content to quenching timescales \citep[e.g.,][]{Kelson2010}, making uncertainties in TP-AGB prescriptions in SPS models particularly consequential for interpreting NIR light and inferred dust masses. Early empirical work suggested substantial TP-AGB dust production in poststarburst galaxies \citep{Kelson2010}, but more recent models demonstrate that predicted dust yields are highly sensitive to poorly constrained parameters, such as metallicity, pulsation-driven wind dynamics, dredge-up efficiency, and the adopted mass-loss prescription \citep[e.g.,][]{FerrarottiGail2006,Nanni2013,Nanni2014,Nanni2018}. As a result, TP-AGB dust production remains one of the least certain components of dust-evolution modeling, limiting our ability to interpret the dust and chemical enrichment histories of quiescent galaxies during the first $\sim$1–2 Gyr after quenching.

\section{QUIESCENT POPULATION DEMOGRAPHICS}
\label{sec:supplement_demographics}

In this supplemental section, we compile and homogenize measurements of the number densities of quiescent and young poststarburst galaxies from a broad set of photometric and spectroscopic studies. Because these datasets span different stellar mass limits, selection strategies, and modeling assumptions, careful normalization is required to compare results on a consistent basis and to place them in the context of theoretical predictions. Below we describe the compilation, stellar mass homogenization, normalization procedures, and the implications for the growth of the quiescent population.

\subsection{Number Density Evolution of Quiescent Galaxies}
\label{subsec:SMnumberdensity}

We compile measurements of the number densities of both young and total quiescent galaxy populations from photometric \citep{Whitaker2012,Wild2016,Forrest2018,Belli2019,Carnall2023b,Clausen2024,Baker2025a,Baker2025b} and spectroscopic datasets \citep{Setton2023,Park2024,Zhang2026} (see \textbf{\autoref{tbl:Ndensity}}). The characteristic stellar mass completeness thresholds for these surveys span $\log(M_{\star}/M_{\odot}) > 10$–11.2, which translates into substantial differences in number densities because of the steep exponential decline at the high mass end of the stellar mass function \citep[SMF; e.g.,][]{Schechter1976}. Reported mass thresholds are also included in \textbf{\autoref{tbl:Ndensity}}. In some studies, only total number densities are reported, even if many  galaxies in the sample would likely be classified as poststarburst; this is particularly relevant at high redshift.

We note that even the definition of stellar mass is sensitive to systematic uncertainties associated with assumed star formation histories \citep[e.g.,][]{Leja2019b}; for self-consistency, we shift stellar masses (including SMFs) derived using non-parametric SFHs downward by 0.2 dex \citep{Leja2019b} for consistency with those derived from parametric models -- these are reported in  \textbf{\autoref{tbl:Ndensity}}. In \textbf{\mainfig{fig:numberdensities}} of the main article, 
we normalize all datasets to $>10^{10.5}M_{\odot}$ in order to most directly compare to simulations; there, each data point is divided by a normalization factor determined from the ratio of the number density implied by the $z=1$ \cite{Leja2020} SMF above $10^{10.5}M_{\odot}$ to that above the survey mass limit, where $z\sim1$ is chosen as a reference epoch because the quiescent population is both well established and well constrained observationally. Because the \cite{Leja2020} SMF includes both star forming and quiescent galaxies, we additionally fit the quiescent fraction as a function of redshift \citep{Muzzin2013,Weaver2023a} with a third-order polynomial and multiply the continuous \citep{Leja2020} SMF by that function. To reduce normalization uncertainties, 
we restrict the comparison to datasets with similar mass limits ($\Delta\log M_{\star} < 0.5$ dex), excluding \citep{Setton2023}. Their Dark Energy Spectroscopic Instrument Luminous Red Galaxy sample targets extremely massive galaxies ($\log (M_{\star, \mathrm{Prospector}}/M_{\odot})>11.2$), requiring a substantial correction factor of $\sim6$; even this is likely uncertain due to the low–redshift cosmic variance at the rare high-mass end of the \citet{Leja2020} SMF.

\begin{figure}
    \centering
    \includegraphics[width=\linewidth]{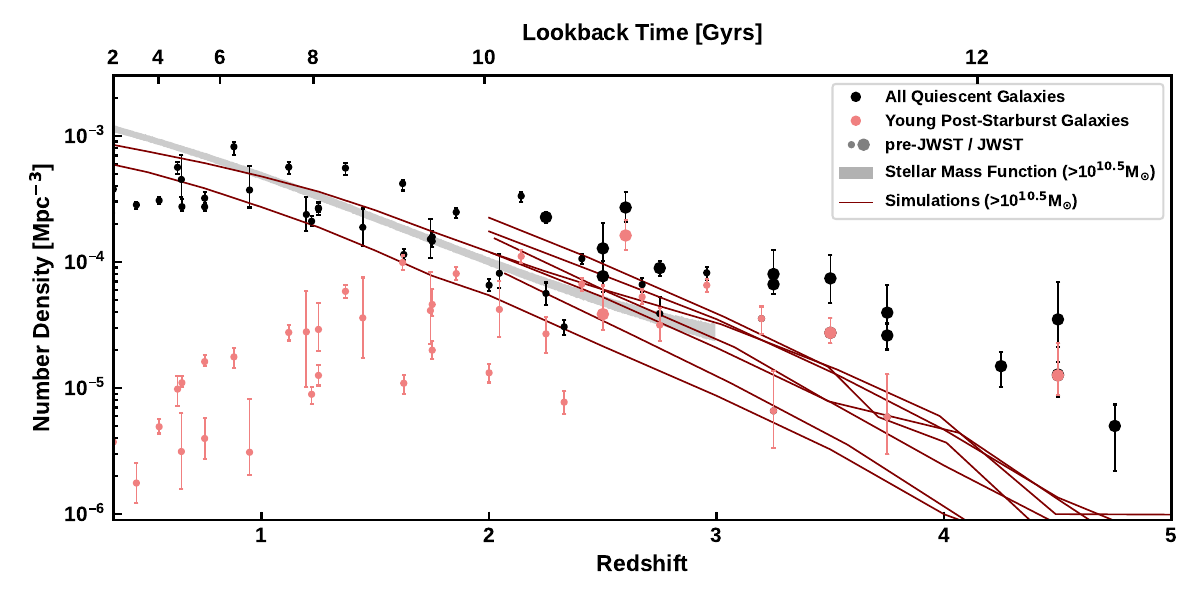}
    \caption{The number density of massive quiescent galaxies, shown without normalizations to account for different mass limits between samples, measured from a host of photometric and spectroscopic surveys \citep[\emph{filled symbols};][and references therein]{Zhang2026}, with the quiescent SMF \citep[\emph{gray band};][]{Leja2020}, and predictions from simulations \citep[\emph{maroon lines};][and references therein]{Lagos25}. In contrast with \textbf{\mainfig{fig:numberdensities}} 
    of the main article (see also \textbf{\autoref{fig:number_density_evolution}}), both the young (\emph{pink}) and total (\emph{black}) populations exhibit substantial scatter at any given redshift. This variation arises naturally from the combination of different survey mass limits and the steep decline of the SMF at high stellar masses. Surveys probing lower stellar mass limits report higher number densities, whereas surveys limited to more massive galaxies yield correspondingly lower values. In the main text, we account for these differences by normalizing each measurement using the \cite{Leja2020} SMF, providing a consistent basis for comparison across heterogeneous datasets. Abbreviations: JWST, James Webb Space Telescope; SMF, stellar mass function.}
    \label{fig:ndens_unnorm}
\end{figure}

For completeness, \textbf{\autoref{fig:ndens_unnorm}} presents the full unnormalized results. Both cases reveal a dramatic growth in the number density of massive quiescent galaxies over the past $\sim12$ Gyr -- nearly two orders of magnitude since $z\sim5$. Young poststarburst galaxies comprise a significant fraction of the quiescent population until $z\sim2$, after which their number density plateaus and declines below $z\sim1$. At the same time, the total number density of massive quiescent galaxies also levels out below $z\sim1$.  Despite this demographic flattening, their structural evolution continues through inside-out growth via minor mergers and the accretion of lower-mass satellites (see \autoref{subsec:mergers} in the main article), building extended low-surface-brightness envelopes over time \citep[e.g.,][]{vanDokkum2010,Patel2013b,Hill2017} -- a process that alters galaxy sizes but less significantly increasing their masses. Even so, survey-to-survey scatter remains large (up to $\sim$1 dex in many cases), and observed number densities exceed theoretical predictions until $z\sim2$ -- a tension first made striking by early spectroscopic surveys finding massive old galaxies at redshifts then considered surprisingly high \citep[e.g.,][]{Cimatti2004}. Although this tension is reduced when accounting for mass-limit discrepancies, significant divergence remains at $z\gtrsim3$.

\subsection{Observed Growth Rates of Recently Quenched Galaxies}

\begin{figure}
    \centering
    \includegraphics[width=\linewidth]{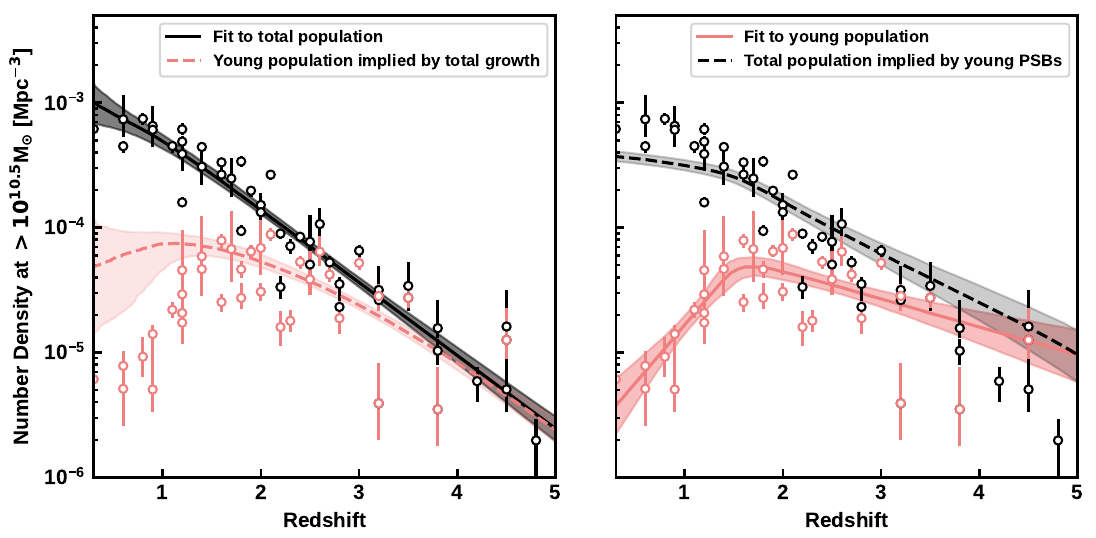}
    \caption{The implied number density evolution of the total massive quiescent galaxy population (\emph{black symbols}) assuming that young massive quiescent galaxies (\emph{pink symbols}) remain identifiable in the poststarburst phase for $\sim500$ Myr (using the same data as \textbf{\autoref{fig:ndens_unnorm}} and \textbf{\autoref{fig:number_density_evolution}}, 
    normalized to $10^{10.5}M_{\odot}$). In each panel, we fit a simple broken log-linear relation to either the total quiescent population (\emph{solid black}, \emph{left}) or the young poststarburst population (\emph{solid pink}, \emph{right}). The pink dashed line (\emph{left}) shows the required number density of young quiescent galaxies to account for the observed growth of the total quiescent population. The dashed line (\emph{right}) shows the implied growth of the total quiescent population as young poststarburst galaxies age into quiescent systems under the same continuity assumption. At $z\gtrsim1$, the two populations are reasonably consistent, implying that poststarburst galaxies -- and therefore rapid quenching -- can plausibly explain the early build-up of the massive quiescent population. At later times ($z<1$), however, the number density of massive quiescent galaxies continues to increase, outpacing the diminishing poststarburst population. This divergence requires an additional pathway, likely a form of slow quenching that imprints more subtle spectral signatures and is therefore not captured in the young poststarburst samples. Abbreviations: PSB, poststarburst.}
    \label{fig:number_density_evolution}
\end{figure}

To test whether the growth of the total quiescent population is consistent with the inflow of newly  quenched systems, we perform a simple continuity-style experiment, illustrated in \textbf{\autoref{fig:number_density_evolution}}. Following \citet{Whitaker2012}, we assume that young poststarburst galaxies remain identifiable for a characteristic timescale of $\sim500$ Myr, corresponding to the brief period during which their spectra are dominated by the light of A-type stars. However, this poststarburst visibility timescale depends on intrinsic properties and selection criteria (e.g., spectroscopic versus photometric definitions), and likely spans $\sim100$ Myr to $\sim1$ Gyr. Once this light fades, these systems rapidly migrate onto the red sequence and become indistinguishable from the older quiescent population, barring any event that triggers renewed star formation.

Using this physical picture, we model the growth of the massive quiescent population by assuming that all galaxies pass through a short-lived poststarburst phase. In each panel of \textbf{\autoref{fig:number_density_evolution}}, we fit the number density evolution of one population (left: total population, right: young population) with a broken linear relation in $\log(\mathrm{number\ density})$–$(1+z)$ space (solid lines with dark-shaded confidence regions). In the left panel, we show the number densities that would be required, under a simple continuity argument, to reproduce the observed growth of the total quiescent population (dashed pink line and shaded band). Although the required growth rate slows with time, it would still demand a larger population of recently quenched (young) galaxies than observed. A similar inconsistency appears in the opposite direction in the right panel, in which the observed young population can roughly account for the growth of the total quiescent population at $z\gtrsim1.5$, but falls short below $z\lesssim1-1.5$. 

We note that young quiescent galaxies in these studies are identified by some combination of photometric and spectroscopic poststarburst features and/or colors, which primarily selects systems undergoing rapid quenching. Slower quenching channels imprint more subtle, transitional spectral signatures that are difficult to identify with current data. A natural explanation is that these previously unaccounted-for slowly quenched galaxies supply the remaining growth of the quiescent population at late times. Testing this picture will likely require substantially larger spectroscopic datasets, which will be enabled by upcoming surveys using massively multiplexed optical-NIR spectrographs \citep{Maiolino2020,Greene2022}.\\

\textbf{Supplementary Table: Number Densities of Quiescent Galaxies}

\begin{longtable}{c@{\hskip 4pt}c@{\hskip 4pt}c@{\hskip 6pt}c@{\hskip 6pt}c@{\hskip 2pt}c@{\hskip 2pt}c@{\hskip 2pt}c@{\hskip 2pt}c}
\toprule
Reference & Subsample & Mass & $z$ & N [Mpc$^{-3}$]  & Raw N [Mpc$^{-3}$] & Norm.\ & Type & Provenance \\
 &  &  &  &  & (Norm. to $10^{10.5}M_{\odot}$) &  & factor &  \\
\midrule
\endfirsthead

\multicolumn{9}{c}%
{{\bfseries Supplementary Table (continued)}} \\
\toprule
Reference & Subsample & Mass & $z$ & N [Mpc$^{-3}$]  & Raw N [Mpc$^{-3}$] & Norm.\ & Type & Provenance \\
\midrule
\endhead

\midrule \multicolumn{9}{r}{{Continued on next page}} \\
\endfoot

\bottomrule
\endlastfoot

\citet{Baker2025a} &  total & 10.0 & 2.5 & ($5.1^{+3.0}_{-1.0})\times10^{-5}$ & ($1.3^{+0.8}_{-0.3})\times10^{-4}$ & 2.53 & phot & JWST \\
--- &  total & 10.0 & 3.2 & ($3.2^{+1.7}_{-0.6})\times10^{-5}$ & ($8.0^{+4.4}_{-1.5})\times10^{-5}$ & 2.53 & phot & JWST \\
--- &  total & 10.0 & 3.8 & ($1.6^{+1.0}_{-0.3})\times10^{-5}$ & ($4.0^{+2.6}_{-0.7})\times10^{-5}$ & 2.53 & phot & JWST \\
--- &  total & 10.0 & 4.5 & ($5.1^{+3.4}_{-1.7})\times10^{-6}$ & ($1.3^{+0.8}_{-0.4})\times10^{-5}$ & 2.53 & phot & JWST \\
\citet{Baker2025b} &  total & 10.0 & 2.2 & ($8.9^{+0.8}_{-0.8})\times10^{-5}$ & ($2.3^{+0.2}_{-0.2})\times10^{-4}$ & 2.53 & phot & JWST \\
--- &  total & 10.0 & 2.8 & ($3.5^{+0.5}_{-0.5})\times10^{-5}$ & ($8.9^{+1.2}_{-1.2})\times10^{-5}$ & 2.53 & phot & JWST \\
--- &  total & 10.0 & 3.2 & ($2.6^{+0.4}_{-0.4})\times10^{-5}$ & ($6.6^{+1.0}_{-1.0})\times10^{-5}$ & 2.53 & phot & JWST \\
--- &  total & 10.0 & 3.8 & ($1.0^{+0.2}_{-0.2})\times10^{-5}$ & ($2.6^{+0.6}_{-0.6})\times10^{-5}$ & 2.53 & phot & JWST \\
--- &  total & 10.0 & 4.2 & ($5.9^{+1.7}_{-1.9})\times10^{-6}$ & ($1.5^{+0.4}_{-0.5})\times10^{-5}$ & 2.53 & phot & JWST \\
--- &  total & 10.0 & 4.8 & ($2.0^{+0.9}_{-1.1})\times10^{-6}$ & ($5.0^{+2.4}_{-2.8})\times10^{-6}$ & 2.53 & phot & JWST \\
--- &  total & 10.0 & 5.5 & ($5.5^{+3.6}_{-3.2})\times10^{-7}$ & ($1.4^{+0.9}_{-0.8})\times10^{-6}$ & 2.53 & phot & JWST \\
--- &  total & 10.0 & 6.8 & ($1.2^{+2.0}_{-1.2})\times10^{-7}$ & ($3.0^{+5.0}_{-3.0})\times10^{-7}$ & 2.53 & phot & JWST \\
\citet{Park2024} &  total & 10.0 & 2.6 & ($1.1^{+0.4}_{-0.2})\times10^{-4}$ & ($2.7^{+0.9}_{-0.6})\times10^{-4}$ & 2.53 & spec & JWST \\
--- &  young & 10.0 & 2.6 & ($6.4^{+2.1}_{-1.5})\times10^{-5}$ & ($1.6^{+0.5}_{-0.4})\times10^{-4}$ & 2.53 & spec & JWST \\
\citet{Carnall2023b} &  total & 10.1 & 3.5 & ($3.4^{+1.8}_{-1.2})\times10^{-5}$ & ($7.4^{+4.0}_{-2.7})\times10^{-5}$ & 2.17 & phot & JWST \\
---  &  total & 10.1 & 4.5 & ($1.6^{+1.6}_{-0.9})\times10^{-5}$ & ($3.5^{+3.4}_{-1.9})\times10^{-5}$ & 2.17 & phot & JWST \\
\citet{Forrest2018} &  total & 10.25 & 1.2 & ($1.6^{+0.2}_{-0.1})\times10^{-4}$ & ($2.7^{+0.3}_{-0.2})\times10^{-4}$ & 1.68 & phot & pre-JWST \\
---  &  total & 10.25 & 1.8 & ($9.4^{+1.1}_{-0.9})\times10^{-5}$ & ($1.6^{+0.2}_{-0.1})\times10^{-4}$ & 1.68 & phot & pre-JWST \\
---  &  total & 10.25 & 2.2 & ($3.3^{+0.7}_{-0.7})\times10^{-5}$ & ($5.6^{+1.2}_{-1.1})\times10^{-5}$ & 1.68 & phot & pre-JWST \\
---  &  total & 10.25 & 2.8 & ($2.3^{+0.8}_{-0.5})\times10^{-5}$ & ($3.9^{+1.4}_{-0.9})\times10^{-5}$ & 1.68 & phot & pre-JWST \\
---  &  total & 10.25 & 3.2 & ($3.9^{+4.3}_{-1.9})\times10^{-6}$ & ($6.6^{+7.2}_{-3.2})\times10^{-6}$ & 1.68 & phot & pre-JWST \\
---  &  total & 10.25 & 3.8 & ($3.5^{+4.1}_{-1.7})\times10^{-6}$ & ($5.9^{+7.0}_{-2.9})\times10^{-6}$ & 1.68 & phot & pre-JWST \\
---  &  young & 10.25 & 1.2 & ($1.7^{+1.1}_{-0.6})\times10^{-5}$ & ($2.9^{+1.8}_{-0.9})\times10^{-5}$ & 1.68 & phot & pre-JWST \\
---  &  young & 10.25 & 1.8 & ($2.7^{+0.9}_{-0.5})\times10^{-5}$ & ($4.6^{+1.5}_{-0.8})\times10^{-5}$ & 1.68 & phot & pre-JWST \\
\citet{Forrest2018} &  young & 10.25 & 2.2 & ($1.6^{+0.6}_{-0.5})\times10^{-5}$ & ($2.7^{+0.9}_{-0.8})\times10^{-5}$ & 1.68 & phot & pre-JWST \\
\citet{Forrest2018} &  young & 10.25 & 2.8 & ($1.9^{+0.7}_{-0.5})\times10^{-5}$ & ($3.2^{+1.2}_{-0.8})\times10^{-5}$ & 1.68 & phot & pre-JWST \\
---  &  young & 10.25 & 3.2 & ($3.9^{+4.3}_{-1.9})\times10^{-6}$ & ($6.6^{+7.2}_{-3.2})\times10^{-6}$ & 1.68 & phot & pre-JWST \\
---  &  young & 10.25 & 3.8 & ($3.5^{+4.1}_{-1.7})\times10^{-6}$ & ($5.9^{+7.0}_{-2.9})\times10^{-6}$ & 1.68 & phot & pre-JWST \\
\citet{Clausen2024} &  total & 10.4 & 0.6 & ($4.5^{+0.5}_{-0.5})\times10^{-4}$ & ($5.6^{+0.6}_{-0.7})\times10^{-4}$ & 1.26 & phot & pre-JWST \\
---  &  total & 10.4 & 0.9 & ($6.5^{+0.6}_{-0.9})\times10^{-4}$ & ($8.2^{+0.8}_{-1.1})\times10^{-4}$ & 1.26 & phot & pre-JWST \\
---  &  total & 10.4 & 1.1 & ($4.5^{+0.5}_{-0.6})\times10^{-4}$ & ($5.6^{+0.6}_{-0.7})\times10^{-4}$ & 1.26 & phot & pre-JWST \\
---  &  total & 10.4 & 1.4 & ($4.4^{+0.4}_{-0.5})\times10^{-4}$ & ($5.5^{+0.6}_{-0.7})\times10^{-4}$ & 1.26 & phot & pre-JWST \\
---  &  total & 10.4 & 1.6 & ($3.3^{+0.2}_{-0.4})\times10^{-4}$ & ($4.2^{+0.3}_{-0.5})\times10^{-4}$ & 1.26 & phot & pre-JWST \\
---  &  total & 10.4 & 1.9 & ($2.0^{+0.2}_{-0.2})\times10^{-4}$ & ($2.5^{+0.2}_{-0.3})\times10^{-4}$ & 1.26 & phot & pre-JWST \\
---  &  total & 10.4 & 2.1 & ($2.6^{+0.2}_{-0.3})\times10^{-4}$ & ($3.3^{+0.2}_{-0.3})\times10^{-4}$ & 1.26 & phot & pre-JWST \\
---  &  total & 10.4 & 2.4 & ($8.4^{+0.7}_{-0.8})\times10^{-5}$ & ($1.1^{+0.1}_{-0.1})\times10^{-4}$ & 1.26 & phot & pre-JWST \\
---  &  total & 10.4 & 2.7 & ($5.3^{+0.7}_{-0.6})\times10^{-5}$ & ($6.6^{+0.8}_{-0.8})\times10^{-5}$ & 1.26 & phot & pre-JWST \\
---  &  total & 10.4 & 3.0 & ($6.5^{+0.7}_{-0.7})\times10^{-5}$ & ($8.2^{+0.9}_{-0.9})\times10^{-5}$ & 1.26 & phot & pre-JWST \\
---  &  total & 10.4 & 3.2 & ($2.8^{+0.7}_{-0.7})\times10^{-5}$ & ($3.5^{+0.9}_{-0.9})\times10^{-5}$ & 1.26 & phot & pre-JWST \\
---  &  young & 10.4 & 0.6 & ($7.8^{+2.1}_{-2.1})\times10^{-6}$ & ($9.8^{+2.7}_{-2.7})\times10^{-6}$ & 1.26 & phot & pre-JWST \\
---  &  young & 10.4 & 0.9 & ($1.4^{+0.3}_{-0.3})\times10^{-5}$ & ($1.8^{+0.3}_{-0.3})\times10^{-5}$ & 1.26 & phot & pre-JWST \\
---  &  young & 10.4 & 1.1 & ($2.2^{+0.3}_{-0.3})\times10^{-5}$ & ($2.8^{+0.4}_{-0.4})\times10^{-5}$ & 1.26 & phot & pre-JWST \\
---  &  young & 10.4 & 1.4 & ($4.7^{+0.5}_{-0.5})\times10^{-5}$ & ($5.8^{+0.7}_{-0.7})\times10^{-5}$ & 1.26 & phot & pre-JWST \\
---  &  young & 10.4 & 1.6 & ($7.9^{+1.1}_{-1.0})\times10^{-5}$ & ($9.9^{+1.3}_{-1.2})\times10^{-5}$ & 1.26 & phot & pre-JWST \\
--- &  young & 10.4 & 1.9 & ($6.4^{+0.8}_{-0.7})\times10^{-5}$ & ($8.1^{+1.1}_{-0.8})\times10^{-5}$ & 1.26 & phot & pre-JWST \\
---  &  young & 10.4 & 2.1 & ($8.8^{+1.1}_{-1.0})\times10^{-5}$ & ($1.1^{+0.1}_{-0.1})\times10^{-4}$ & 1.26 & phot & pre-JWST \\
---  &  young & 10.4 & 2.4 & ($5.3^{+0.6}_{-0.6})\times10^{-5}$ & ($6.7^{+0.8}_{-0.8})\times10^{-5}$ & 1.26 & phot & pre-JWST \\
--- &  young & 10.4 & 2.7 & ($4.2^{+0.6}_{-0.5})\times10^{-5}$ & ($5.3^{+0.8}_{-0.7})\times10^{-5}$ & 1.26 & phot & pre-JWST \\
--- &  young & 10.4 & 3.0 & ($5.2^{+0.6}_{-0.6})\times10^{-5}$ & ($6.5^{+0.8}_{-0.8})\times10^{-5}$ & 1.26 & phot & pre-JWST \\
---  &  young & 10.4 & 3.2 & ($2.8^{+0.7}_{-0.7})\times10^{-5}$ & ($3.5^{+0.9}_{-0.9})\times10^{-5}$ & 1.26 & phot & pre-JWST \\
\citet{Zhang2026} &  total & 10.5 & 2.5 & ($7.7^{+4.8}_{-1.9})\times10^{-5}$ & ($7.7^{+4.8}_{-1.9})\times10^{-5}$ & 1.00 & spec & JWST \\
---  &  total & 10.5 & 3.5 & ($2.7^{+0.9}_{-0.5})\times10^{-5}$ & ($2.7^{+0.9}_{-0.5})\times10^{-5}$ & 1.00 & spec & JWST \\
---  &  total & 10.5 & 4.5 & ($1.3^{+1.0}_{-0.4})\times10^{-5}$ & ($1.3^{+1.0}_{-0.4})\times10^{-5}$ & 1.00 & spec & JWST \\
---  &  young & 10.5 & 2.5 & ($3.9^{+2.6}_{-0.9})\times10^{-5}$ & ($3.9^{+2.6}_{-0.9})\times10^{-5}$ & 1.00 & spec & JWST \\
---  &  young & 10.5 & 3.5 & ($2.7^{+0.9}_{-0.5})\times10^{-5}$ & ($2.7^{+0.9}_{-0.5})\times10^{-5}$ & 1.00 & spec & JWST \\
---  &  young & 10.5 & 4.5 & ($1.3^{+1.0}_{-0.4})\times10^{-5}$ & ($1.3^{+1.0}_{-0.4})\times10^{-5}$ & 1.00 & spec & JWST \\
\citet{Whitaker2012} &  total & 10.7 & 0.3 & ($6.2^{+6.4}_{-2.2})\times10^{-4}$ & ($3.8^{+3.9}_{-1.3})\times10^{-4}$ & 0.61 & phot & pre-JWST \\
---  &  total & 10.7 & 0.6 & ($7.4^{+4.1}_{-2.0})\times10^{-4}$ & ($4.5^{+2.5}_{-1.2})\times10^{-4}$ & 0.61 & phot & pre-JWST \\
--- &  total & 10.7 & 0.9 & ($6.1^{+3.4}_{-1.7})\times10^{-4}$ & ($3.7^{+2.1}_{-1.0})\times10^{-4}$ & 0.61 & phot & pre-JWST \\
--- &  total & 10.7 & 1.2 & ($3.9^{+1.4}_{-1.0})\times10^{-4}$ & ($2.4^{+0.9}_{-0.6})\times10^{-4}$ & 0.61 & phot & pre-JWST \\
--- &  total & 10.7 & 1.4 & ($3.1^{+1.3}_{-0.9})\times10^{-4}$ & ($1.9^{+0.8}_{-0.5})\times10^{-4}$ & 0.61 & phot & pre-JWST \\
--- &  total & 10.7 & 1.7 & ($2.5^{+1.1}_{-0.7})\times10^{-4}$ & ($1.5^{+0.7}_{-0.4})\times10^{-4}$ & 0.61 & phot & pre-JWST \\
--- &  total & 10.7 & 2.0 & ($1.3^{+0.6}_{-0.3})\times10^{-4}$ & ($8.1^{+3.5}_{-1.9})\times10^{-5}$ & 0.61 & phot & pre-JWST \\
--- &  young & 10.7 & 0.3 & ($6.1^{+12.5}_{-4.4})\times10^{-6}$ & ($3.7^{+7.6}_{-2.7})\times10^{-6}$ & 0.61 & phot & pre-JWST \\
--- &  young & 10.7 & 0.6 & ($5.1^{+5.2}_{-2.5})\times10^{-6}$ & ($3.1^{+3.2}_{-1.6})\times10^{-6}$ & 0.61 & phot & pre-JWST \\
--- &  young & 10.7 & 0.9 & ($5.1^{+8.4}_{-1.7})\times10^{-6}$ & ($3.1^{+5.1}_{-1.0})\times10^{-6}$ & 0.61 & phot & pre-JWST \\
--- &  young & 10.7 & 1.2 & ($4.6^{+5.0}_{-2.9})\times10^{-5}$ & ($2.8^{+3.0}_{-1.8})\times10^{-5}$ & 0.61 & phot & pre-JWST \\
--- &  young & 10.7 & 1.4 & ($5.9^{+6.4}_{-3.0})\times10^{-5}$ & ($3.6^{+3.9}_{-1.9})\times10^{-5}$ & 0.61 & phot & pre-JWST \\
--- &  young & 10.7 & 1.7 & ($6.7^{+7.0}_{-3.1})\times10^{-5}$ & ($4.1^{+4.3}_{-1.9})\times10^{-5}$ & 0.61 & phot & pre-JWST \\
--- &  young & 10.7 & 2.0 & ($6.9^{+4.6}_{-2.7})\times10^{-5}$ & ($4.2^{+2.8}_{-1.7})\times10^{-5}$ & 0.61 & phot & pre-JWST \\
\citet{Belli2019} &  total & 10.8 & 1.2 & ($4.9^{+0.5}_{-0.4})\times10^{-4}$ & ($2.1^{+0.2}_{-0.2})\times10^{-4}$ & 0.43 & phot & pre-JWST \\
--- &  total & 10.8 & 1.6 & ($2.7^{+0.3}_{-0.3})\times10^{-4}$ & ($1.1^{+0.1}_{-0.1})\times10^{-4}$ & 0.43 & phot & pre-JWST \\
--- &  total & 10.8 & 2.0 & ($1.5^{+0.2}_{-0.2})\times10^{-4}$ & ($6.5^{+0.8}_{-0.7})\times10^{-5}$ & 0.43 & phot & pre-JWST \\
---  &  total & 10.8 & 2.3 & ($7.1^{+1.0}_{-1.0})\times10^{-5}$ & ($3.1^{+0.4}_{-0.4})\times10^{-5}$ & 0.43 & phot & pre-JWST \\
---  &  young & 10.8 & 1.2 & ($2.1^{+0.3}_{-0.3})\times10^{-5}$ & ($8.9^{+1.3}_{-1.5})\times10^{-6}$ & 0.43 & phot & pre-JWST \\
---  &  young & 10.8 & 1.6 & ($2.5^{+0.4}_{-0.4})\times10^{-5}$ & ($1.1^{+0.2}_{-0.2})\times10^{-5}$ & 0.43 & phot & pre-JWST \\
--- &  young & 10.8 & 2.0 & ($3.1^{+0.5}_{-0.5})\times10^{-5}$ & ($1.3^{+0.2}_{-0.2})\times10^{-5}$ & 0.43 & phot & pre-JWST \\
--- &  young & 10.8 & 2.3 & ($1.8^{+0.4}_{-0.3})\times10^{-5}$ & ($7.7^{+1.7}_{-1.5})\times10^{-6}$ & 0.43 & phot & pre-JWST \\
\citet{Wild2016}  &  total & 10.8 & 0.8 & ($7.4^{+0.9}_{-0.8})\times10^{-4}$ & ($3.2^{+0.4}_{-0.3})\times10^{-4}$ & 0.43 & phot & pre-JWST \\
--- &  total & 10.8 & 1.2 & ($6.1^{+0.7}_{-0.6})\times10^{-4}$ & ($2.6^{+0.3}_{-0.3})\times10^{-4}$ & 0.43 & phot & pre-JWST \\
--- &  total & 10.8 & 1.8 & ($3.4^{+0.4}_{-0.3})\times10^{-4}$ & ($1.5^{+0.2}_{-0.1})\times10^{-4}$ & 0.43 & phot & pre-JWST \\
--- &  young & 10.8 & 0.8 & ($9.2^{+4.1}_{-2.8})\times10^{-6}$ & ($4.0^{+1.8}_{-1.2})\times10^{-6}$ & 0.43 & phot & pre-JWST \\
--- &  young & 10.8 & 1.2 & ($2.9^{+0.6}_{-0.5})\times10^{-5}$ & ($1.3^{+0.3}_{-0.2})\times10^{-5}$ & 0.43 & phot & pre-JWST \\
--- &  young & 10.8 & 1.8 & ($4.6^{+0.8}_{-0.7})\times10^{-5}$ & ($2.0^{+0.3}_{-0.3})\times10^{-5}$ & 0.43 & phot & pre-JWST \\
\citet{Setton2023} &  total & 11.0 & 0.5 & ($1.7^{+0.1}_{-0.1})\times10^{-3}$ & ($2.8^{+0.1}_{-0.2})\times10^{-4}$ & 0.17 & spec & pre-JWST \\
--- &  total & 11.0 & 0.6 & ($1.8^{+0.1}_{-0.1})\times10^{-3}$ & ($3.1^{+0.2}_{-0.2})\times10^{-4}$ & 0.17 & spec & pre-JWST \\
--- &  total & 11.0 & 0.7 & ($1.6^{+0.2}_{-0.1})\times10^{-3}$ & ($2.7^{+0.4}_{-0.2})\times10^{-4}$ & 0.17 & spec & pre-JWST \\
--- &  total & 11.0 & 0.8 & ($1.6^{+0.1}_{-0.1})\times10^{-3}$ & ($2.7^{+0.2}_{-0.2})\times10^{-4}$ & 0.17 & spec & pre-JWST \\
--- &  young & 11.0 & 0.5 & ($1.0^{+0.5}_{-0.3})\times10^{-5}$ & ($1.8^{+0.8}_{-0.5})\times10^{-6}$ & 0.17 & spec & pre-JWST \\
---  &  young & 11.0 & 0.6 & ($2.9^{+0.4}_{-0.3})\times10^{-5}$ & ($4.9^{+0.8}_{-0.6})\times10^{-6}$ & 0.17 & spec & pre-JWST \\
---  &  young & 11.0 & 0.7 & ($6.5^{+0.8}_{-0.5})\times10^{-5}$ & ($1.1^{+0.1}_{-0.1})\times10^{-5}$ & 0.17 & spec & pre-JWST \\
---  &  young & 11.0 & 0.8 & ($9.5^{+1.2}_{-0.8})\times10^{-5}$ & ($1.6^{+0.2}_{-0.1})\times10^{-5}$ & 0.17 & spec & pre-JWST \\

\label{tbl:Ndensity}
\end{longtable}

\section{QUENCHING IN AN EVOLUTIONARY FRAMEWORK}
\label{sec:sm_evolutionary_framework}

\begin{figure}
    \centering
    \includegraphics[width=\linewidth]{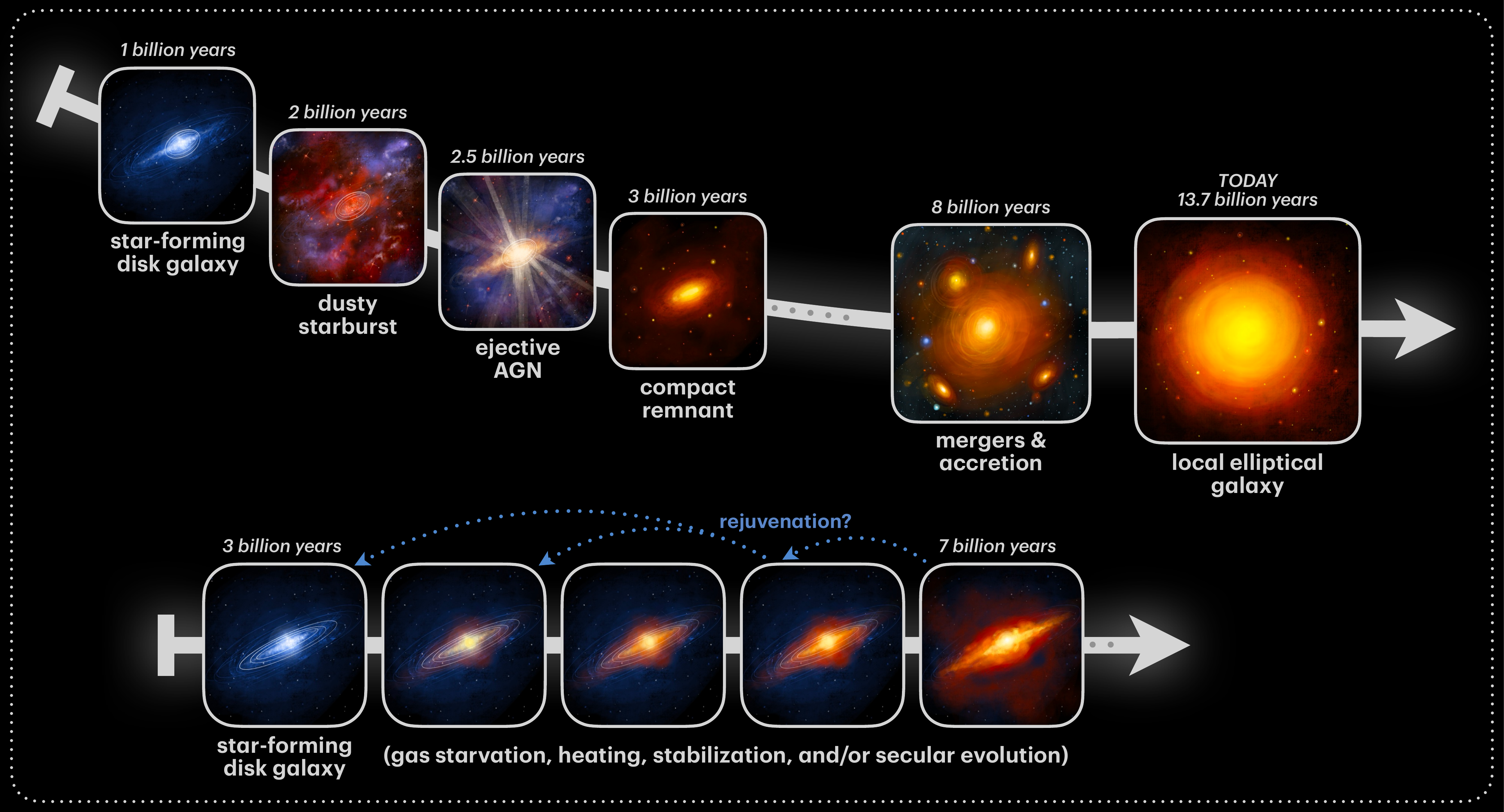}
    \caption{Illustrated evolutionary pathways for galaxy quenching at $M_\star \gtrsim 10^{10}$ M$_\odot$, contrasting rapid (top) and slow (bottom) modes. Galaxies evolve from a star formation phase, through quenching, and a late-stage structural assembly phase post-quenching. The spatial location of cold gas is shown with white contours, with central concentrations for rapid pathways and extended for slow. AGN-driven outflows dominate rapid shutdown at high redshift, whereas extended gas disks enable slow quenching via gradual heating, stabilization, or other long timescale processes. In general, all quenched galaxies grow structurally at late times through minor mergers and accretion. Artwork by Carrington Bryan is available at: \href{https://zenodo.org/records/17123339}{doi:10.5281/zenodo.17123339}. Inspired by \citet{Toft2014}.}
    \label{fig:sm_evolutionary_framework}
\end{figure}

The main text of this review article frames current observations of quiescent galaxies within an evolutionary framework that separates quenching mechanisms into rapid and slower processes. Here, we provide an alternate version of this figure with most text annotations removed, which may be preferable for presentations (\textbf{\autoref{fig:sm_evolutionary_framework}}). We also provide versions with transparent and white backgrounds, as well as the individual evolutionary-stage illustrations by Carrington Bryan (2025 Physics \& Astronomy Artist in Residence at the University of Pittsburgh) in high resolution. All files are available at: \href{https://zenodo.org/records/17123339}{doi:10.5281/zenodo.17123339}. Inspired by \citet{Toft2014}.

%








\end{bibunit}
\end{document}